\documentclass{article}[13pt]
\usepackage{amssymb,amsmath}

\newcommand{\R}{\mbox{${\rm I\!R}$}}

\newcounter{mathe}[section]

\newtheorem{thm}{Theorem}[section]
\newtheorem{cor}[thm]{Corollary}
\newtheorem{lem}[thm]{Lemma}

\begin{document}

\title{Flux-Across-Surfaces Theorem for a Dirac
Particle}
\author{D.~D\"urr, P.~Pickl \\
Fakult\"at f\"ur Mathematik\\ Universit\"at M\"unchen,
     Theresienstr. 39, 80333 M\"unchen, Germany}

\date{March 2002}
\maketitle

\begin{abstract}
 We consider the asymptotic evolution of a relativistic spin-$\frac{1}{2}$ particle. i.e.
  a particle whose wavefunction satisfies the Dirac equation with external static potential.
We prove that the probability for the particle crossing  a
(detector) surface converges to the probability, that the
direction of the momentum of the particle lies within the solid
angle defined by the (detector) surface, as the distance of the
surface goes to infinity. This generalizes earlier non
relativistic results, known as flux across surfaces theorems,  to
the relativistic regime.
\end{abstract}

\section{Introduction}

In scattering experiments the scattered particles are measured at
a macroscopic distance, but the computations of scattering cross
sections are based on the distribution of the wavefunction in
momentum space. Therefore a relationship between the crossing
probability through a far distant detector surface and the shape
of the wavefunction in momentum space is needed.

This relationship is given by the flux-across-surfaces theorem,
which - as a problem in mathematical physics - has been formulated
by Combes, Newton and Shtokhamer \cite{CNS}, see also
\cite{DDGZ2,SCAT}. For scattering states (material on scattering
states for the Dirac equation is in \cite{thaller}) the theorem
asserts that the probability of crossing a far distant surface
(physical interaction with the detector is neglected) subtended by
a solid angle is equal to the probability that the scattered
particle will, in the distant future, have a momentum, whose
direction lies in that same solid angle. Moreover, the
probability, that the particle will cross the detector within a
certain area is given by the integral of the flux over that area
and time. This has been proven for Schr\"odinger evolutions in
great generality, see for instance
\cite{JMP,Amrein,AmreinPearson,thesis,Panati, AntPan}.

 We consider here wavefunctions $\psi_{t}\in
 L^{2}(\mathbb{R}^{3})\bigotimes\mathbb{C}^{4}$ which satisfy
  the Dirac-equation (conveniently setting
$c=\hbar=1$)
\begin{equation}\label{dirac}
  i\frac{\partial\psi_{t}}{\partial t}=-i\sum_{l=1}^{3}\alpha_{l}\partial_{l}\psi_{t}+A\hspace{-0.2cm}/\psi_{t}+\beta m\psi_{t}\equiv H\psi_{t}
\end{equation} where
\begin{eqnarray}\label{alphas}
\alpha_{l}=
\begin{pmatrix}
  _{0} & _{\sigma_{l}}\\
  _{\sigma_{l}} & _{0}
\end{pmatrix}; \beta=
\begin{pmatrix}
  _{\mathbf{1}} & _{0}\\
  _{0} & _{-\mathbf{1}}
\end{pmatrix}; l=1,2,3
\end{eqnarray} $\sigma_{l}$ being the Pauli matrices:
\begin{eqnarray*}
 \sigma_{1}=\begin{pmatrix}
  _{1} & _{0} \\
  _{0} & _{-1}
\end{pmatrix}; \sigma_{2}=\begin{pmatrix}
  _{0} & _{1} \\
  _{1} & _{0}
\end{pmatrix}; \sigma_{3}=\begin{pmatrix}
  _{0} & _{-i} \\
  _{i} & _{0}\;
\end{pmatrix}
\end{eqnarray*}
$\mathbf{1}$ the $2\times2$-unit matrix and $A\hspace{-0.2cm}/$
the 4-potential in the form
$$A\hspace{-0.2cm}/:=A_{0}+\mathbf{A}\cdot\boldsymbol{\alpha}$$ with
$\boldsymbol{\alpha}:=(\alpha_1,\alpha_2,\alpha_3)$ . In the
following we will always denote solutions of the Dirac equation by
$\psi_{t}$ and by $\psi_{0}$ the "time zero" wavefunction.

$A\hspace{-0.2cm}/$ is an external static four-potential, which
satisfies
 condition A (see \ref{potcond}), which concerns smoothness and is for the
sake of simplicity taken stronger than needed:

\begin{equation}\label{potcond}
\hspace{-2.1cm}\text{\bf{Condition A}}
\hspace{2cm}A\hspace{-0.2cm}/(\mathbf{x})\in
C^{\infty}\hspace{1cm}\exists M,\xi>0:\hspace{0.5cm}\mid
A\hspace{-0.2cm}/(\mathbf{x})\mid\leq M\langle
x\rangle^{4+\xi}\vspace{0.5cm}\;.
 \end{equation}
The norm $\mid\cdot\mid$ is defined as:

$$\mid B\mid:=\sup_{\parallel\varphi\parallel_{s}=1}\parallel
B\varphi\parallel_{s}$$ where
$$\parallel\varphi\parallel_{s}:=\langle\varphi,\varphi\rangle^{\frac{1}{2}}$$
with the inner product in spin space

  $$\langle\cdot ,\cdot \rangle: \mathbb{C}\parallel\varphi\parallel_{s}(\mathbf{x})
  ^{4}\bigotimes \mathbb{C}
  ^{4}\rightarrow
  \mathbb{C}\hspace{1cm}\langle\varphi,\chi\rangle:=
  \sum_{l=1}^{4}\overline{\varphi_{l}}\chi_{l}\,\,.$$
  Often we have spinors depending on $\mathbf{x}$, in that case we
  have $\parallel\varphi\parallel_{s}(\mathbf{x})$.

The continuity equation involving the quantum flux of a
relativistic spin $\frac{1}{2}$ particle reads
\begin{equation}\label{cont}
\frac{\partial}{\partial t}\overline\psi_{t}\psi_{t} =
\nabla\cdot\mathbf{j} ,
\end{equation}
whereas the 4-flux is defined for any $\varphi\in
L^{2}(\mathbb{R}^{3})\bigotimes\mathbb{C}^{4}$ by
\begin{equation}\label{relflux}
\underline{j} = \begin{pmatrix}
  _{j_{0}} \\
  _{\mathbf{j}}
\end{pmatrix}=\langle\varphi,\underline{\alpha}\varphi\rangle\;,
\end{equation}

with $\underline{\alpha}=\begin{pmatrix}
  _{1} \\
  _{\boldsymbol{\alpha}}
\end{pmatrix}.$

For notational convenience we sometimes omit the dependence on
$\mathbf{x}.$ Furthermore we have the usual $L^2$-Norm on the
space of $4-$spinors given by
$$\parallel\varphi\parallel=\left(\int
\parallel\varphi\parallel_{s}^2d^3x\right)^{\frac{1}{2}}\,\,.$$ We
introduce the Fouriertransform of $\varphi(\mathbf{x})$ as
representation in the generalized basis (\ref{basis}) of the free
Hamiltonian ,i.e.

\begin{equation}\label{Four}
\widehat{\varphi}_s(\mathbf{k})= \int(2\pi)^{-\frac{3}{2}}
 \langle\varphi_{\mathbf{k}}^s(\mathbf{x}),\varphi(\mathbf{x})\rangle
 d^{3}x\hspace{1cm}
 \widehat{\varphi}(\mathbf{k}):=\sum_{s=1}^{2}s_{\mathbf{k}}^{s}\widehat{\varphi}_s(\mathbf{k})\,.
\end{equation}
We denote by $x$ the euclidian length of $\mathbf{x}$.

We assume that asymptotic completeness holds, i.e. that the wave
operators exist on the spectral subspace ${\cal H}_{ac}$ of the
continuous positive spectrum ("scattering state") of the
Dirac-Hamiltonian: Let $\psi_{\text{out}}$ denote the wavefunction
of the free asymptotic of a scattering state $\psi$ then

  $$\lim_{t\rightarrow\infty}\parallel
  e^{-iH_{0}t}\psi_{\text{out}}-e^{iHt}\psi\parallel=0\,\,.$$
$\psi_{\text{out}}$ is given by the wave operator:

  $$\Omega_{+}=\lim_{t\rightarrow\infty}e^{iHt}e^{-iH_{0}t}\hspace{1cm}\psi=\Omega_{+}\psi_{\text{out}}\,\,.$$
The existence of the wave operators and asymptotic completeness
has been proven for short range potentials. See e.g. Thaller
\cite{thaller}.

We remark (see Lemma \ref{properties}(d))(\ref{her}), that  the
Fourier transform $\widehat{\psi}_{out,s}(\mathbf{k})$ of
$\psi_{\text{out}}$ equals the generalized Fourier transform
$\psi^{\ddagger}_s$ of $\psi$ in the generalized eigen-basis of
the Dirac hamiltonian with potential.

In general, we do not have much information about scattering
states. One can prove the flux across surface theorem with
conditions merely on the ``out''-states, where the corresponding
 properties
of the scattering states are hidden in the mapping properties of
the wave operators, or, better, in the smoothness properties of
the generalized eigenfunctions. On the other hand, one would like
to be sure, that such conditions are not too restrictive on the
set of scattering states.

We introduce the set $\cal{G}$ of functions
$\widehat{\psi}_{\text{out}}$, for which the flux across surfaces
can naturally be proven:

\begin{equation}\label{definitionG}
f(\mathbf{k})\in\cal{G}\Longleftrightarrow\bigg{\{}
\begin{tabular}{llll}
 $\exists M\in\mathbb{R}:$
& $\parallel\partial_{k}^{j}f(\mathbf{k})\parallel_{s}\leq
M\langle k\rangle^{-n}$ & for $j=0,1,2;$ $n\in\mathbb{N}$
\\
 $\forall\mathbf{k}\neq0:$
& $\parallel k^{\mid\gamma\mid-1}
D_{\mathbf{k}}^{\gamma}f(\mathbf{k})\parallel_{s}\leq M\langle
k\rangle^{-n}$ & for $n\in\mathbb{N}$ \\
\end{tabular}
\end{equation}
where $\gamma=(\gamma_{1};\gamma_{2};\gamma_{3})$ is a multi-index
with$\mid\gamma\mid\leq2$,
$D_{\mathbf{k}}^{\gamma}:=\partial_{k_{1}}^{\gamma_{1}}\partial_{k_{2}}^{\gamma_{2}}\partial_{k_{3}}^{\gamma_{3}}$.

This set maps under the wave operator to a dense set in the set of
scattering states. After the theorem we shall give under more
restrictive conditions more detailed information on the set of
scattering states for which the theorem holds.

The paper is organized as follows: In the next section we shall
state the theorem.   We shall also give its formulation in
covariant form, but we shall prove the theorem using the rest
frame of the dectector and the potential.

The following sections contain the proof of the theorem.
 We first prove the
statement for the free case ($A\hspace{-0.2cm}/=0$) and then for
the case of nonzero potential. Both are done in section 3. The
proof relies almost entirely on the stationary phase method, which
we need to  adapt to our purposes. The main lemma is lemma
(\ref{statphas}), whose lengthy technical proof is put in the
Appendix \ref{appendix1}.

The difficulty we have to face and which makes this paper not a
simple generalization of the results in the Schr\"odinger
situation is, that the time evolution with the Dirac hamiltonian
is not of a "nice" form for the stationary phase method to be
easily applied to. The Schr\"odinger case is easier. On the other
hand, the expression for the flux needs no differentiability of
the wavefunction and one might be lead to believe, that to
describe scattering in the relativistic regime is simpler---in
particular less restrictive theorems should result. One may even
get the idea, that asymptotic completeness and the flux across
surfaces theorem become more or less equivalent statements in the
relativistic regime. But we are far from that. Nevertheless, that
we require smoothness and good decay on the potential may well be
due to our method of proof.

We also need information about the generalized eigenfunctions of
the Dirac hamiltonian with external potential, see Lemma
\ref{properties}, whose proof is also put into the Appendix
\ref{appendix3}.  The appendix, which in fact is almost half of
the paper, contains other tedious technical details.
\bigskip

\noindent\textit{\textbf{Acknowledgement:}} Very helpful
discussions with Stefan Teufel are gratefully acknowledged. We
would also like to thank the referee for a very critical reading
of the manuscript, which let to a clear improvement of the paper.

\section{The Theorem}

The flux-across-surfaces theorem deals with the flux $\mathbf j$
integrated over a spherical surface at a far distance and asserts
that\begin{enumerate} \item the absolute value of the flux and the
flux itself yield the same asymptotics, allowing to interpret the
flux integral as crossing probability \cite{DDGZ2,Daumer},
\item the crossing probability equals the probability for the
momentum to lie within the cone defined by the surface.
\end{enumerate}

\begin{thm} \label{frflsu} Let $\psi$ be a scattering state with outgoing free
asymptotic $\widehat{\psi}_{\text{out}}$, whose Fourier transform
$\widehat{\psi}_{\text{out}}$ lies in $\mathcal{G}$(cf.
\ref{definitionG}). Let $R^{2}d\Omega$ be the surface element at
distance R with solid angle differential $d\Omega$ and let
$\mathbf{n}$ denote the outward normal of the surface element.
Furthermore let $S$ be a subset of the unit sphere. Then for all
$t_{i}\in\mathrm{R}$:

\begin{eqnarray}\label{limes}
\lim_{R\rightarrow\infty}\int_{S}\int_{t_{i}}^{\infty}j(\mathbf{R},t)dt
    R^{2}d\Omega & = &\lim_{R\rightarrow\infty}\int_{S}\int_{t_{i}}^{\infty}\mathbf{j}(\mathbf{R},t)\cdot\mathbf{n}dt
    R^{2}d\Omega \nonumber\\& = & \int_{S}\int_{0}^{\infty}\langle\widehat{\psi}_{\text{out}}(\mathbf{k}),
    \widehat{\psi}_{\text{out}}(\mathbf{k})\rangle k^{2}dkd\Omega\ \, ,
\end{eqnarray}
\end{thm}
Observing that
$\parallel\widehat{\psi}_{\text{out}}\parallel(\mathbf{k})$ does
not depend on time, we can choose a coordinate system
$t^{\prime}=t-t_{i}$, so that we may for definiteness always put
$t_{i}=0$ in (\ref{limes}).

The conditions on $\psi_{\text{out}}$ can be translated into more
detailed  conditions on the scattering states under more
restrictive conditions on the potential: Let

\begin{eqnarray}\label{potcond2}
\hspace{-1.8cm}\text{\bf Condition
B}\hspace{1.5cm}\mid\partial_{x}^{n}A\hspace{-0.2cm}/(\mathbf{x})\mid\in
L^{2}(\mathbb{R}^3)\;\forall n\in\{0,1,2...\}\hspace{1cm}\exists M
|A\hspace{-0.2cm}/(\mathbf{x})|\leq
 M\langle x\rangle^{-6}\nonumber\,\,.
\end{eqnarray}
Then (for the proof see Appendix \ref{appendix4})

\begin{lem}\label{equiv}

\begin{equation}\label{equivalence}
\widehat{\psi}_{\text{out}}(\mathbf{k})\in\mathcal{G}\Leftrightarrow\psi(\mathbf{x})\in\widehat{\mathcal{G}}
\end{equation}
where $\widehat{\mathcal{G}}$ is the space of functions
$\psi(\mathbf{x})\in\mathcal{H}_{ac}$ with
$x^{j}\nabla\hspace{-0.25cm}/\hspace{0.1cm}^{n}\psi(\mathbf{x})\in
L^{2}$ for all $j=0,1,2$; $n\in\mathbb{N}_{0}$, where
$\nabla\hspace{-0.25cm}/:=-i\sum_{l=1}^{3}\alpha_{l}\partial_{l}$.

\end{lem}

\subsection{Covariant form of the theorem}

As we deal with a relativistic regime, it might be of interest to
have also a covariant formulation of the theorem. As
$\langle\widehat{\psi}_{out},\widehat{\psi}_{out}\rangle$ is not
conserved under Lorentz function we use

$$\widehat{\psi}^{LI}_{\text{out}}(\underline{k})=(k^{2}+m^{2})^{\frac{1}{4}}\widehat{\psi}_{\text{out}}(\underline{k})\;,$$
of which it is known, that
$\langle\widehat{\psi}_{out}^{LI},\widehat{\psi}_{out}^{LI}\rangle$
is a Lorentz-scalar (see for instance \cite{haag}). Then the
flux-across-surfaces theorem reads in a general and covariant way:

\begin{thm} \label{relfast} Let the conditions of Theorem \ref{frflsu} be satisfied. Let
$$\underline{x}\diamond\underline{y}:=x_{0}y_{0}-\sum_{j=1}^{3}x_{j}y_{j}$$
be the Minkowski scalar product. Then for any subspace
$Z\subseteq\{\underline{x}\mid
\underline{x}\diamond\underline{x}=m^{2}\}\subset\mathbb{R}^{4}$
and any smooth scalar function
$\eta(\underline{x})$ bounded away from zero:

\begin{eqnarray} \label{relativistic}
\lim_{\lambda\rightarrow\infty}\int_{\widetilde{Z}(\lambda)}
\underline{j}(\mathbf{x})\diamond\underline{n}\widetilde{d\sigma}
=\int_{Z}\langle\widehat{\psi}^{LI}_{\text{out}}(\underline{k}),
\widehat{\psi}^{LI}_{\text{out}}(\underline{k})\rangle d\sigma \;
.
\end{eqnarray}
where $$\widetilde{Z}(\lambda):=\{\underline{y}\mid\exists
\underline{x}\in Z:
\underline{y}=\lambda\eta(\underline{x})\underline{x}\}\subset\mathbb{R}^{4}$$
\end{thm}
and $d\sigma$ is the invariant measure on $Z$,
$\widetilde{d\sigma}$ the invariant measure on $\widetilde{Z}$ and
$\underline{n}$ is the vector orthogonal on $\widetilde{Z}$ with
Lorentz length one.

This formulation may perhaps not be directly guessed, but once one
understands its basics like (\ref{jscat0}), this formulation
becomes clear: The arbitrariness of the scalar function $\eta$
follows directly from (\ref{jscat0}), observing that
$$\lim_{\lambda\to\infty}\psi(\lambda \underline{k})=\lim_{\lambda\to\infty}\psi(\lambda \eta(\underline{k})\underline{k}).$$

Physically this is related to the fact, that (on big scales) it is
possible to "catch" any part of the wave-function in different
ways (for example by using a detector which is "close" and catches
the wavefunction at an "early" time or one uses a far detector at
a later time-interval).

Let us explain how (\ref{limes}) follows  from
(\ref{relativistic}). We choose a set $Z$ whose projection on the
 $t=0$-subspace is a cone with angular distribution S:

$$Z=\{\underline{k}\mid\frac{\mathbf{k}}{k}\in S\}\cap\{\underline{k}\diamond\underline{k}=m^{2}\}\;.$$

The invariant measure on the mass hyperboloid
$d\sigma=\frac{d^{3}k}{\sqrt{k^{2}+m^{2}}}$ we get for the right
hand side of (\ref{relativistic})
\begin{equation}\label{rechteseite}
\int_{Z}\langle\widehat{\psi}^{LI}_{\text{out}}(\underline{k}),
\widehat{\psi}^{LI}_{\text{out}}(\underline{k})\rangle
d\sigma=\int_{S}\int_{0}^{\infty}\langle\widehat{\psi}_{\text{out}}(\mathbf{k}),
    \widehat{\psi}_{\text{out}}(\mathbf{k})\rangle
    k^{2}dkd\Omega\;.
\end{equation}
For the left hand side of (\ref{relativistic}) we take:

$$\eta(\underline{x}):=\frac{1}{x}\hspace{1cm}x\neq0\;.$$
As both integrands in (\ref{relativistic}) are bounded, a small
neighborhood of $\mathbf{x}=0$ can be neglected.
For constant $\lambda$, $\widetilde{Z}$ represents a radial
surface with arbitrary time $t\geq0$. So we have:

\begin{equation}\label{linkeseite}
\lim_{\lambda\rightarrow\infty}\int_{\widetilde{Z}(\lambda)}\underline{j}(\mathbf{x})\diamond\underline{n}\widetilde{d\sigma}=\lim_{R\rightarrow\infty}\int_{S}\int_{t_{i}}^{\infty}j(\mathbf{R},t)dt
    R^{2}d\Omega\;.
\end{equation}

\section{The proof}

\subsection{Scattering into cones heuristics}

The flux-across-surfaces theorem is based on an asymptotic
connection between the shape of the wavefunction in momentum space
and in ordinary space. This is often referred to as the scattering
into cones theorem, which has been proven for non-relativistic
particles by Dollard \cite{dollard}. For that one chooses a
certain parameterization of $\mathbb{R}^{4}$ and evaluates the
wavefunction, as the parameter of the parameterizations goes to
infinity. In the non-relativistic case, it is easiest to choose
time as the parameter of the parameterization. In the relativistic
case it is simplest to have lorentz-invariant three-dimensional
subspaces of the time-like part of $\mathbb{R}^{4}$ as leaves of
the parametrization\footnote{We only parameterize the time-like
region, as for big time-scales the main part of our wavefunction
will be in this region.}. This can easily be done, by choosing a
lorentz-vector as argument of $\psi$, i.e. a vector
$\underline{x}$ with
$\underline{x}\diamond\underline{x}=x_{0}^{2}-\mathbf{x}\cdot\mathbf{x}=\lambda
m^{2}$. Set
$\psi(\lambda\underline{k})=\psi(\mathbf{x}=\lambda\mathbf{k},t=\lambda\sqrt{k^{2}+m^{2}})$.
We denote the two different eigenstates of momentum $\mathbf{k}$
of the free Hamiltonian with positive energy by
$\varphi_{\mathbf{k}}^{s}$, whereas the s labels the two different
spins our electron may have. In the standard representation these
eigenstates can be written as:

\begin{equation}\label{basis}
\varphi_{\mathbf{k}}^{s}=e^{i\mathbf{k}\cdot\mathbf{x}}s^{s}_{\mathbf{k}},
\end{equation}
where the $s_{\mathbf{k}}^{s}$ are:

\begin{eqnarray*}
  s^{1}_{\mathbf{k}}=(2E_{k}\widehat{E}_{k})^{-\frac{1}{2}}\begin{pmatrix}
    _{\widehat{E}_{k}} \nonumber \\
    _{0} \nonumber \\
    _{k_{1}} \nonumber \\
    _{k^{+}}
  \end{pmatrix}\hspace{1cm}
  s^{2}_{\mathbf{k}}=(2E_{k}\widehat{E}_{k})^{-\frac{1}{2}}\begin{pmatrix}
    _{0} \nonumber \\
    _{\widehat{E}_{k}} \nonumber \\
    _{k^{-}} \nonumber \\
    _{-k_{1}}
  \end{pmatrix}    \,,
\end{eqnarray*}
where $$k^{\pm}=k_{2}\pm
 ik_{3}\hspace{1cm}\widehat{E}_{k}=E_{k}+m\hspace{1cm}E_{k}=\sqrt{k^{2}+m^{2}}\,.$$
(For a detailed calculation of these spinors see (\cite{thaller}))

 The
asymptotics result from a stationary phase analysis:

\begin{eqnarray*}
  \psi(\lambda\underline{k}) & = & U(t=\lambda\sqrt{k^{2}+m^{2}})\psi(\lambda\mathbf{k},0)\nonumber
  \\&=&\sum_{s=1}^{2}e^{-iH\lambda\sqrt{k^{2}+m^{2}}}\int(2\pi)^{-\frac{3}{2}}
  \varphi_{\mathbf{k^{\prime}}}^{s}(\lambda\mathbf{k})\widehat{\psi}_{s}(\mathbf{k^{\prime}})d^{3}k^{\prime}\nonumber
  \\&=&\sum_{s=1}^{2} e^{-iH\lambda\sqrt{k^{2}+m^{2}}}\int(2\pi)^{-\frac{3}{2}}
  e^{i\mathbf{k^{\prime}}\cdot\lambda
  \mathbf{k}}s^{s}_{\mathbf{k^{\prime}}}\widehat{\psi}_{s}(\mathbf{k^{\prime}})d^{3}k^{\prime}\nonumber
  \,.
\end{eqnarray*}
For convenience we define:

$$\widehat{\psi}(\mathbf{k^{\prime}})=\sum_{s=1}^{2}s^{s}_{\mathbf{k^{\prime}}}\widehat{\psi}_{s}(\mathbf{k^{\prime}})\,.$$
This leads to:

\begin{eqnarray}\label{ps}
  \psi(\lambda\underline{k}) & = & e^{-iH\lambda\sqrt{k^{2}+m^{2}}}\int(2\pi)^{-\frac{3}{2}}
  e^{i\mathbf{k^{\prime}}\cdot\lambda
  \mathbf{k}}\widehat{\psi}(\mathbf{k^{\prime}})d^{3}k^{\prime}\nonumber\\
  &=& \int(2\pi)^{-\frac{3}{2}}
  e^{-i\lambda(\sqrt{k^{\prime 2}+m^{2}}\sqrt{k^{2}+m^{2}}-\mathbf{k^{\prime}}\cdot\mathbf{k})}\widehat{\psi}(\mathbf{k^{\prime}})d^{3}k^{\prime}\;.
\end{eqnarray}
In view of the stationary phase method, in the limit
$\lambda\rightarrow\infty $ only a small neighborhood of the
stationary point of the phase function

$$h(\mathbf{k}^{\prime}):=(\sqrt{k^{\prime
2}+m^{2}}\sqrt{k^{2}+m^{2}}-\mathbf{k^{\prime}}\cdot\mathbf{k})$$
will be relevant for the integral. The stationary point is given
by:

\begin{equation}\label{point}
\nabla_{\mathbf{k}^{\prime}}h(\mathbf{k}_{stat})=0 \Rightarrow
\mathbf{k}_{stat}=\mathbf{k}
\end{equation}
Without loss of generality we can set $k_{2}=k_{3}=0$. Near the
stationary point the phase is to second order:

$$
  -i\lambda(\sqrt{k^{\prime
  2}+m^{2}}\sqrt{k^{2}+m^{2}}-\mathbf{k^{\prime}}\cdot\mathbf{k})\approx-i\lambda(m^{2}+\frac{m^{2}}{2(k^{2}+m^{2})}(k_{1}^{\prime}-k)^{2}+\frac{1}{2}(k_{2}^{\prime2}+k_{3}^{\prime2}))
$$ This in equation(\ref{ps}) leads to

\begin{eqnarray*}
  \psi(\lambda\underline{k}) & \approx & \int(2\pi)^{-\frac{3}{2}}
  e^{-i\lambda(m^{2}+\frac{m^{2}}{2(k^{2}+m^{2})}(k_{1}^{\prime}-k)^{2}+\frac{1}{2}(k_{2}^{\prime2}+k_{3}^{\prime2}))}\widehat{\psi}(\mathbf{k^{\prime}})d^{3}k^{\prime}\;,
\end{eqnarray*}
and replacing $\widehat{\psi}(\mathbf{k}^{\prime})$ by
$\widehat{\psi}(\mathbf{k})$ we obtain by integrating the
gaussian

$$
  \psi(\lambda\underline{k})\approx\frac{e^{-i\lambda m^{2}}}{(i\lambda)^{\frac{3}{2}}}\widehat{\psi}(\mathbf{k})\sqrt{\frac{k^{2}}{m^{2}}+1}
$$ We shall state now the stationary phase result in a some what
more general setting, to cover also applications to the potential
case considered later:

\subsection{The stationary phase}

\begin{lem} \label{statphas}

                    Let $\widetilde{\chi}$ be in $\cal{G}$
                    (see(\ref{definitionG})) and let the ``phase function'' $g$ be

$$g(\mathbf{k}^{\prime})=\sqrt{k^{\prime2}+m^{2}}+a\mid
k^{\prime}\mid-\mathbf{y}\cdot\mathbf{k}^{\prime}.$$ Let
$\mathbf{k}_{stat}$ be the stationary point of the phase-function:

$$\nabla g(\mathbf{k}_{stat})=0\, .$$ Then there exist
 $C_{1},C_{2},C_{3}\in\mathbb{R}$ so, that for all $\chi$ with
$\parallel\partial_{k}^{j}\chi\parallel_{s}\leq\parallel\partial_{k}^{j}\widetilde{\chi}\parallel_{s}$
for $j=0,1,2$, $\mathbf{y}\in\mathbb{R}^{3}$ and $a\geq 0$

\begin{equation}\label{hoerm}
\parallel\int e^{-i\mu
g(\mathbf{k}^{\prime})}\chi(\mathbf{k}^{\prime})d^{3}k^{\prime}-C_{1}
\mu^{-\frac{3}{2}}\chi(\mathbf{k}_{stat})
\parallel_{s}< C_{2}\mu^{-2}+C_{3}
\frac{k_{stat}^{\frac{1}{2}}}{\mu}\chi(\mathbf{k}_{stat})\, .
\end{equation}
For phase functions without stationary point $C_{1}=C_{3}=0$.
Moreover the $C_{j}$ are uniformly bounded for all $\chi$, $a$ and
$\mathbf{y}$. For $a=0$  we can choose
$C_{1}=(-2\pi i)^{\frac{3}{2}}e^{-i\mu
g(\mathbf{k}_{stat})}\frac{(k_{stat}^{2}+m^{2})^{\frac{5}{4}}}{m}$
and $C_{3}=0 $ .
 \end{lem}

One may be disturbed about the nature of the inequality
(\ref{hoerm}) when $C_{3}\neq0$. The point here is that our
estimate is uniform in $k_{stat}$ and then later we shall use
(\ref{hoerm}) for $C\neq0$ such that $k_{stat}$ will be of order
$\mu^{-1}$, so for $C_{3}\neq0$ the last term will be part of the
leading term in our estimation.

This statement is a slight adaptation to our situation of a
 theorem of H\"ormander \cite{hoer}, and its proof in
 the appendix \ref{appendix1}.

\subsection{Scattering into cones for a free particle}

Applying Lemma \ref{statphas} to (\ref{ps}) we choose:

$$\mu=\lambda\sqrt{k^{2}+m^{2}} ;\; a=0 ;\;
\mathbf{y}=\frac{\mathbf{k}}{\sqrt{k^{2}+m^{2}}} ;\;
\chi(\mathbf{k}^{\prime})=(2\pi)^{-\frac{3}{2}}\widehat{\psi}(\mathbf{k}^{\prime})$$
and calculate the stationary point $k_{stat}$:

\begin{eqnarray*}
\frac{\mathbf{k_{stat}}}{\sqrt{k_{stat}^{2}+m^{2}}}-\mathbf{y}&=&0
\\k_{stat}^{2}&=&y^{2}(k_{stat}^{2}+m^{2})
\\\\k_{stat}&=&\frac{ym}{\sqrt{1-y^{2}}}
\end{eqnarray*}
obtaining

\begin{cor}\label{scintocones} ("Scattering into cones") There exists a constant $C<\infty $ so that
 for all $ \mathbf{k}\in\R^{3} $
$$
\parallel\psi(\lambda\underline{k})-\frac{e^{-i\lambda m^{2}}}{(i\lambda)^{\frac{3}{2}}}\widehat{\psi}(\mathbf{k})\sqrt{\frac{k^{2}}{m^{2}}+1}\parallel_{s}
   \leq
  C\lambda^{-2}\;.
$$
\end{cor}
Note, that this implies

\begin{eqnarray}\label{scatlim}
\lim_{\lambda\rightarrow\infty}\sup_{\mathbf{k}}(\parallel\sqrt{\lambda}^{3}\psi(\lambda\underline{k})\parallel_{s}-\parallel
\widehat{\psi}(\mathbf{k})\sqrt{\frac{k^{2}}{m^{2}}+1}\parallel_{s}
)=0\;.
\end{eqnarray}

For the flux-across-surfaces theorem we need the asymptotics of
the relativistic quantum flux (\ref{relflux}) of the particle.
Since all the $\alpha_{l}$ are bounded matrices and $\widehat{\psi}\in
{\cal G}$, we obtain from (\ref{relflux}) and (\ref{scatlim}) for
the flux:

\begin{eqnarray}\label{jscat0}
\lim_{\lambda\rightarrow\infty}\sup_{\mathbf{k}}\mid\lambda^{3}j_{\hspace{0.01cm}l}(\lambda\underline{k})-\langle\widehat{\psi}(\mathbf{k}),
\alpha_{l}\widehat{\psi}(\mathbf{k})\rangle(\frac{k^{2}}{m^{2}}+1)\mid=0\;.
\end{eqnarray}
Next observe (see the appendix \ref{appendix2}), that:

\begin{equation}\label{ridofalpha}
 \langle\widehat{\psi}(\mathbf{k}),\mathbf{\alpha}\widehat{\psi}(\mathbf{k})\rangle=\frac{\mathbf{k}}{\sqrt{k^{2}+m^{2}}}\langle\widehat{\psi}(\mathbf{k}),\widehat{\psi}(\mathbf{k})\rangle\;.
\end{equation}
Thus we get the uniform bound:

\begin{cor}
\begin{eqnarray}\label{jscat}
\forall\varepsilon>0\hspace{1cm}\exists\lambda\in\mathbb{R}:\hspace{2.1cm}\nonumber\\
\sup_{\mathbf{k}}\mid\lambda^{3}\mathbf{j}(\lambda\underline{k})-\langle\widehat{\psi}(\mathbf{k}),\widehat{\psi}(\mathbf{k})\rangle
  \frac{\mathbf{k}}{m^{2}}\sqrt{k^{2}+m^{2}}\mid<\varepsilon\;.
\end{eqnarray}
 \end{cor}
Observe, that after a long time of propagation, the flux at
$\mathbf{x}=\lambda\mathbf{k}$ will always be parallel to
$\mathbf{k}$. So in the limit $t\rightarrow\infty$ it will always
point away from the origin of the coordinate system.

\subsection{Flux across surfaces for a free particle}

Theorem \ref{frflsu} reads in this case

\begin{eqnarray}\label{limes2}
\lim_{R\rightarrow\infty}\int_{S}\int_{0}^{\infty}\mathbf{j}(\mathbf{R},t)\cdot\mathbf{n}dt
    R^{2}d\Omega-\int_{S}\int_{0}^{\infty}\langle\widehat{\psi}(\mathbf{k}),\widehat{\psi}(\mathbf{k})\rangle k^{2}dkd\Omega=0\
\end{eqnarray}
and

\begin{eqnarray}\label{limes3}
\lim_{R\rightarrow\infty}\int_{S}\int_{0}^{\infty}j(\mathbf{R},t)dt
    R^{2}d\Omega-\int_{S}\int_{0}^{\infty}\langle\widehat{\psi}(\mathbf{k}),\widehat{\psi}(\mathbf{k})\rangle
    k^{2}dkd\Omega=0\;.
\end{eqnarray}
In the following, we will prove (\ref{limes2}) by inserting the
longtime asymptotic (\ref{jscat}) for $\mathbf{j}$ and showing,
that the integral of the error we get by this approximation tends
to zero in the limit $R\rightarrow\infty$.

Now, the long time asymptotic of $\mathbf{j}$ is parallel to the
normal $\mathbf{n}$ of the radial surface. Therefore the longtime
asymptotic of $j$ is equal to the longtime asymptotic of
$\mathbf{j}\cdot\mathbf{n}$. More detailed, one sees that using
the approximation (\ref{jscat}) for $\mathbf{j}$ in (\ref{limes2})
and (\ref{limes3}), the bound on the error terms in (\ref{limes2})
and (\ref{limes3}) arising from (\ref{jscat}) are equal.

So the proof of (\ref{limes3}) is essentially the same as for
(\ref{limes2}) and we shall concentrate only on showing
(\ref{limes2}).

The left side of (\ref{limes2}) includes an integral over t,
whereas the right hand side is integrated over k. We therefore
substitute for t in the first term, to get integration over k,
too. Since $\lambda$ plays the role of a time parameter it is
natural to substitute:

 $$\mathbf{k}=\frac{R\mathbf{n}}{\lambda}$$
with

$$\lambda=\frac{\sqrt{t^{2}-R^{2}}}{m}\;.$$ But this substitution
is only possible in the time-like region ($t\geq R$). So we first
handle the integral starting at $t=R$, later we deal with the
space-like part of the integral. Then, substituting t by k, we
obtain

\begin{eqnarray}\label{subst1}
 \int_{S}\int_{R}^{\infty}\mathbf{j}(\mathbf{R},t)\cdot\mathbf{n}dt
 R^{2}d\Omega & = & \int_{S}\int_{0}^{\infty}\mathbf{j}(\mathbf{R},\frac{R}{k}\sqrt{k^{2}+m^{2}})\cdot\mathbf{n}\frac{m^{2}}{\sqrt{k^{2}+m^{2}}}\frac{R^{3}}{k^{2}}dk
 d\Omega \nonumber \\
 & = & \int_{S}\int_{0}^{\infty}\mathbf{j}(\lambda(k)
 k,\lambda(k)\sqrt{k^{2}+m^{2}})\cdot\mathbf{n}\frac{m^{2}}{\sqrt{k^{2}+m^{2}}}k\lambda(k)^{3}dkd\Omega\nonumber\;.
\end{eqnarray}
The integrand is now in the form that we can replace it by the
asymptotic in (\ref{jscat}).

It turns out however, that the error in the integrand will be
$\sim\frac{k}{\sqrt{k^{2}+m^{2}}}$ which is not integrable,
therefore the replacement is not straight forward. We separate
large momenta $k>X$ and small momenta $k<X$. In the following we
choose $X>m$. Given X and $R_{0}=\lambda_{0}X$

$$k\leq X\Leftrightarrow\frac{R_{0}}{k}=\lambda(k)\geq\lambda_{0}=\frac{R_{0}}{X}.$$
Then by (\ref{jscat}) for small momenta ($k\leq X\Leftrightarrow
t\geq R\sqrt{1+\frac{m^{2}}{X^{2}}}$):

$$ \forall\varepsilon>0\hspace{1cm}\exists
R_{0}\in\mathrm{R}\hspace{1cm}\forall R\geq R_{0}\:$$
\begin{eqnarray}\label{insert}
\lefteqn{\hspace{-1cm}
\mid\int_{S}\int_{R\sqrt{1+\frac{m^{2}}{X^{2}}}}^{\infty}\big(\mathbf{j}(\mathbf{R},t)\cdot\mathbf{n}dt
 R^{2}-\int_{0}^{X}\langle\widehat{\psi}(\mathbf{k}),\widehat{\psi}(\mathbf{k})\rangle k^{2}\big)dkd\Omega\mid}\\ &=&
    \mid\int_{S}\int_{0}^{X}\mathbf{j}(\lambda
 k,\lambda\sqrt{k^{2}+m^{2}})\cdot\mathbf{n}\frac{m^{2}}{\sqrt{k^{2}+m^{2}}}k\lambda^{3}-\langle\widehat{\psi}(\mathbf{k}),\widehat{\psi}(\mathbf{k})\rangle k^{2}dkd\Omega\mid\nonumber \\
    &\leq&\int_{S}\int_{0}^{X}\frac{km^{2}\varepsilon}{\sqrt{k^{2}+m^{2}}}
    dkd\Omega=:\chi(X)\varepsilon\nonumber
\end{eqnarray}
where

$$\chi(X):=4\pi\int_{0}^{X}\frac{km^{2}}{\sqrt{k^{2}+m^{2}}}dk\;.$$
Given $X$ we can take $\varepsilon$ arbitrarily small, choosing
$R_{0}$ large enough, so that the r.h.s. of (\ref{insert}) goes to
zero. Thus

\begin{equation}\label{in}
  \lim_{X\rightarrow\infty}\lim_{R\rightarrow\infty}\mid\int_{S}\int_{R\sqrt{1+\frac{m^{2}}{X^{2}}}}^{\infty}\mathbf{j}(\mathbf{R},t)\cdot\mathbf{n}dt
 R^{2}d\Omega-\int_{S}\int_{0}^{X}\langle\widehat{\psi}(\mathbf{k}),\widehat{\psi}(\mathbf{k})\rangle
 k^{2}dkd\Omega\mid=0\;.
\end{equation}
For the large momenta note that by virtue of
$\widehat{\psi}\in\mathcal{G}$:

\begin{equation}\label{bigX1}
    \lim_{X\rightarrow\infty}\int_{S}\int_{X}^{\infty}\langle\widehat{\psi}(\mathbf{k}),\widehat{\psi}(\mathbf{k})\rangle
    k^{2}dkd\Omega=0
\end{equation}
and all it remains to show is that

\begin{equation}\label{bigXX}
     \lim_{X\rightarrow\infty}\lim_{R\rightarrow\infty}\int_{S}\int_{0}^{R\sqrt{1+\frac{m^{2}}{X^{2}}}}\mathbf{j}(\mathbf{R},t)\cdot\mathbf{n}dt
    R^{2}d\Omega=0
\end{equation}
where we also included the time integration outside the light
cone, which we excluded in the substitution.

We first estimate the part of the integral (\ref{bigXX}) that lies
in the space-like region (more precisely: $t\in [0,R]$) then we
estimate the time-like part near the light cone ( $t\in
[R,R\sqrt{1+\frac{m^{2}}{X^{2}}}]$). That is, we first show that

\begin{equation}\label{othertimeb}
    \lim_{R\rightarrow\infty}
    \int\int_{0}^{R}\mathbf{j}(\mathbf{R},t)\cdot\mathbf{n}dtR^{2}d\Omega=0\;.
\end{equation}
That this holds is physically related to the fact, that a particle
moves slower than light, so for big time and space scales the main
part of the wavefunction will be inside the light cone. This
follows from a straightforward application of the stationary phase
method, outside of the stationary point. Two partial integrations
lead to:

\begin{eqnarray*}
\parallel\psi(\mathbf{x},\eta x)\parallel_{s}&=&\parallel\int
(2\pi)^{-\frac{3}{2}}e^{-ix(\sqrt{k^{2}+m^{2}}\eta-k_{1})
}\widehat{\psi}(\mathbf{k})d^{3}k\parallel_{s}
\\&=&\parallel\int
(2\pi)^{-\frac{3}{2}}e^{-ixg
}\widehat{\psi}(\mathbf{k})d^{3}k\parallel_{s}
\\&\leq&\frac{1}{x^{2}}\int\parallel(2\pi)^{-\frac{3}{2}}
(\frac{\widehat{\psi}^{\prime\prime}}{g^{\prime2}}-\frac{3\widehat{\psi}^{\prime}g^{\prime\prime}}{g^{\prime
3}}+\frac{3\widehat{\psi}
g^{\prime\prime2}}{g^{\prime4}}-\frac{\widehat{\psi}
g^{\prime\prime\prime}}{g^{\prime3}})\parallel_{s}d^{3}k
\end{eqnarray*}
where

$$g:=(\sqrt{k^{2}+m^{2}}\eta-k_{1})\hspace{1cm}f^{\prime}:=\partial_{k_{1}}f\;.$$
Since

$$-g^{\prime}=1-\frac{k_{1}\eta}{\sqrt{k^{2}+m^{2}}}\geq
1-\frac{\mid k_{1}\mid}{\sqrt{k^{2}+m^{2}}}>0$$ it follows:

\begin{equation}\label{spacelike}
\parallel\psi(\mathbf{x},\eta x)\parallel_{s}
\leq(2\pi)^{-\frac{3}{2}}\frac{C_{2}}{x^{2}}
\end{equation}
uniform in $\eta\leq 1$. Hence

\begin{eqnarray*}
\lefteqn{\hspace{-1cm}\lim_{R\rightarrow\infty}\int\int_{0}^{R}\mathbf{j}(\mathbf{R},t)\cdot\mathbf{n}dtR^{2}d\Omega}\\&\leq&
4\pi\lim_{R\rightarrow\infty}\int_{0}^{R}\parallel\psi(\mathbf{x},t)\parallel_{s}^{2}
dtR^{2}\leq\frac{1}{2\pi^{2}}
C_{2}^{2}\lim_{R\rightarrow\infty}R^{3}\frac{1}{R^{4}}=0
\end{eqnarray*}
It is left to prove that the second part of the integral in
(\ref{bigXX}) goes to zero, i.e. that

$$\lim_{X\rightarrow\infty}\lim_{R\rightarrow\infty}\int_{S}\int_{R}^{R\sqrt{1+\frac{m^{2}}{X^{2}}}}\mathbf{j}(R,t)\cdot\mathbf{n}dtR^{2}d\Omega=0\;.$$
The scalar norm of $\psi(\mathbf{x},t)$ is:

\begin{eqnarray}\label{norm}
    \parallel\psi(\mathbf{x},t)\parallel_{s}&=&\parallel\int (2\pi)^{-\frac{3}{2}}e^{-i\sqrt{k^{2}+m^{2}}t+i\mathbf{k}\cdot
    \mathbf{x}}\widehat{\psi}d^{3}k\parallel_{s}\\&=&\parallel\int (2\pi)^{-\frac{3}{2}}e^{-i(\sqrt{k^{2}+m^{2}}-\mathbf{k}\cdot
    \mathbf{r})t}\widehat{\psi}d^{3}k\parallel_{s}\;.
\end{eqnarray}
Applying Lemma \ref{statphas} with

$$\mu=t;\; a=0 ;\; \mathbf{y}=\mathbf{r} ;\;
\chi(\mathbf{k}^{\prime})=(2\pi)^{-\frac{3}{2}}\widehat{\psi}(\mathbf{k}^{\prime})$$
we have by (\ref{hoerm}), that:

\begin{eqnarray*}
\parallel\int
e^{-iE_{k}t+i\mathbf{k}\cdot\mathbf{x}}\widehat{\psi}(\mathbf{x})d^{3}k-C_{1}t^{-\frac{3}{2}}\widehat{\psi}(k_{stat})\parallel_{s}<C_{2}t^{-2}\;.
\end{eqnarray*}
As $\widehat{\psi}$ is bounded, we have:

\begin{eqnarray}\label{wehave}
\nonumber\exists M\in\R:\forall t>R
\hspace{1cm}\parallel\psi(\mathbf{x},t)\parallel_{s}=\parallel\int
e^{-iE_{k}t+i\mathbf{k}\cdot\mathbf{x}}\widehat{\psi}(\mathbf{k})d^{3}k\parallel_{s}\leq
Mt^{-\frac{3}{2}}\;.
\end{eqnarray}
So

$$\mid\int_{S}\mathbf{j}(\mathbf{R},t)\cdot\mathbf{n}R^{2}d\Omega\mid\leq4\pi\frac{MR^{2}}{t^{3}}\;.$$
So we can write:

\begin{eqnarray*}
\lefteqn{\hspace{-1cm}\mid\int_{S}\int_{R}^{R\sqrt{1+\frac{m^{2}}{X^{2}}}}\mathbf{j}(\mathbf{R},t)\cdot\mathbf{n}dt
    R^{2}d\Omega\mid}\\&\leq&2\pi
    MR^{2}(R^{-2}-R^{-2}\big(1+\frac{m^{2}}{X^{2}})^{-1}\big)=2\pi
    M\big(1-(1+\frac{m^{2}}{X^{2}})^{-1}\big)\;.
\end{eqnarray*}
This term goes to zero as $X\rightarrow\infty$

\subsection{The flux-across-surfaces theorem  with potential}

\subsubsection{Generalized Eigenfunctions for the Dirac equation with potential}

For the proof of the free flux-across-surfaces theorem we used the
$\varphi_{\mathbf{k}}^{s}$ as basis of the Hilbert space. In the
potential case we adopt a new basis for doing calculations.

Like in the free case, we again get four linear independent
eigenfunctions for each $\mathbf{k}$, two of them have positive
energy-eigenvalue $E_{k}^{eig}=E_{k}=\sqrt{k^{2}+m^{2}}$, two of
them have negative energy-eigenvalue $E_{k}^{eig}=-E_{k}$. We
denote by $\widetilde{\varphi}^{s}_{\mathbf{k}}(\mathbf{x})$ the
eigenfunctions with $s\in\{1,2\}$:

\begin{equation}\label{dgmp}
E_{k}\widetilde{\varphi}^{s}_{\mathbf{k}}(\mathbf{x})=(H_{0}+A\hspace{-0.2cm}/)\widetilde{\varphi}^{s}_{\mathbf{k}}(\mathbf{x})\;.
\end{equation}
The corresponding Lipmann Schwinger equation reads:

\begin{equation}\label{EH}
 \widetilde{\varphi}^{s}_{\mathbf{k}}(\mathbf{x})=\varphi_{\mathbf{k}}^{s}(\mathbf{x})+(E_{k}-H_{0})^{-1}A\hspace{-0.2cm}/\widetilde{\varphi}_{\mathbf{k}}^{s}(\mathbf{x})\;.
\end{equation}
We replace the
formal expression $(E_{k}-H_{0})^{-1}$ by the integral kernel
$G^{+}_{k}$:

\begin{equation}\label{Green}
 (E_{k}-H_{0})G^{+}_{k}(\mathbf{x}-\mathbf{x^{\prime}})=\delta(\mathbf{x}-\mathbf{x^{\prime}})\;.
\end{equation}
The explicit form for $G^{+}_{k}(\mathbf{x}-\mathbf{x^{\prime}})$
can be found in \cite{thaller}:

\begin{equation}\label{kernel}
G^{+}_{k}(\mathbf{x})=\frac{1}{4\pi}e^{ikx}\left(-x^{-1}(E_{k}+\sum_{j=1}^{3}\alpha_{j}k\frac{x_{j}}{x}+\beta
m)
+x^{-2}\sum_{j=1}^{3}\alpha_{j}\frac{x_{j}}{x}\right)=:\frac{e^{ikx}}{x}S_{k}^{+}(\mathbf{x})\;.
\end{equation}
Thus:

\begin{equation}\label{LSE}
 \widetilde{\varphi}^{s}_{\mathbf{k}}(\mathbf{x})=\varphi_{\mathbf{k}}^{s}(\mathbf{x})-\int
 A\hspace{-0.2cm}/(\mathbf{x^{\prime}})G^{+}_{k}(\mathbf{x}-\mathbf{x^{\prime}})
 \widetilde{\varphi}_{\mathbf{k}}^{s}(\mathbf{x^{\prime}})d^{3}x^{\prime}\;.
\end{equation}
For $S^{+}_{\mathbf{k}}$, defined in (\ref{kernel}), we have:

\begin{eqnarray*}
\mid\partial_{k}^{j}S^{+}_{\mathbf{k}}\mid&=&\mid\frac{1}{4\pi}\partial_{k}^{j}(-E_{k}-\sum_{j=1}^{3}\alpha_{j}k\frac{x_{j}}{x}-\beta
m+x^{-1}\sum_{j=1}^{3}\alpha_{j}\frac{x_{j}}{x})\mid\\
&=&\mid\frac{1}{4\pi}\partial_{k}^{j}(E_{k}+\sum_{j=1}^{3}\alpha_{j}(k\frac{x_{j}}{x}-\frac{x_{j}}{x^{2}})+\beta
m\mid
\end{eqnarray*}
for $j=0,1,2$. Choosing $x\geq 1$ we have

$$\frac{x_{j}}{x}\leq1 \hspace{1cm}\frac{x_{j}}{x^{2}}\leq1$$
and it follows, that

$$\mid\partial_{k}^{j}S^{+}_{\mathbf{k}}\mid\leq\mid\frac{1}{4\pi}\partial_{k}^{j}(E_{k}+\sum_{j=1}^{3}\alpha_{j}(k+1)+\beta
m)\mid$$
Thus with
\begin{equation}\label{sgreen}
\widetilde{S}_{\mathbf{k}}^{+}:=\frac{1}{4\pi}(E_{k}+\sum_{j=1}^{3}\alpha_{j}(k+1)+\beta
m)
\end{equation}
we have:

\begin{equation}\label{sbound}
\mid\partial_{k}^{j}S^{+}_{\mathbf{k}}\mid\leq\mid\partial_{k}^{j}\widetilde{S}_{\mathbf{k}}^{+}\mid
\end{equation}
for $j=0,1,2$, $x\geq 1$.

For the next steps we need some properties of the generalized
eigenfunctions. We summarize these properties in the following
Lemma which is proven in the Appendix \ref{appendix3}:

\begin{lem} \label{properties}

Let $A\hspace{-0.2cm}/$ satisfy Condition A (\ref{potcond}). Then
there exist unique solutions
$\widetilde{\varphi}^{s}_{\mathbf{k}}(\mathbf{x})$ of (\ref{LSE})
for all $k\in \mathbb{R}^{3}$, such that:

\begin{description}
     \item[(a)]For any $\mathbf{k}\in\mathbb{R}^{3}$, $s=1,2$ the
functions $\widetilde{\varphi}^{s}_{\mathbf{k}}(\mathbf{x})$ are
H\"older continuous of degree 1 in $\mathbf{x}$
\item[(b)]Any
$\widetilde{\varphi}^{s}_{\mathbf{k}}(\mathbf{x})$ which is a
solution of (\ref{LSE}) automatically satisfies (\ref{dgmp}).
    \item[(c)]The functions
\begin{equation}\label{zeta}
\zeta^{s}_{\mathbf{k}}(\mathbf{x}):=\widetilde{\varphi}^{s}_{\mathbf{k}}(\mathbf{x})-\varphi^{s}_{\mathbf{k}}(\mathbf{x})
\end{equation}
are infinitely often continuously differentiable with respect to
$k$, furthermore we have for $j\in\mathbb{N}$ and any multi-index
$\gamma$ with $\mid\gamma\mid\leq2$ :

\begin{eqnarray*}
i)&&\sup_{\mathbf{x}\in\mathbb{R}^{3}}\parallel x
\zeta^{s}_{\mathbf{k}}(\mathbf{x})\parallel_{s}<\infty
\\ii)&&\sup_{\mathbf{x}\in\mathbb{R}^{3}}\parallel\partial_{k}^{j}\frac{\zeta^{s}_{\mathbf{k}}(\mathbf{x})}{\mid x+1\mid^{j-1}}\parallel_{s}<\infty
\\iii)&&\sup_{\mathbf{x}\in\mathbb{R}^{3}}\parallel k^{\mid\gamma\mid-1}D_{\mathbf{k}}^{\gamma}\frac{\zeta^{s}_{\mathbf{k}}(\mathbf{x})}{\mid
x+1\mid^{j-1}}\parallel_{s}<\infty\;.
\end{eqnarray*}

\item[(d)]The $\widetilde{\varphi}^{s}_{\mathbf{k}}(\mathbf{x})$
    form a basis of the space of scattering states, i.e. for scattering states $\psi(\mathbf{x},t)$:
\begin{equation}\label{hin}
    \psi(\mathbf{x},t)=\sum_{s=1}^{2}\int(2\pi)^{-\frac{3}{2}}
  e^{-i\sqrt{k^{
  2}+m^{2}}t}\widetilde{\varphi}^{s}_{\mathbf{k}}(\mathbf{x})\widehat{\psi}_{out,s}(\mathbf{k})d^{3}k
 \end{equation}
\begin{equation}\label{her}
\widehat{\psi}_{out,s}(\mathbf{k})=\int(2\pi)^{-\frac{3}{2}}\langle\widetilde{\varphi}^{s}_{\mathbf{k}}(\mathbf{x}),\psi(\mathbf{x})\rangle
d^{3}x
\end{equation}
   where $\widehat{\psi}_{out,s}(\mathbf{k})$ is the
   fourier transform of $\psi_{\text{out}}=\Omega_{+}\psi$.
  \end{description}
\end{lem}

\subsubsection{Flux-across-surfaces for the Dirac-equation with potential}

We prove now Theorem \ref{frflsu}. As in the free case only the
equality of the second and third integral is shown. From the
nature of the estimates in the proof it will become evident, that
essentially by the same argument as in the free case, the first
equality can be established, and we do not say anything more to
that.

 We again split our flux integral into two
parts, one inside the light-cone (from $R$ to $\infty$) and one
outside the light-cone (from $0$ to $R$), where the main
contribution comes from the times $t>R$, i.e. we prove that

\begin{eqnarray}\label{newsplit}
i)&&\lim_{R\rightarrow\infty}\mid\int_{S}\int_{R}^{\infty}\mathbf{j}(\mathbf{R},t)\cdot\mathbf{n}dt
 R^{2}d\Omega-\int_{S}\int_{0}^{\infty}\langle\widehat{\psi}_{\text{out}}(\mathbf{k}),\widehat{\psi}_{\text{out}}(\mathbf{k})\rangle
 k^{2}dkd\Omega\mid=0 \nonumber \\
ii)&&\lim_{R\rightarrow\infty}
\int\int_{0}^{R}\mathbf{j}(\mathbf{R},t)\cdot\mathbf{n}dtR^{2}d\Omega=0
\end{eqnarray}
We start with i):

According to (\ref{hin})

$$\psi(\mathbf{x},t)=\sum_{s=1}^{2}\int(2\pi)^{-\frac{3}{2}}
  e^{-i\sqrt{k^{
  2}+m^{2}}t}\widetilde{\varphi}^{s}_{\mathbf{k}}(\mathbf{x})\widehat{\psi}_{out,s}(\mathbf{k})d^{3}k\;.$$
Setting
$$\widehat{\psi}_{\text{out}}(\mathbf{k^{\prime}})=\sum_{s=1}^{2}s^{s}_{\mathbf{k^{\prime}}}\widehat{\psi}_{out,s}(\mathbf{k^{\prime}})$$
and using (\ref{LSE}) with (\ref{zeta}) we get:

\begin{eqnarray}\label{sum}
\psi(\mathbf{x},t) & = &
  \int(2\pi)^{-\frac{3}{2}}
  e^{-i\sqrt{k^{
  2}+m^{2}}t}e^{i\mathbf{k}\cdot\mathbf{x}}\widehat{\psi}_{\text{out}}(\mathbf{k})d^{3}k\nonumber\\
&  & -\int(2\pi)^{-\frac{3}{2}}
  e^{-i\sqrt{k^{
  2}+m^{2}}t}\int
\frac{e^{ik\mid\mathbf{x}-\mathbf{x}^{\prime}\mid}}{\mid\mathbf{x}-\mathbf{x}^{\prime}\mid}S^{+}_{k}(\mathbf{x}-\mathbf{x}^{\prime})A\hspace{-0.2cm}/(\mathbf{x}^{\prime})e^{i\mathbf{k}\cdot\mathbf{x}^{\prime}}d^{3}x^{\prime}\widehat{\psi}_{\text{out}}(\mathbf{k})d^{3}k
 \nonumber\\
 & & -\sum_{s=1}^{2}\int(2\pi)^{-\frac{3}{2}}
  e^{-i\sqrt{k^{
  2}+m^{2}}t}\int
\frac{e^{ik\mid\mathbf{x}-\mathbf{x}^{\prime}\mid}}{\mid\mathbf{x}-\mathbf{x}^{\prime}\mid}S^{+}_{k}(\mathbf{x}-\mathbf{x}^{\prime})A\hspace{-0.2cm}/(\mathbf{x}^{\prime})\zeta^{s}_{\mathbf{k}}(\mathbf{x}^{\prime})d^{3}x^{\prime}\widehat{\psi}_{out,s}(\mathbf{k})d^{3}k
  \nonumber\\
 &=:& S_{0}-S_{1}-S_{2}\;.
\end{eqnarray}
$S_{0}$ is the propagation of the free outgoing state. The free
Flux-Across-Surfaces-Theorem yields therefore:

$$\lim_{R\rightarrow\infty}\mid\int_{S}\int_{R}^{\infty}\langle
S_{0},\mathbf{\alpha}S_{0}\rangle\cdot\mathbf{n}dt
 R^{2}d\Omega-\int_{S}\int_{0}^{\infty}\langle\widehat{\psi}_{\text{out}}(\mathbf{k}),\widehat{\psi}_{\text{out}}(\mathbf{k})\rangle
 k^{2}dkd\Omega\mid=0\;.$$
Hence for (\ref{newsplit})(i) it remains to show, that (using
\ref{relflux}):

\begin{eqnarray*}
\lefteqn{\hspace{-1cm}
\lim_{R\rightarrow\infty}\int_{S}\int_{R}^{\infty}(\mathbf{j}(R,t)-\langle
S_{0},\mathbf{\boldsymbol{\alpha}}S_{0}\rangle)\cdot\mathbf{n}dt
 R^{2}d\Omega}
\\&=&\lim_{R\rightarrow\infty}\int_{S}\int_{R}^{\infty}(\langle
\sum_{j=0}^{2}S_{j},\mathbf{\boldsymbol{\alpha}}\sum_{j=0}^{2}S_{j}\rangle-\langle
S_{0},\mathbf{\boldsymbol{\alpha}}S_{0}\rangle)\cdot\mathbf{n}dt
 R^{2}d\Omega
 \\&=&\lim_{R\rightarrow\infty}\int_{S}\int_{R}^{\infty}(\langle
\psi,\mathbf{\boldsymbol{\alpha}}\sum_{j=1}^{2}S_{j}\rangle+\langle
\sum_{j=1}^{2}S_{j},\mathbf{\boldsymbol{\alpha}}\psi\rangle)\cdot\mathbf{n}dt
 R^{2}d\Omega=0\;.
\end{eqnarray*}
By Schwartz-inequality we need only show:

\begin{equation}\label{summesa}
\lim_{R\rightarrow\infty}\int_{S}\int_{R}^{\infty}
\parallel\psi\parallel_{s} \sum_{j=1}^{2}\parallel
S_{j}\parallel_{s}dt R^{2}d\Omega=0\;.
\end{equation}
We first want to estimate $\parallel S_{1}\parallel_{s}$.
Recalling (\ref{sum}) we get by Fubinis theorem:

\begin{eqnarray*}
S_{1}&=&\int(2\pi)^{-\frac{3}{2}}
  e^{-i\sqrt{k^{
  2}+m^{2}}t}\int
\frac{e^{ik\mid\mathbf{x}-\mathbf{x}^{\prime}\mid}}{\mid\mathbf{x}-\mathbf{x}^{\prime}\mid}S^{+}_{k}(\mathbf{x}-\mathbf{x}^{\prime})A\hspace{-0.2cm}/(\mathbf{x}^{\prime})e^{i\mathbf{k}\cdot\mathbf{x}^{\prime}}d^{3}x^{\prime}\widehat{\psi}_{\text{out}}(\mathbf{k})d^{3}k
\\&=&\int\int(2\pi)^{-\frac{3}{2}}
  e^{-i\sqrt{k^{
  2}+m^{2}}t}
\frac{e^{ik\mid\mathbf{x}-\mathbf{x}^{\prime}\mid}}{\mid\mathbf{x}-\mathbf{x}^{\prime}\mid}S^{+}_{k}(\mathbf{x}-\mathbf{x}^{\prime})A\hspace{-0.2cm}/(\mathbf{x}^{\prime})e^{i\mathbf{k}\cdot\mathbf{x}^{\prime}}d^{3}x^{\prime}\widehat{\psi}_{\text{out}}(\mathbf{k})d^{3}k
\\&=:&\int(2\pi)^{-\frac{3}{2}}
\frac{1}{\mid\mathbf{x}-\mathbf{x}^{\prime}\mid}\widetilde{S}_{1}(\mathbf{x},\mathbf{x}^{\prime})A\hspace{-0.2cm}/(\mathbf{x}^{\prime})d^{3}x^{\prime}
\end{eqnarray*}
where

\begin{equation}\label{innen}
\widetilde{S}_{1}(\mathbf{x},\mathbf{x}^{\prime})=\int(2\pi)^{-\frac{3}{2}}
  e^{-i\sqrt{k^{
  2}+m^{2}}t}
e^{ik\mid\mathbf{x}-\mathbf{x}^{\prime}\mid}S^{+}_{k}(\mathbf{x}-\mathbf{x}^{\prime})e^{i\mathbf{k}\cdot\mathbf{x}^{\prime}}d^{3}x^{\prime}\widehat{\psi}_{\text{out}}(\mathbf{k})d^{3}k\;.
\end{equation}
Next we use Lemma \ref{statphas}, setting:

$$\mu=t;\; a=t^{-1}\mid\mathbf{x}-\mathbf{x}^{\prime}\mid ;\;
\mathbf{y}=t^{-1}\mathbf{x}^{\prime} ;\; k^{\prime}=k;\;
\chi(\mathbf{k}^{\prime})=(2\pi)^{-\frac{3}{2}}S^{+}_{\mathbf{k}}(\mathbf{x}-\mathbf{x}^{\prime})\widehat{\psi}(\mathbf{k}^{\prime})\;.$$
With regard to (\ref{sbound}), the function

$$\widetilde{\chi}(\mathbf{k})=(2\pi)^{-\frac{3}{2}}\widetilde{S}_{\mathbf{k}}^{+}\widehat{\psi}(\mathbf{k}^{\prime})$$
satisfies the properties we need in (\ref{hoerm}). Furthermore we
observe that for the stationary point:

\begin{eqnarray*}
\frac{k_{stat}}{\sqrt{k_{stat}^{2}+m^{2}}}+a-y&=&0\nonumber\\
 k_{stat}&=&
\sqrt{k_{stat}^{2}+m^{2}}(y-a)\;.
\end{eqnarray*}
So we can estimate $k_{stat}$ by:

\begin{eqnarray}\label{kstat}
k_{stat}&=&\sqrt{k_{stat}^{2}+m^{2}}t^{-1}(x^{\prime}-\mid
\mathbf{x}-\mathbf{x}^{\prime}\mid) \leq
\sqrt{k_{stat}^{2}+m^{2}}xt^{-1}\;.
\end{eqnarray}
Hence by (\ref{hoerm}) we obtain for (\ref{innen}) that there
exists $M_{1}<\infty$, bounding in particular
$\sqrt{k_{stat}^{2}+m^{2}}\widehat{\chi}(\mathbf{k}_{stat)}$,
which is bounded by the choice of
$\widehat{\psi}_{\text{out}}\in\mathcal{G}$ and incorporating also
the constants $C_{1}$ and $C_{2}$, uniformly in $\mathbf{y}$ and $a$
so that:

\begin{eqnarray}\label{seins}
\parallel S_{1}\parallel_{s}&\leq&\parallel M_{1}t^{-\frac{3}{2}}(1+x^{\frac{1}{2}})\int
\frac{1}{\mid\mathbf{x}-\mathbf{x}^{\prime}\mid}
A\hspace{-0.2cm}/(\mathbf{x}^{\prime})d^{3}x^{\prime}\parallel_{s}
\nonumber\\&=&M_{1}t^{-\frac{3}{2}}P_{1}(\mathbf{x})\rightarrow_{x\rightarrow\infty}0\;.
\end{eqnarray}
That the function $P_{1}$ goes to zero in the limit
$x\rightarrow\infty$ may be seen as follows:\newline For any
function $f(\mathbf{x})\in L^{1}$ with
$\limsup_{x\rightarrow\infty}\mid x^{3}f(\mathbf{x})\mid<\infty$
we have:

\begin{eqnarray}\label{decayofint}
&&\lim_{x\rightarrow\infty}x\mid\int\frac{1}{\mid\mathbf{x}-\mathbf{x}^{\prime}\mid}f(\mathbf{x}^{\prime})d^{3}x^{\prime}\mid\nonumber\\&\leq&
\lim_{x\rightarrow\infty}x\int\mid\frac{1}{x^{\prime}}f(\mathbf{x}-\mathbf{x}^{\prime})\mid
d^{3}x^{\prime}
\nonumber\\&=&\lim_{x\rightarrow\infty}x\big(\int_{B(0,\frac{x}{2})}\mid\frac{1}{x^{\prime}}f(\mathbf{x}-\mathbf{x}^{\prime})\mid
d^{3}x^{\prime}+\int_{\mathbb{R}^{3}\backslash
B(0,\frac{x}{2})}\mid\frac{1}{x^{\prime}}f(\mathbf{x}-\mathbf{x}^{\prime})\mid
d^{3}x^{\prime}\big)
\nonumber\\&\leq&\lim_{x\rightarrow\infty}x\big(
\sup_{\widetilde{x}\geq\frac{x}{2}}\{\mid
f(\mathbf{\widetilde{x}})\mid\}\int_{B(0,\frac{x}{2})}\frac{1}{x^{\prime}}
d^{3}x^{\prime}+\frac{2}{x}\int_{\mathbb{R}^{3}\backslash
B(0,\frac{x}{2})}\mid f(\mathbf{x}-\mathbf{x}^{\prime})\mid
d^{3}x^{\prime}\big)
\nonumber\\&\leq&\lim_{x\rightarrow\infty}\frac{1}{8}x^{3}\sup_{\widetilde{x}\geq\frac{x}{2}}\{\mid
f(\mathbf{\widetilde{x}})\mid\}
+\lim_{x\rightarrow\infty}2\int_{\mathbb{R}^{3}\backslash
B(0,\frac{x}{2})}\mid f(\mathbf{x}-\mathbf{x}^{\prime})\mid
d^{3}x^{\prime}<\infty
\end{eqnarray}
where $B(\mathbf{a},r)$ means the ball with center $\mathbf{a}$
and radius r.

Next we estimate $\parallel S_{2}\parallel_{s}$. According to
(\ref{sum}) we can write it down as:

$$S_{2}=\sum_{s=1}^{2}\int(2\pi)^{-\frac{3}{2}}
  e^{-i\sqrt{k^{
  2}+m^{2}}t}\int
\frac{e^{ik\mid\mathbf{x}-\mathbf{x}^{\prime}\mid}}{\mid\mathbf{x}-\mathbf{x}^{\prime}\mid}S^{+}_{k}(\mathbf{x}-\mathbf{x}^{\prime})(x^{\prime}+1)A\hspace{-0.2cm}/(\mathbf{x}^{\prime})\frac{\zeta^{s}_{\mathbf{k}}(\mathbf{x}^{\prime})}{x^{\prime}+1}d^{3}x^{\prime}\widehat{\psi}_{out,s}(\mathbf{k})d^{3}k\;.$$
Therefore we again use Lemma \ref{statphas}, setting:

$$\mu=t ;\; a=t^{-1}(\mid\mathbf{x}-\mathbf{x}^{\prime}\mid) ;\;
\mathbf{y}=0 ;\; k^{\prime}=k;\;
\chi(\mathbf{k}^{\prime})=(2\pi)^{-\frac{3}{2}}\sum_{s=1}^{2}\frac{\zeta^{s}_{\mathbf{k}}(\mathbf{x}^{\prime}
)}{x^{\prime}+1}S^{+}_{\mathbf{k}}(\mathbf{x}-\mathbf{x}^{\prime})\widehat{\psi}_{out,s}(\mathbf{k}^{\prime})\;.$$
With regard to (\ref{sbound}) and Lemma \ref{properties}(c) there
exists a $M_{2}<\infty$, so that the function

$$\widetilde{\chi}=(2\pi)^{-\frac{3}{2}}M_{2}\widetilde{S}_{\mathbf{k}}^{+}\widehat{\psi}(\mathbf{k}^{\prime})$$
satisfies the properties we need in (\ref{hoerm}).

Since our phase function has no stationary point we get with
(\ref{hoerm}):

\begin{eqnarray}\label{szwei}
\parallel S_{2}\parallel_{s}\leq M_{2}t^{-2}\mid\int
\frac{1}{\mid\mathbf{x}-\mathbf{x}^{\prime}\mid}
(x^{\prime}+1)A\hspace{-0.2cm}/(\mathbf{x}^{\prime})d^{3}x^{\prime}\mid
\nonumber=M_{2}t^{-2}P_{2}(\mathbf{x})_{\overrightarrow{x\rightarrow\infty}}0\;.
\end{eqnarray}
Choosing $(x^{\prime}+1)A\hspace{-0.2cm}/(\mathbf{x}^{\prime})$
for $f$ in the most left side of (\ref{decayofint}), one can see,
that $xP_{2}(\mathbf{x})$ is bounded, so $P_{2}$ goes to zero in
the limit $x\rightarrow\infty$. Since $S_{0}$ is the analogue of
the freely evolving wavefunction, we have by Corollary
\ref{scintocones}:

\begin{equation}\label{snull}
\parallel S_{0}\parallel_{s}\leq
M_{0}t^{-\frac{3}{2}}\;.
\end{equation}
We use the estimates (\ref{snull}), (\ref{seins}) and
(\ref{szwei}) in the right side of (\ref{summesa}) and get,
defining $M:=M_{0}+M_{1}+M_{2}$:

\begin{eqnarray*}
\lefteqn{\hspace{-1cm}\lim_{R\rightarrow\infty}\mid\int_{S}\int_{R}^{\infty}
(\parallel\psi\parallel_{s}\parallel\sum_{j=1}^{2}S_{j}\parallel_{s})dt
R^{2}d\Omega\mid}\\&\leq&\lim_{R\rightarrow\infty}\int_{R}^{\infty}M^{2}(P_{1}(\mathbf{R})+P_{2}(\mathbf{R}))t^{-3}dt
R^{2}\leq\lim_{R\rightarrow\infty}3M^{2}(P_{1}(\mathbf{R})+P_{2}(\mathbf{R}))=0
\end{eqnarray*}
and (\ref{summesa}) is proved.

Like in the free case, (\ref{newsplit} ii) follows directly from
an analogous argument which used equation (\ref{spacelike}), thus we prove (\ref{spacelike}) for the case at hand. Since in
(\ref{newsplit} ii) we need only estimates of the wavefunction for
times $t\leq x$ we have in view of (\ref{sum}), setting $t=\eta x$
with $0\leq\eta\leq 1$ and using Fubinis Theorem:

\begin{eqnarray*}
\psi(\mathbf{x},\eta x)&=&\int(2\pi)^{-\frac{3}{2}}
e^{-i\sqrt{k^{2}+m^{2}}\eta
x+i\mathbf{k}\cdot\mathbf{x}}\widehat{\psi}_{\text{out}}(\mathbf{k})d^{3}k\\&&-\int\int
e^{-i\sqrt{k^{2}+m^{2}}\eta
x+ik\mid\mathbf{x}-\mathbf{x}^{\prime}\mid+i\mathbf{k}\cdot\mathbf{x}^{\prime}}\frac{A\hspace{-0.2cm}/(\mathbf{x}^{\prime})S^{+}_{\mathbf{k}}(\mathbf{x}-\mathbf{x}^{\prime})\widehat{\psi}_{\text{out}}(\mathbf{k})}{(2\pi)^{\frac{3}{2}}\mid\mathbf{x}-\mathbf{x}^{\prime}\mid}d^{3}kd^{3}x^{\prime}
\\&&-\sum_{s=1}^{2}\int\int e^{-i\sqrt{k^{2}+m^{2}}\eta
x+ik\mid\mathbf{x}-\mathbf{x}^{\prime}\mid}\frac{A\hspace{-0.2cm}/(\mathbf{x}^{\prime})\zeta^{s}_{\mathbf{k}}(\mathbf{x}^{\prime})S^{+}_{\mathbf{k}}(\mathbf{x}-\mathbf{x}^{\prime})\widehat{\psi}_{out,s}(\mathbf{k})}{(2\pi)^{\frac{3}{2}}\mid\mathbf{x}-\mathbf{x}^{\prime}\mid}d^{3}kd^{3}x^{\prime}
\\&=:&S_{0}-S_{1}-S_{2}\;.
\end{eqnarray*}
For $S_{0}$ we have (\ref{spacelike}), for the other summands we define:
\begin{eqnarray*}
\widetilde{S}_{1}&:=&\int(2\pi)^{-\frac{3}{2}}
e^{-i\sqrt{k^{2}+m^{2}}\eta
x+ik\mid\mathbf{x}-\mathbf{x}^{\prime}\mid+i\mathbf{k}\cdot\mathbf{x}^{\prime}}S^{+}_{\mathbf{k}}(\mathbf{x}-\mathbf{x}^{\prime})\widehat{\psi}_{\text{out}}(\mathbf{k})d^{3}k
\\\widetilde{S}_{2}&:=&\sum_{s=1}^{2}\int(2\pi)^{-\frac{3}{2}} e^{-i\sqrt{k^{2}+m^{2}}\eta
x+ik\mid\mathbf{x}-\mathbf{x}^{\prime}\mid+i\mathbf{k}\cdot\mathbf{x}^{\prime}}e^{-i\mathbf{k}\cdot\mathbf{x}^{\prime}}\zeta^{s}_{\mathbf{k}}(\mathbf{x}^{\prime})S^{+}_{\mathbf{k}}(\mathbf{x}-\mathbf{x}^{\prime})\widehat{\psi}_{out,s}(\mathbf{k})d^{3}k\;.
\end{eqnarray*}
So we have for $S_{j}$, j=1;2:

$$
S_{j}=\int\widetilde{S}_{j}\frac{A\hspace{-0.2cm}/(\mathbf{x}^{\prime})}{\mid\mathbf{x}-\mathbf{x}^{\prime}\mid}d^{3}x^{\prime}\;.$$
We can estimate the $\widetilde{S}_{j}$ by two partial
integrations. One can easily see, that the phase functions of
$\widetilde{S}_{j}$ have no stationary point. This leads to:

\begin{eqnarray*}
\parallel\widetilde{S}_{j}\parallel_{s}&=&\parallel\int(2\pi)^{-\frac{3}{2}}
e^{-ixg(\mathbf{k})}\chi_{j}(\mathbf{x},\mathbf{x}^{\prime},\mathbf{k})d^{3}k\parallel_{s}
\\&=&\frac{1}{x^{2}}\parallel\int(2\pi)^{-\frac{3}{2}}
e^{-ixg(\mathbf{k})}\partial_{k_{1}}(\frac{1}{g^{\prime}}\partial_{k_{1}}\frac{\chi_{j}}{g^{\prime}})d^{3}k\parallel_{s}
\\&=&\frac{1}{x^{2}}\parallel\int(2\pi)^{-\frac{3}{2}}
(\frac{\chi_{j}^{\prime\prime}}{g^{\prime2}}-\frac{3\chi_{j}^{\prime}g^{\prime\prime}}{g^{\prime
3}}+\frac{3\chi_{j}
g^{\prime\prime2}}{g^{\prime4}})d^{3}k\parallel_{s}
\end{eqnarray*}
where

\begin{eqnarray*}
g(\mathbf{k})&:=&\sqrt{k^{2}+m^{2}}\eta
-k\frac{\mid\mathbf{x}-\mathbf{x}^{\prime}\mid}{x}-\mathbf{k}\cdot\frac{\mathbf{x}^{\prime}}{x}\\
\chi_{1}(\mathbf{x},\mathbf{x}^{\prime},\mathbf{k})&:=&S^{+}_{\mathbf{k}}(\mathbf{x}-\mathbf{x}^{\prime})\widehat{\psi}_{\text{out}}(\mathbf{k})
\\\chi_{2}(\mathbf{x},\mathbf{x}^{\prime},\mathbf{k})&:=&\sum_{s=1}^{2}e^{-i\mathbf{k}\cdot\mathbf{x}^{\prime}}\zeta^{s}_{\mathbf{k}}(\mathbf{x}^{\prime})S^{+}_{\mathbf{k}}(\mathbf{x}-\mathbf{x}^{\prime})\widehat{\psi}_{out,s}(\mathbf{k})
\\g^{\prime}&:=&\partial_{k_{1}}g\;.
\end{eqnarray*}
Since

\begin{eqnarray*}
\mid
g^{\prime}\mid&=&\frac{x^{\prime}}{x}+\frac{k_{1}\mid\mathbf{x}-\mathbf{x}^{\prime}\mid}{kx}-\frac{k_{1}\eta}{\sqrt{k^{2}+m^{2}}}\geq\frac{k_{1}}{k}(\frac{x^{\prime}}{x}+\frac{\mid\mathbf{x}-\mathbf{x}^{\prime}\mid}{x}-\frac{k}{\sqrt{k^{2}+m^{2}}})
\\&\geq&\frac{k_{1}}{k}(1-\frac{k}{\sqrt{k^{2}+m^{2}}})>0\;.
\end{eqnarray*}
$g^{\prime\prime}$ is bounded and due to Lemma \ref{properties}
the $\chi_{j}$ are bounded, we can find $C_{2}<\infty$ with:

$$\sum_{j=1}^{2}\widetilde{S}_{j}\leq \frac{C_{2}}{x^{2}}\;.$$ So
$x^{2}\sum_{j=1}^{2}S_{j}$ is bounded (see \ref{decayofint}) and the analogue of
(\ref{spacelike}) is proved.

\newpage

\section{Appendix}

\subsection{Proof of Lemma(\ref{statphas})}\label{appendix1}

We consider for a family of phase functions g, which we should
think of being indexed by $a\geq0,\mathbf{y}$:

$$g(\mathbf{k})=\sqrt{k^{2}+m^{2}}+a\mid
k\mid-\mathbf{y}\cdot\mathbf{k}$$ the integral

$$I:=\int e^{-i\mu g(\mathbf{k})}\chi(\mathbf{k})d^{3}k$$ where
$\chi\in\mathcal{G}$ (see\ref{definitionG}).

We shall find its asymptotic behavior as a function of $\mu$. In
major parts we will recall the proof of theorem 7.7.5 in the book
of H\"ormander \cite{hoer}, which unfortunately is formulated for
compactly supported $\chi$ and which moreover does not give
uniformity over the family, i.e. uniformity in $a,\mathbf{y}$
which we need. The compactness can easily be handled but for the
uniformity we must invoke the special form of the family of phase
functions $g$ and we shall give the argument here.

The stationary points of the phase functions are given by:

\begin{eqnarray}\label{kstat2}
g^{\prime}({\mathbf{k}}_{stat})&=&\frac{\mathbf{k}_{stat}}{\sqrt{k_{stat}^{2}+m^{2}}}
+a\frac{\mathbf{k}_{stat}}{k_{stat}}-\mathbf{y}=0 \nonumber\\
k^{2}_{stat}&=&(k_{stat}^{2}+m^{2})(y-a)^{2} \nonumber\\
k_{stat}&=&\frac{m(y-a)}{\sqrt{1-(y-a)^{2}}}
\nonumber\\\mathbf{k}_{stat}&\parallel&\mathbf{y}\;.
\end{eqnarray}
Since $k_{stat}$ is a function of $a$ and $\mathbf{y}$, we sometimes use the phrase: uniform in $k_{stat}$ to express uniformity in $a$ and $\mathbf{y}$.
\bigskip
(I) For $y\geq a+1$ there is no stationary point and for $y=a$ the
stationary point is at $k_{stat}=0$.

First we handle the family where $y \in [a+\frac{1}{2};a+1[$.
These phase-functions do exactly have one stationary point bounded
away from zero:

\begin{equation}\label{ungleichungk}
k_{stat}=\frac{m(y-a)}{\sqrt{1-(y-a)^{2}}}\geq\frac{m}{\sqrt{3}}
\end{equation}
Later we will handle phase functions, where the stationary point
is close to zero and phase-functions without stationary point.

We choose a coordinate system, where the $k_{1}$-direction is
parallel to $\mathbf{y}$. So the stationary points will have the
coordinates $(k_{stat},0,0)$. To estimate the integral, we
separate from the integral the contribution coming from near the
stationary point. This part of integral includes the leading term.
Therefore we define a  smooth function $\rho_{k_{stat}}$ which is
one near the stationary points and zero away from the stationary
point. (We shall omit further on for ease of notation the index
$k_{stat}$).

More precisely we define the compact set Q by:

$$\mathbf{k}\in Q\Leftrightarrow
k_{1}\in[\frac{k_{stat}}{2},2k_{stat}] \wedge
k_{2},k_{3}\in[-1,1]$$ and choose

\begin{equation}\label{rho}
\rho(\mathbf{k}):=1 \hspace{1cm}\forall \mathbf{k}\in Q
\end{equation}
falling quickly off to zero outside of Q, lets say

\begin{equation}\label{qepsilon}
\rho(\mathbf{k}):=0 \hspace{1cm}\forall k\notin Q_{\varepsilon}
\end{equation}
where $Q_{\varepsilon}$ is some $\varepsilon$-neighborhood of $Q$ for some $\varepsilon>0$.
With the help of $\rho$ we can split $\chi=\chi_{1}+\chi_{2}$ by
defining:

\begin{equation}\label{chi}
\begin{tabular}{ll}
  $\chi_{1}:=\rho\chi\nonumber$ & \hspace{1cm}$\chi_{2}:=(1-\rho)\chi$ \\
  $I_{1}:=\int e^{-i\mu
 g(\mathbf{k})}\chi_{1}(\mathbf{k})d^{3}k\nonumber$ & \hspace{1cm}$I_{2}:=\int e^{-i\mu
 g(\mathbf{k})}\chi_{2}(\mathbf{k})d^{3}k\;.$ \\
\end{tabular}
\end{equation}
This split has the following advantages:

The compactly supported $\chi_{1}$ includes the stationary point,
so $I_{1}$ can be estimated the same way as in H\"ormanders
theorem, but with focus on the uniformity of the estimates.
$\chi_{2}$ is zero near the stationary point, so $I_{2}$ can be
easily estimated by partial integrations. $\rho$ has been defined
in such a way, that we may estimate the terms we get by the
partial integrations uniform in $k_{stat}.$

We start with $I_{1}$. We move the stationary point to the center
of our coordinate system setting
$\mathbf{k}^{\prime}:=\mathbf{k}-\mathbf{k}_{stat}$, i.e.
$g(\mathbf{k})$ becomes
$\widetilde{g}(\mathbf{k}^{\prime})=g(\mathbf{k}^{\prime}+\mathbf{k}_{stat})$.
Slightly abusing notation we simply write $g(\mathbf{k}^{\prime})$
for $\widetilde{g}$. By Taylor's formula we obtain a function
$f$:

\begin{equation}\label{taylorg}
g(\mathbf{k}^{\prime})=g(\mathbf{k}^{\prime}=0)+\sum_{\mid\gamma\mid=2}\frac{D_{\mathbf{k}^{\prime}}^{\gamma}g(\mathbf{k}^{\prime}
=0)\mathbf{k}^{\prime\gamma}}{\gamma!}+f(\mathbf{k}^{\prime})\;,
\end{equation}
where $\frac{f(\mathbf{k}^{\prime})}{k^{\prime 3}}$ bounded.

Computing the second-order terms of $g(\mathbf{k}^{\prime})$ we
find that only diagonal terms survive at $(k_{stat},0,0)$ and

\begin{eqnarray}\label{partialjj}
\partial_{k^{\prime}_{j}}^{2}g(\mathbf{k}^{\prime}=0)
&=&\partial_{k_{j}}^{2}g(\mathbf{k}=\mathbf{k}_{stat})
\nonumber\\&=&(\partial_{k_{j}}(\frac{k_{j}}{\sqrt{k^{2}+m^{2}}}
+a\frac{k_{j}}{k}-y_{l}))\mid_{\mathbf{k}=\mathbf{k}_{stat}}
\nonumber\\&=&(\frac{k^{2}-k_{j}^{2}+m^{2}}{\sqrt{k^{2}+m^{2}}^{3}}
+a\frac{k^{2}-k_{j}^{2}}{k^{3}})\mid_{\mathbf{k}=\mathbf{k}_{stat}}
\end{eqnarray}
so that
\begin{eqnarray}\label{d2g}
\partial_{k^{\prime}_{j}}^{2}g(\mathbf{k}^{\prime}=0)
&=&\frac{k_{stat}^{2}+m^{2}}{\sqrt{k_{stat}^{2}+m^{2}}^{3}}
+a\frac{1}{k_{stat}}\,\,\mbox{for}\,  j=2,3 \nonumber\\
\partial_{k^{\prime}_{1}}^{2}g(\mathbf{k}^{\prime}=0)
&=&\frac{m^{2}}{\sqrt{k_{stat}^{2}+m^{2}}^{3}}\;.
\end{eqnarray}
We define:
\begin{equation}\label{triangle}
g_{2}(\vartheta,\theta):=
\frac{\sum_{j=1}^{3}\partial_{k^{\prime}_{j}}^{2}g(\mathbf{k}^{\prime}=0
)k^{\prime2}_{j}}{k^{\prime2}}\;.
\end{equation}
By this definition, $g_{2}$ does only depend on the angular, not
on the radial coordinate of $\mathbf{k}^{\prime}$. Using
(\ref{triangle}) in (\ref{taylorg}), we may write

\begin{equation}\label{gkstrich}
g(\mathbf{k}^{\prime})=g(0)+\frac{1}{2}k^{\prime2}g_{2}(\vartheta,\theta)
+f(\mathbf{k'})\;.
\end{equation}
Furthermore for $s\in[0,1]$ set:

\begin{equation}\label{gs}
g_{s}:=g(0)+\frac{1}{2}k^{\prime2}g_{2}(\vartheta,\theta)
+sf(\mathbf{k'})
\end{equation}
and

$$I(s)=\int e^{-i\mu
g_{s}(\mathbf{k}^{\prime})}\chi_{1}(\mathbf{k}^{\prime})d^{3}k^{\prime}\;.$$
Note that $g=g_{1}$, $I_{1}=I(1)$. By Taylor's Formula there exits
$\xi\leq 1$ so that:

\begin{equation}\label{xx}
I_{1}=I(1)=I(0)+\partial_{s}I(s)\mid_{\xi}\;.
\end{equation}
We begin with $I(0)$, introducing spherical coordinates. With
slight abuse of notation: (leaving the notation for the functions
unchanged)

$$I(0)=\int
e^{-i\mu(g(0)+\frac{1}{2}k^{\prime2}g_{2}(\vartheta,\theta))
}\chi_{1}(k^{\prime},\vartheta,\theta)k^{\prime2}dk^{\prime}d\Omega\;.$$
Writing $\chi_{1}=\chi(k^{\prime}=0)+\widetilde{\chi}$ the
integral splits into:

\begin{eqnarray}\label{xxx}
I(0)&=&\int
e^{-i\mu(g(0)+\frac{1}{2}k^{\prime2}g_{2}(\vartheta,\theta) )
}\chi(k^{\prime}=0)k^{\prime2}dk^{\prime}d\Omega\nonumber\\&+&\int
e^{-i\mu(g(0)+\frac{1}{2}k^{\prime2}g_{2}(\vartheta,\theta))
}\widetilde{\chi}(k^{\prime},\vartheta,\theta)k^{\prime2}dk^{\prime}d\Omega=:I_{1}^{1}+I_{1}^{2}\;.
\end{eqnarray}

The integral $I_{1}^{1}$ is a gaussian integral, which includes
the leading term:

\begin{eqnarray}\label{I11}
I_{1}^{1}&=&\int
e^{-i\mu(g(0)+\frac{1}{2}k^{\prime2}g_{2}(\vartheta,\theta) )
}\chi(k^{\prime}=0)k^{\prime2}dk^{\prime}d\Omega
\nonumber\\&=&\int
e^{-i\mu\sum_{j=1}^{3}\frac{1}{2}\partial_{k^{\prime2}_{j}}g(\mathbf{k}^{\prime}=0))k_{j}^{2}}e^{-i\mu
g(0) }\chi(\mathbf{k}^{\prime}=0)k^{\prime2}d^{3}k^{\prime}
\nonumber\\&=&(2\pi)\frac{3}{2}\mu^{-\frac{3}{2}}e^{-i\mu g(0)
}(\prod_{j=1}^{3}\partial^{2}_{k^{\prime}_{j}}g(\mathbf{k}^{\prime}=0))^{-\frac{1}{2}}\chi(\mathbf{k}_{stat})\;.
\end{eqnarray}
For $a=0$ the $\partial_{k^{\prime2}_{j}}g(\mathbf{k}^{\prime}=0)$
terms can be easily calculated. We get:

\begin{eqnarray*}
\partial^{2}_{k^{\prime}_{j}}g(\mathbf{k}^{\prime}=0)&=&\partial^{2}_{k^{\prime}_{j}}g(\mathbf{k}=\mathbf{k}_{stat})
\\&=&\partial_{k^{\prime}_{j}}\frac{k_{j}}{\sqrt{k^{2}+m^{2}}}\mid_{\mathbf{k}=\mathbf{k}_{stat}}=\frac{k^{2}+m^{2}-k_{j}^{2}}{\sqrt{k^{2}+m^{2}}^{3}}\mid_{\mathbf{k}=\mathbf{k}_{stat}}\;.
\end{eqnarray*}
So we get:

$$\prod_{j=1}^{3}\partial^{2}_{k^{\prime}_{j}}g(\mathbf{k}^{\prime}=0)=\frac{m^{2}(k_{stat}^{2}+m^{2})^{2}}{\sqrt{k_{stat}^{2}+m^{2}}^{9}}=\frac{m^{2}}{\sqrt{k_{stat}^{2}+m^{2}}^{5}}\;.$$
$I_{1}^{1}$ is the leading term of our integral. For $a=0$ we get
the desired value for $C_{1}$ (\ref{statphas}).

For $I_{1}^{2}$ put:

$$\phi(k^{\prime},\vartheta,\theta):=\widetilde{\chi}(k^{\prime},\vartheta,\theta)k^{\prime-1}$$
which is bounded and smooth.

\begin{equation}\label{ieinszwei}
I_{1}^{2}=\int
e^{-i\mu(g(0)+\frac{1}{2}k^{\prime2}g_{2}(\vartheta,\theta) )
}\phi(k^{\prime},\vartheta,\theta)k^{\prime3}dk^{\prime}d\Omega\;.
\end{equation}
One partial integration leads to:
\begin{eqnarray*}
\parallel I_{1}^{2}\parallel_{s}&=&\mu^{-1}\parallel\int
e^{-i\mu\frac{1}{2}k^{\prime2}g_{2}(\vartheta,\theta)
}\partial_{k^{\prime}}\frac{\phi(k^{\prime},\vartheta,\theta)k^{\prime3}}{k^{\prime}g_{2}(\vartheta,\theta)
}dk^{\prime}d\Omega\parallel_{s}\\&=&\mu^{-1}\parallel\int
e^{-i\mu\frac{1}{2}k^{\prime2}g_{2}(\vartheta,\theta)
}\frac{\partial_{k^{\prime}}\phi(k^{\prime},\vartheta,\theta)k^{\prime2}+2\phi(k^{\prime},\vartheta,\theta)k^{\prime}}{g_{2}(\vartheta,\theta)
}dk^{\prime}d\Omega\parallel_{s}\;.
\end{eqnarray*}
So another partial integration is possible:

\begin{eqnarray}\label{i12}
\parallel I_{1}^{2}\parallel_{s}&=&\mu^{-2}\parallel\int
e^{-i\mu\frac{1}{2}k^{\prime2}g_{2}(\vartheta,\theta)
}\partial_{k^{\prime}}(\frac{\partial_{k^{\prime}}\phi(k^{\prime},\vartheta,\theta)k^{\prime2}+2\phi(k^{\prime},\vartheta,\theta)k^{\prime}}{
k^{\prime}(g_{2}(\vartheta,\theta))^{2}})dk^{\prime}d\Omega\parallel_{s}\nonumber\\&=&\mu^{-2}\parallel\int
e^{-i\mu\frac{1}{2}k^{\prime2}g_{2}(\vartheta,\theta)
}\partial_{k^{\prime}}(\frac{\partial_{k^{\prime}}\phi(k^{\prime},\vartheta,\theta)k^{\prime}+2\phi(k^{\prime},\vartheta,\theta)}{
(g_{2}(\vartheta,\theta))^{2}})dk^{\prime}d\Omega\parallel_{s}
\nonumber\\&\leq&\mu^{-2}\parallel\int
\partial_{k^{\prime}}(\frac{\partial_{k^{\prime}}\phi(k^{\prime},\vartheta,\theta)k^{\prime}+2\phi(k^{\prime},\vartheta,\theta)}{
(g_{2}(\vartheta,\theta))^{2}})dk^{\prime}d\Omega\parallel_{s}\;.
\end{eqnarray}
With our definition of Q, the support of the integrand increases
and $g_{2}(\vartheta,\theta)$ decreases polynomially with
$k_{stat}$ (see (\ref{d2g}) and (\ref{triangle})). While the
support moves away from the center of our coordinate system. But
$\widetilde{\chi}=\chi-\chi(\mathbf{k}_{stat})$ and its
derivatives decay faster in $k_{stat}$ than any power, so we get a
constant C uniform in $k_{stat}$ with:

$$I_{1}^{2}\leq \mu^{-2}C\;.$$ For $I_{1}$ it is left to estimate
$\partial_{s}I(s)\mid_{\xi}$:

\begin{equation}\label{partialsi}
\partial_{s}I(s)\mid_{\xi}=\int -i\mu
f(k^{\prime},\vartheta,\theta) e^{-i\mu
g_{\xi}(k^{\prime},\vartheta,\theta)}\chi_{1}(k^{\prime},\vartheta,\theta)k^{\prime2}dk^{\prime}d\Omega\;.
\end{equation}
By Taylor's formula we can define:

\begin{eqnarray}\label{g-tilde}
\widetilde{f}(k^{\prime},\vartheta,\theta):=f(k^{\prime},\vartheta,\theta)k^{\prime-3}
\nonumber\hspace{1cm}\widetilde{g}(k^{\prime},\vartheta,\theta):=k^{\prime-1}\partial_{k^{\prime}}g_{\xi}(k^{\prime},\vartheta,\theta)
\end{eqnarray}
and thus:

\begin{equation}\label{50a}
\partial_{s}I(s)\mid_{\xi}=\int -i\mu
\widetilde{f}(k^{\prime},\vartheta,\theta) e^{-i\mu
g_{\xi}(k^{\prime},\vartheta,\theta)}\chi_{1}(k^{\prime},\vartheta,\theta)k^{\prime5}dk^{\prime}d\Omega\;.
\end{equation}
On $Q_{\varepsilon}$ (see below (\ref{qepsilon}), g is infinitely
often differentiable. So these functions are well defined and
bounded on $Q_{\varepsilon}$.

To estimate the integral by partial integrations we have to
assure, that $g_{\xi}$ has only one stationary point, which is
$k_{stat}=0$ as one easily sees from (\ref{next}).



By (\ref{gs}):

\begin{equation}\label{next}
g_{\xi}=g(k^{\prime}=0)+\frac{1}{2}k^{\prime2}g_{2}(\vartheta,\theta)
+\xi f(\mathbf{k'})=\xi
g+(1-\xi)\left(g(k^{\prime}=0)+\frac{1}{2}k^{\prime2}g_{2}(\vartheta,\theta)\right)\;.
\end{equation}
Looking at

$$\partial_{k^{\prime}}^{2}g_{\xi}=\xi\partial_{k^{\prime}}^{2}g+(1-\xi)\partial_{k^{\prime}}^{2}\frac{1}{2}k^{\prime2}g_{2}$$
we
observe, that

\begin{eqnarray*}
\partial_{k^{\prime}}^{2}g&=&\partial_{k^{\prime}}^{2}(\sqrt{k^{\prime2}-2k^{\prime}k_{stat}\cos(\vartheta)+k_{stat}^{2}+m^{2}}+a\sqrt{k^{\prime2}-2k^{\prime}k_{stat}\cos(\vartheta)+k_{stat}^{2}}-\mathbf{y}\cdot\mathbf{k}^{\prime})
\\&=&\partial_{k^{\prime}}(\frac{k^{\prime}-k_{stat}\cos(\vartheta)}{\sqrt{k^{\prime2}-2k^{\prime}k_{stat}\cos(\vartheta)+k_{stat}^{2}+m^{2}}}+a\frac{k^{\prime}-k_{stat}\cos(\vartheta)}{\sqrt{k^{\prime2}-2k^{\prime}k_{stat}\cos(\vartheta)+k_{stat}^{2}}})
\\&=&\frac{(1-\cos(\vartheta)^{2})k_{stat}^{2}+m^{2}}{\sqrt{k^{\prime2}-2k^{\prime}k_{stat}\cos(\vartheta)+k_{stat}^{2}+m^{2}}^{3}}+a\frac{(1-\cos(\vartheta)^{2})k_{stat}^{2}}{\sqrt{k^{\prime2}-2k^{\prime}k_{stat}\cos(\vartheta)+k_{stat}^{2}}^{3}}>0\;.
\end{eqnarray*}
And for $\mathbf{k}\in Q_{\varepsilon}$, $k_{1}$ is positive, so
the angular component
$\vartheta\in]-\frac{\pi}{2},\frac{\pi}{2}[$, we also have, that
on $Q_{\varepsilon}$ also $g_{2}$ is positive. Since $\xi\in[0;1]$
it follows, that $\partial_{k^{\prime}}^{2}g_{\xi}$ is positive,
so $\partial_{k^{\prime}}g_{\xi}$ is strictly monotonous on
$Q_{\varepsilon}$ and has only one stationary point. Recalling the
definition of $\widetilde{g}$ (see \ref{g-tilde}) we see, that
$\widetilde{g}$ is bounded away from zero.

Now we can estimate the integral (\ref{50a}). By partial
integration

\begin{eqnarray*}
\partial_{s}I(s)\mid_{\xi}&=&\int e^{-i\mu
g_{\xi}(k^{\prime},\vartheta,\theta)}
\partial_{k^{\prime}}\frac{\widetilde{f}(k^{\prime},\vartheta,\theta)
\chi_{1}(k^{\prime},\vartheta,\theta)k^{\prime4}}{\widetilde{g}(k^{\prime},\vartheta,\theta)}dk^{\prime}d\Omega\\&=&
\int e^{-i\mu g_{\xi}(k^{\prime},\vartheta,\theta)}(
\partial_{k^{\prime}}\frac{\widetilde{f}(k^{\prime},\vartheta,\theta)
\chi_{1}(k^{\prime},\vartheta,\theta)}{\widetilde{g}(k^{\prime},\vartheta,\theta)}k^{\prime4}+4\frac{\widetilde{f}(k^{\prime},\vartheta,\theta)
\chi_{1}(k^{\prime},\vartheta,\theta)}{\widetilde{g}(k^{\prime},\vartheta,\theta)})k^{\prime3}dk^{\prime}d\Omega\;.
\end{eqnarray*}
Setting

\begin{equation}\label{psitilde}
\widetilde{\psi}(k^{\prime},\vartheta,\theta):=\partial_{k^{\prime}}\frac{\widetilde{f}(k^{\prime},\vartheta,\theta)
\chi_{1}(k^{\prime},\vartheta,\theta)}{\widetilde{g}(k^{\prime},\vartheta,\theta)}k^{\prime}+4\frac{\widetilde{f}(k^{\prime},\vartheta,\theta)
\chi_{1}(k^{\prime},\vartheta,\theta)}{\widetilde{g}(k^{\prime},\vartheta,\theta)}\;.
\end{equation}
Hence

$$\partial_{s}I(s)\mid_{\xi}=\int e^{-i\mu
g_{\xi}(k^{\prime},\vartheta,\theta)}\widetilde{\psi}(k^{\prime},\vartheta,\theta)k^{\prime3}dk^{\prime}d\Omega\;.$$
This term is similar to (\ref{ieinszwei}). The only differences
are, that we have $\widetilde{\psi}$ instead of $\phi$ and
$g_{\xi}$ instead of $g_{0}$.

So with the same estimate as in (\ref{ieinszwei}) we get:

\begin{equation}\label{dsi}
\parallel\partial_{s}I(s)\mid_{\xi}\parallel_{s}\leq\mu^{-2}\parallel\int
\partial_{k^{\prime}}\frac{\partial_{k^{\prime}}\widetilde{\psi}(k^{\prime},\vartheta,\theta)k^{\prime}+2\widetilde{\psi}(k^{\prime},\vartheta,\theta)}{
(\widetilde{g}(\mathbf{k}^{\prime},\vartheta,\theta))^{2}}dk^{\prime}d\Omega\parallel_{s}\;.
\end{equation}
This term again has uniform bound in $k_{stat}$, as its support
moves away from the center of the coordinate system. So we get a
constant C uniform in $k_{stat}$ with:

$$\parallel\partial_{s}I(s)\mid_{\xi}\parallel_{s}\leq
\mu^{-2}C\;.$$

Now we estimate $I(2)$ (\ref{chi}). As this integral includes no
stationary point, two partial integrations are possible without
any problem, but we have to assure, that we can estimate the
factors we get by these partial integrations uniform in
$k_{stat}$. To be able to find an uniform estimate, we estimate
the areas of $\chi$ separately.

So we again split our integral:

\begin{eqnarray*}
I_{2}&=&\int_{k_{1}<\frac{k_{stat}}{2}} e^{-i\mu
g(\mathbf{k})}\chi_{2}(\mathbf{k})d^{3}k+\int_{k_{1}>2k_{stat}}
e^{-i\mu
g(\mathbf{k})}\chi_{2}(\mathbf{k})d^{3}k\\&&+\int_{k_{1}\in B;\mid
k_{2}\mid>1} e^{-i\mu
g(\mathbf{k})}\chi_{2}(\mathbf{k})d^{3}k+\int_{k_{1}\in B;\mid
k_{2}\mid<1;\mid k_{3}\mid>1} e^{-i\mu
g(\mathbf{k})}\chi_{2}(\mathbf{k})d^{3}k\\&=:&I_{2}^{1}+I_{2}^{2}+I_{2}^{3}+I_{2}^{4}
\end{eqnarray*}
where $B:=[\frac{k_{stat}}{2};2k_{stat}]$.

\smallskip

The integrals $I_{2}^{1}$ and $I_{2}^{2}$ we estimate by two
partial integrations under the $k_{1}$-integral. This leads to:

\begin{eqnarray}\label{kurz}
\parallel
I_{2}^{1}\parallel_{s}&\leq&\mu^{-2}\int_{k_{1}<\frac{k_{stat}}{2}}\parallel\partial_{k_{1}}(\frac{1}{\dot{g}(\mathbf{k})}\partial_{k_{1}}\frac{\chi_{2}(\mathbf{k})}{\dot{g}(\mathbf{k})})\parallel_{s}
d^{3}k
\nonumber\\&=&\mu^{-2}\int_{k_{1}<\frac{k_{stat}}{2}}\parallel3\frac{\ddot{\chi_{2}}}{\dot{g}^{2}}+3\frac{\chi_{2}\ddot{g}^{2}}{\dot{g}^{4}}-3\frac{\dot{\chi_{2}}\ddot{g}}{\dot{g}^{3}}\parallel_{s}
d^{3}k\nonumber\\ \parallel
I_{2}^{2}\parallel_{s}&\leq&\mu^{-2}\int_{k_{1}>2k_{stat}}\parallel\partial_{k}(\frac{1}{g^{\prime}(\mathbf{k})}\partial_{k}\frac{\chi_{2}(\mathbf{k})}{g^{\prime}(\mathbf{k})})\parallel_{s}
d^{3}k
\nonumber\\&=&\mu^{-2}\int_{k_{1}>2k_{stat}}\parallel3\frac{\chi_{2}^{\prime\prime}}{g^{\prime2}}+3\frac{\chi_{2}g^{\prime\prime2}}{g^{\prime4}}-3\frac{\chi_{2}^{\prime}g^{\prime\prime}}{g^{\prime3}}\parallel_{s}
d^{3}k
\end{eqnarray}
where $\dot{g}(\mathbf{k}):=\partial_{k_{1}}g(\mathbf{k})$;
$g^{\prime}(\mathbf{k}):=\partial_{k}g(\mathbf{k})$

At first sight these estimates do not seem to be uniform in a and
$\mathbf{y}$. In fact

$$\ddot{g}(\mathbf{k})=\frac{m^{2}}{\sqrt{k^{2}+m^{2}}^{3}}+a\frac{k_{2}^{2}+k_{3}^{2}}{k^{3}}$$
and

$$g^{\prime\prime}(\mathbf{k})=\frac{m^{2}}{\sqrt{k^{2}+m^{2}}^{3}}$$
are
bounded on the area of integration. So it is left to show, that we
can find functions $h_{j}$ with j=1;2, which do not depend on a
and $\mathbf{y}$ and which is bounded away from zero on
$\R^{3}\backslash Q$ with

\begin{eqnarray*}
h_{1}(\mathbf{k})\leq g^{\prime}(\mathbf{k})
\hspace{1cm}h_{2}(\mathbf{k})\leq\dot{g}(\mathbf{k})
\end{eqnarray*}
for all $a$, $\mathbf{y}$, $\mathbf{k}$.

For this we estimate $\dot{g}$ for $k_{1}\leq \frac{k_{stat}}{2}$.
As $\ddot{g}>0$, it follows, that (see (\ref{xxx}))

$$\mid
\dot{g}(\mathbf{k})\mid=y-\frac{k_{1}}{\sqrt{k^{2}+m^{2}}}-a\frac{k_{1}}{k}\geq\frac{1}{2}\;.
$$ For $k_{1}<0$ and by virtue $y\geq
a+\frac{1}{2}\geq\frac{1}{2}$.

For $k_{1}>0$ we estimate, using that
$y-a-\frac{k_{stat}}{\sqrt{k_{stat}^{2}+m^{2}}}=0$:

\begin{eqnarray*}
\mid
\dot{g}(\mathbf{k})\mid&=&y-\frac{k_{1}}{\sqrt{k^{2}+m^{2}}}-a\frac{k_{1}}{k}
\geq y-a-\frac{k_{1}}{\sqrt{k^{2}+m^{2}}}
\\&\geq&y-a-\frac{k_{stat}}{\sqrt{k_{stat}^{2}+m^{2}}}+\frac{k_{stat}}{\sqrt{k_{stat}^{2}+m^{2}}}-\frac{k_{1}}{\sqrt{k^{2}+m^{2}}}
\\&=&\frac{k_{stat}}{\sqrt{k_{stat}^{2}+m^{2}}}-\frac{k_{1}}{\sqrt{k^{2}+m^{2}}}
\\&\geq&\frac{k_{stat}\sqrt{k^{2}+m^{2}}-k_{1}\sqrt{k_{stat}^{2}+m^{2}}}{\sqrt{k^{2}+m^{2}}\sqrt{k_{stat}^{2}+m^{2}}}
\\&=&\frac{k_{stat}^{2}(k^{2}+m^{2})-k_{1}^{2}(k_{stat}^{2}+m^{2})}{\left(k_{stat}\sqrt{k^{2}+m^{2}}+k_{1}\sqrt{k_{stat}^{2}+m^{2}}\right)\sqrt{k_{1}^{2}+m^{2}}\sqrt{k_{stat}^{2}+m^{2}}}\;.
\end{eqnarray*}
Recalling $k\in[0;\frac{k_{stat}}{2}]$

\begin{eqnarray*}
\mid
\dot{g}(\mathbf{k})\mid&\geq&\frac{\frac{3}{4}k_{stat}^{2}m^{2}}{\left(k_{stat}\sqrt{k^{2}+m^{2}}+k_{1}\sqrt{k_{stat}^{2}+m^{2}}\right)\sqrt{k_{1}^{2}+m^{2}}\sqrt{k_{stat}^{2}+m^{2}}}
\\&=&\frac{3m^{2}}{4\left(\sqrt{k^{2}+m^{2}}+k_{1}\sqrt{1+(\frac{m}{k_{stat}})^{2}}\right)\sqrt{k_{1}^{2}+m^{2}}\sqrt{1+(\frac{m}{k_{stat}})^{2}}}\;.
\end{eqnarray*}
As $k_{stat}\geq\frac{m}{\sqrt{3}}$ (see \ref{ungleichungk}) it
follows:

$$\mid
\dot{g}(\mathbf{k})\mid\geq\frac{3m^{2}}{8\left(\sqrt{k^{2}+m^{2}}+2k_{1}\right)\sqrt{k_{1}^{2}+m^{2}}}=:h_{1}\;.
$$ For $k_{1}\geq 2k_{stat}$,  $g^{\prime}$ is positive. Therefore
similar as before:

\begin{eqnarray*}
\mid
g^{\prime}(\mathbf{k})\mid&=&\frac{k}{\sqrt{k^{2}+m^{2}}}+a-y\cos(\vartheta)
\\&\geq&\frac{k}{\sqrt{k^{2}+m^{2}}}-\frac{k_{stat}}{\sqrt{k_{stat}^{2}+m^{2}}}+\frac{k_{stat}}{\sqrt{k_{stat}^{2}+m^{2}}}+a-y
\\&=&\frac{k}{\sqrt{k^{2}+m^{2}}}-\frac{k_{stat}}{\sqrt{k_{stat}^{2}+m^{2}}}
=\frac{k\sqrt{k_{stat}^{2}+m^{2}}-k_{stat}\sqrt{k^{2}+m^{2}}}{\sqrt{k^{2}+m^{2}}\sqrt{k_{stat}^{2}+m^{2}}}
\\&=&\frac{k^{2}(k_{stat}^{2}+m^{2})-k_{stat}^{2}(k^{2}+m^{2})}{(k^{2}+m^{2})(k_{stat}^{2}+m^{2})}
\geq\frac{\frac{1}{4}k^{2}m^{2}}{(k^{2}+m^{2})^{2}}=:h_{2}(\mathbf{k})\;.
\end{eqnarray*}
Note, that $h_{1}$ and $h_{2}$ do not depend on $a$ and
$\mathbf{y}$.

We can use this estimate in (\ref{kurz}). As $g^{\prime\prime}$
and $\ddot{g}$ have uniform bounds in $a$ and $\mathbf{y}$ we get
uniform estimates for $I_{2}^{1}$ and $I_{2}^{2}$:

\begin{eqnarray*}
\parallel
I_{2}^{1}\parallel_{s}&\leq&\mu^{-2}\int_{k_{1}<\frac{k_{stat}}{2}}\parallel3\frac{\ddot{\chi_{2}}}{h_{1}^{2}}+3\frac{\chi_{2}\ddot{g}^{2}}{h_{1}^{4}}+3\frac{\dot{\chi_{2}}\ddot{g}}{h_{1}^{3}}\parallel_{s}
d^{3}k\\&\leq&\mu^{-2}\int_{\mathbb{R}^{3}}\parallel3\frac{\ddot{\widetilde{\chi}_{2}}}{h_{1}^{2}}+3\frac{\widetilde{\chi}_{2}\ddot{g}^{2}}{h_{1}^{4}}+3\frac{\dot{\widetilde{\chi}_{2}}\ddot{g}}{h_{1}^{3}}\parallel_{s}
d^{3}k\\ \parallel
I_{2}^{2}\parallel_{s}&\leq&\mu^{-2}\int_{k_{1}>2k_{stat}}\parallel3\frac{\chi_{2}^{\prime\prime}}{h_{2}^{2}}+3\frac{\chi_{2}g^{\prime\prime2}}{h_{2}^{4}}+3\frac{\chi_{2}^{\prime}g^{\prime\prime}}{h_{2}^{3}}\parallel_{s}
d^{3}k\\
&\leq&\mu^{-2}\int_{k_{1}\geq\frac{m}{\sqrt{3}}}\parallel3\frac{\widetilde{\chi}_{2}^{\prime\prime}}{h_{2}^{2}}+3\frac{\widetilde{\chi}_{2}g^{\prime\prime2}}{h_{2}^{4}}+3\frac{\widetilde{\chi}_{2}^{\prime}g^{\prime\prime}}{h_{2}^{3}}\parallel_{s}
d^{3}k\;.
\end{eqnarray*}
Hence

$$\parallel I_{2}^{1}\parallel_{s}+\parallel
I_{2}^{2}\parallel_{s}\leq\mu^{-2}C$$ with a constant C uniform in
$k_{stat}$.

The integrals $I_{2}^{3}$ and $I_{2}^{4}$ can be estimated in a
similar way, partial integration now be done with $k_{2}$ and
$k_{3}$

$$\mid
\partial_{k_{j}}g(\mathbf{k})\mid=\frac{1}{\sqrt{k^{2}+m^{2}}}+\frac{ak_{j}}{k^{2}}\leq\frac{1}{\sqrt{k^{2}+m^{2}}}+\frac{a}{k}\hspace{1cm}\text{
for } j=1;2 $$
which is uniformly bounded away from zero on the
area of integration.

So we have a uniform constant C with:

$$I_{2}\leq \mu^{-2}C\;,$$ and the lemma is proven for
$y\in[a+\frac{1}{2},a+1]$.

\bigskip

(II) Next we prove the Lemma for $y<a+1/2$.

We again have to assure, that all estimates are uniform in $a$ and
$\mathbf{y}$. In the last section the main difficulty we had to
solve was, that $g^{\prime}$ near the stationary point was
increasing with $k_{stat}$ (recall that $\lim_{y\rightarrow
a+1}k_{stat}=\infty$).

So on the first view it seems to be simple to have uniform
estimates for $y<a+1/2$ just by setting $Q=\mathbb{R}^{3}$. But we
have to face a new problem, which is, that the stationary point
may be very close to zero. This is problematical in the
differentiation of $k$ appearing in our estimates.

For $a=0$ this problem does not appear and the lemma is also
proven for $y<\frac{1}{2}$ with $a=0$.

As the divergence only appears for small $k_{stat}$ we can set
$k_{stat}<\frac{1}{2}$ (For $k_{stat}\geq\frac{1}{2}$ the
estimates can be done very closely to the ones of (I), setting
$Q=\mathbb{R}^{3}$).

We solve the problem by first "cutting out" the stationary point.
We split our integral:

$$I=\int_{B(0,\sqrt{k_{stat}})} e^{-i\mu
g(\mathbf{k})}\chi(\mathbf{k})d^{3}k+\int_{\R^{3}\setminus
B(0,\sqrt{k_{stat}})} e^{-i\mu
g(\mathbf{k})}\chi(\mathbf{k})d^{3}k=:I_{1}+I_{2}\;.$$ As
$k_{stat}<\frac{1}{2}$, the stationary point is inside the ball.

We estimate $I_{1}$, writing it in spherical coordinates
"centered" around the stationary point, by one partial
integration:

\begin{eqnarray*}
\parallel I_{1}\parallel_{s}&\leq&\parallel\int_{B(0,\sqrt{k_{stat}})}e^{-i\mu
g(\mathbf{k}^{\prime})}\chi(\mathbf{k}^{\prime})k^{\prime2}dk^{\prime}d\Omega\parallel_{s}
\\&\leq&\parallel\mu^{-1}\int_{B(0,\sqrt{k_{stat}})}(\frac{\chi^{\prime}k^{\prime2}}{g^{\prime}}+\frac{2\chi k^{\prime}}{g^{\prime}}+\frac{\chi
g^{\prime\prime}k^{\prime2}}{g^{\prime2}})dk^{\prime}d\Omega\parallel_{s}\;.
\end{eqnarray*}
As $\chi\in\mathcal{G}$ all these terms are bounded, we have

$$\parallel I_{1}\parallel_{s}\leq M\mu^{-1}\sqrt{k_{stat}}$$ We now estimate $I_{2}$.

The first idea is to estimate this integral by two partial
integrations. But the integrand still comes "very close" to the
stationary point, where $(g^{\prime})^{-1}$ is not bounded. So
this procedure will not yield uniform bound in $a$ and
$\mathbf{y}$.

The trick to get uniform bound is to redo the split (\ref{xx}),
(\ref{xxx}) of (I) into the integral for $a+\frac{1}{2}<y<a+1$.

$$I_{2}=I_{1}^{1}+I_{1}^{2}+\partial_{s}I(s)\mid_{s=\xi}$$ now
using $k=0$ as the center for our Taylor-expansion. So we get:

\begin{eqnarray*}
I_{1}^{1}&=&\int_{\R^{3}\setminus B(0,\sqrt{k_{stat}})}
e^{i(g^{\prime}(0)k+\frac{1}{2}g^{\prime\prime}(0)k^{2})}
\chi(0)d^{3}k
\\I_{1}^{2}&=&\int_{\R^{3}\setminus
B(0,\sqrt{k_{stat}})}e^{i(g^{\prime}(0)k+\frac{1}{2}g^{\prime\prime}(0)k^{2})}(\chi(\mathbf{k})-\chi(0))k^{2}dkd\Omega
\\\partial_{s}I(s)\mid_{s=\xi}&=&\int_{\R^{3}\setminus
B(0,\sqrt{k_{stat}})}\lambda
k^{3}\widetilde{f}(\mathbf{k})e^{g_{\xi}}(\mathbf{k})\chi(k)k^{2}dkd\Omega
\end{eqnarray*}
where

\begin{eqnarray}\label{newtaylor}
g^{\prime}(0)&=&\frac{k}{\sqrt{k^{2}+m^{2}}}+a-y\cos(\vartheta)\mid_{k=0}=a-y\cos(\vartheta)\nonumber\\
g^{\prime\prime}(0)&=&\partial_{k}^{2}g=\frac{m^{2}}{\sqrt{k^{2}+m^{2}}^{3}}\mid_{k=0}=\frac{1}{m}\nonumber\\
\widetilde{f}(\mathbf{k})&=&(g(\mathbf{k})-g(0)-g^{\prime}(0)k-\frac{1}{2}g^{\prime\prime}(0)k^{2})k^{-3}\nonumber\\
g_{\xi}(\mathbf{k})&=&g(0)+g^{\prime}(0)k+\frac{1}{2}g^{\prime\prime}(0)k^{2}+\xi\widetilde{f}(\mathbf{k})\;.
\end{eqnarray}
As by similar argument concerning (\ref{next}) $g_{\xi}$ has only
one stationary point $\widetilde{\mathbf{k}}_{stat}$. One can
easily see, that

$$g^{\prime}(0)+g^{\prime\prime}(0)k=a-y\cos(\vartheta)+\frac{k}{m}\geq
a-y\cos(\vartheta)+\frac{k}{\sqrt{k^{2}+m^{2}}}=g^{\prime}(\mathbf{k})\;.$$
Furthermore we have, that:

$$g^{\prime}_{\xi}(\mathbf{k})=(1-\xi)(g^{\prime}(0)+g^{\prime\prime}(0)k)+\xi
g^{\prime}(\mathbf{k})\;.$$ It follows, that

$$g^{\prime}(0)+g^{\prime\prime}(0)k\geq g^{\prime}_{\xi}\geq
g^{\prime}\;.$$ Therefore at
$\mathbf{k}=\widetilde{\mathbf{k}}_{stat}$ (where by definition
$g^{\prime}_{\xi}(\widetilde{\mathbf{k}}_{stat})=0$) the
$g^{\prime}$ has to be negative. It follows (recalling, that
$g^{\prime}$ increases monotonously on the $k_{1}$-axis), that

$$0\leq\widetilde{k}_{stat}\leq k_{stat}\;.$$ For the same reasons
we have the zero point $\overline{\mathbf{k}}_{stat}$ of
$g^{\prime}(0)+g^{\prime\prime}(0)k$ (i.e.
$\overline{\mathbf{k}}_{stat}=-\frac{g^{\prime}(0)}{g^{\prime\prime}(0)})$:

$$0\leq\overline{k}_{stat}\leq k_{stat}\;.$$ As the second
derivative of $g_{\xi}^{\prime\prime}(\widetilde{k}_{stat})$ is
not equal to zero, we can define a function $\widetilde{g}_{\xi}$
with:

\begin{equation}\label{gtilde}
0<M\leq\widetilde{g}_{\xi}:=\mid\mathbf{k}-\mathbf{\widetilde{k}}_{stat}\mid^{-1}g_{\xi}\;.
\end{equation}

The integral $I_{1}^{1}$ includes the leading term. It can be
estimated like (\ref{I11}). The other terms can be estimated again
by partial integrations. For that we define:

\begin{eqnarray*}
\zeta_{1}:=(\chi(\mathbf{k})-\chi(0))k^{2}=:\widetilde{\zeta}_{1}k^{3}
\hspace{1cm}\zeta_{2}:=\widetilde{f}(\mathbf{k})\chi(k)k^{5}=:\widetilde{\zeta}_{2}k^{5}
\end{eqnarray*}
where $\widetilde{\zeta}_{1,2}$ are bounded
$C^{\infty}$-functions.

We now make two partial integrations in $I_{1}^{2}$ and three
partial integrations in $\partial_{s}I(s)$ to get the estimates

\begin{eqnarray*}
\parallel I_{1}^{2}\parallel_{s}&\leq&\mu^{-2}\parallel\int_{\R^{3}\setminus
B(0,\sqrt{k_{stat}})}\partial_{k}\big(\frac{1}{g^{\prime}(0)+g^{\prime\prime}(0)k}\partial_{k}(\frac{\zeta_{1}}{g^{\prime}(0)+g^{\prime\prime}(0)k})\big)dkd\Omega\parallel_{s}
\\&=&\mu^{-2}\parallel\int_{\R^{3}\setminus
B(0,\sqrt{k_{stat}})}\partial_{k}\big(\frac{\zeta_{1}^{\prime}}{(g^{\prime}(0)+g^{\prime\prime}(0)k)^{2}}-\frac{\zeta_{1}g^{\prime\prime}(0)}{(g^{\prime}(0)+g^{\prime\prime}(0)k)^{3}}\big)dkd\Omega\parallel_{s}
\\\parallel\partial_{s}I(s)\mid_{s=\xi}\parallel_{s}&\leq&\mu^{-2}\parallel\int_{\R^{3}\setminus
B(0,\sqrt{k_{stat}})}\partial_{k}\big(\frac{1}{g_{\xi}^{\prime}}\partial_{k}(\frac{1}{g_{\xi}^{\prime}}\partial_{k}\frac{\zeta_{2}}{g_{\xi}^{\prime}})\big)dkd\Omega\parallel_{s}
\\&=&\mu^{-2}\parallel\int_{\R^{3}\setminus
B(0,\sqrt{k_{stat}})}\partial_{k}\big(\frac{1}{g_{\xi}^{\prime}}\partial_{k}(\frac{\zeta_{2}^{\prime}}{g_{\xi}^{\prime2}}-\frac{\zeta_{2}g_{\xi}^{\prime\prime}}{g_{\xi}^{\prime3}})\big)dkd\Omega\parallel_{s}
\\&=&\mu^{-2}\parallel\int_{\R^{3}\setminus
B(0,\sqrt{k_{stat}})}\partial_{k}\big(\frac{\zeta_{2}^{\prime\prime}}{g_{\xi}^{\prime3}}-3\frac{\zeta_{2}^{\prime}g_{\xi}^{\prime\prime}}{g_{\xi}^{\prime4}}-\frac{\zeta_{2}g_{\xi}^{\prime\prime\prime}}{g_{\xi}^{\prime3}}+3\frac{\zeta_{2}g_{\xi}^{\prime\prime2}}{g_{\xi}^{\prime5}}\big)dkd\Omega\parallel_{s}\;.
\end{eqnarray*}
So we can define functions $f_{j}$, $j=1;...;5$ which are bounded,
with:

\begin{eqnarray*}
\parallel I_{1}^{2}\parallel_{s}&\leq&\mu^{-2}\parallel\int_{\R^{3}\setminus
B(0,\sqrt{k_{stat}})}\partial_{k}(f_{1}q_{1}^{2}+f_{2}q_{1}^{3})dkd\Omega\parallel_{s}
\\\parallel\partial_{s}I(s)\mid_{s=\xi}\parallel_{s}&\leq&\mu^{-2}\parallel\int_{\R^{3}\setminus
B(0,\sqrt{k_{stat}})}\partial_{k}(f_{3}q_{2}^{3}+f_{4}q_{2}^{4}+f_{5}q_{2}^{5})dkd\Omega\parallel_{s}
\end{eqnarray*}
where

\begin{eqnarray*}
q_{1}:=\frac{k}{\mid\mathbf{k}-\overline{\mathbf{k}}_{stat}\mid}
\hspace{1cm}q_{2}:=\frac{k}{\mid\mathbf{k}-\mathbf{\widetilde{k}}_{stat}\mid}\;.
\end{eqnarray*}
So it is only left to show, that $\partial_{k}q_{1}$ and
$\partial_{k}q_{2}$ are bounded on $\R^{3}\setminus
B(0,\sqrt{k_{stat}})$. But this is easy:

\begin{eqnarray*}
\partial_{k}q_{1}=\partial_{k}\frac{1}{\sqrt{1-2\frac{k_{stat}\cos(\vartheta)}{k}+\frac{k_{stat}^{2}}{k^{2}}}}
=\frac{1}{\sqrt{1-2\frac{k_{stat}\cos(\vartheta)}{k}+\frac{k_{stat}^{2}}{k^{2}}}^{3}}
(\frac{k_{stat}^{2}}{k^{3}}-\frac{k_{stat}\cos(\vartheta)}{k^{2}})
\end{eqnarray*}
for $k\geq\sqrt{k_{stat}}$ this term has obviously uniform bound.

The derivative of $q_{2}$ can be estimated in the same way. We
only have to replace $\widetilde{k}_{stat}$ by
$\overline{k}_{stat}$.

\bigskip
(III) For $y>a+1$ we have no stationary point any more. So two
partial integrations are possible without any problem. We again
choose $k_{1}$ parallel to $\mathbf{y}$

\begin{eqnarray*}
\parallel
I_{2}\parallel_{s}&\leq&\mu^{-2}\int\parallel\partial_{k}(\frac{1}{g^{\prime}(\mathbf{k})}\partial_{k}\frac{\chi_{2}(\mathbf{k})}{g^{\prime}(\mathbf{k})})\parallel_{s}
d^{3}k=\mu^{-2}\int\parallel\partial_{k}(\frac{\chi^{\prime}}{g^{\prime2}}-\frac{\chi
g^{\prime\prime} }{g^{\prime3}})\parallel_{s} d^{3}k
\\&=&\mu^{-2}\int\parallel\frac{\chi^{\prime\prime}}{g^{\prime2}}-2\frac{\chi^{\prime}g^{\prime\prime}}{g^{\prime
3}}-\frac{\chi^{\prime} g^{\prime\prime}}{g^{\prime3}}-\frac{\chi
g^{\prime\prime\prime}}{g^{\prime3}}+3\frac{\chi
g^{\prime\prime2}}{g^{\prime4}}\parallel_{s} d^{3}k
\end{eqnarray*}
($f^{\prime}$ means $\partial_{k_{1}}f$).

This integral still depends on $k_{stat}$. To get an estimate
uniform in $k_{stat}$ we use:

 $$\mid
g^{\prime}(\mathbf{k})\mid=y-\frac{k_{1}}{\sqrt{k^{2}+m^{2}}}-a\frac{k_{1}}{k}\geq
1-\frac{k_{1}}{\sqrt{k^{2}+m^{2}}}=:h(\mathbf{k})\;.$$ It follows:

$$\parallel
I_{2}\parallel_{s}\leq\mu^{-2}\int\parallel\frac{\chi^{\prime\prime}}{h^{2}}+3\frac{\chi^{\prime}g^{\prime\prime}}{h^{3}}+3\frac{\chi
g^{\prime\prime2}}{h^{4}}+\frac{\chi
g^{\prime\prime\prime}}{h^{3}}\parallel_{s} d^{3}k=:\mu^{-2}C\;.$$

\subsection{Proof of
equation(\ref{ridofalpha})}\label{appendix2}

For each $\mathbf{k}$ we have two eigenstates for electrons. These
two eigenstates span the two dimensional spinor subspace for
electrons. In the standard representation of the Dirac matrices
these two spinors\footnote{The spinors here are not normalized!}
are:

\begin{eqnarray}\label{spinors}
  s^{1}_{\mathbf{k}}=\begin{pmatrix}
    _{\widehat{E}_{k}} \nonumber \\
    _{0} \nonumber \\
    _{k_{1}} \nonumber \\
    _{k^{+}}
  \end{pmatrix}\hspace{1cm}
  s^{2}_{\mathbf{k}}=\begin{pmatrix}
    _{0} \nonumber \\
    _{\widehat{E}_{k}} \nonumber \\
    _{k^{-}} \nonumber \\
    _{-k_{1}}
  \end{pmatrix}
\end{eqnarray}
where $$k^{\pm}=k_{2}\pm
 ik_{3}\hspace{1cm}\widehat{E}_{k}=E_{k}+m\hspace{1cm}E_{k}=\sqrt{k^{2}+m^{2}}\;.$$

If we now take any linear combination of these spinors
$s_{\mathbf{k}}=a_{\mathbf{k}}s^{1}(\mathbf{k})+b(\mathbf{k})s^{2}_{\mathbf{k}}$
and compute for example $\langle s_{\mathbf{k}}^{*},\alpha_{1}
s_{\mathbf{k}}\rangle$, we get (see(\ref{alphas})):

\begin{eqnarray*}
\langle s_{\mathbf{k}}^{*},\alpha_{1}s_{\mathbf{k}}\rangle &=&
\big\langle(a^{*}(\mathbf{k})s^{1*}_{\mathbf{k}}+b^{*}(\mathbf{k})s^{2*}_{\mathbf{k}}),\alpha_{1}(a(\mathbf{k})s^{1}_{\mathbf{k}}+b(\mathbf{k})s^{2}_{\mathbf{k}})\big\rangle
\\ &=& \big(a^{*}(\mathbf{k})
\begin{pmatrix}
    _{{\widehat{E}_{k}}} \\
    _{0} \\
    _{k_{1}} \\
    _{k^{-}}

   \end{pmatrix}
+b^{*}(\mathbf{k})
\begin{pmatrix}
    _{0} \\
   _{\widehat{E}_{k}}\\
    _{k^{+}} \\
    _{-k_{1}}
  \end{pmatrix}),
(a(\mathbf{k})
\begin{pmatrix}
    _{k_{1}} \\
    _{-k^{+}} \\
    _{\widehat{E}_{k}} \\
    _{0}
\end{pmatrix}
+b(\mathbf{k})
\begin{pmatrix}
    _{k^{-}} \\
    _{k_{1}} \\
    _{0} \\
    _{-\widehat{E}_{k}}
\end{pmatrix}\big)
\\ &=&
  \big(a^{2}(\mathbf{k})+b^{2}
  (\mathbf{k})\big)2\widehat{E}_{k}k_{1}\;.
\end{eqnarray*}
With the normalization factor

\begin{eqnarray*}
\langle
s_{\mathbf{k}}^{*},s_{\mathbf{k}}\rangle&=&(a^{2}(\mathbf{k})+b^{2}(\mathbf{k}))(\widehat{E}_{k}^{2}+k^{2})=(a^{2}(\mathbf{k})+b^{2}(\mathbf{k}))(E_{k}^{2}+2E_{k}m+m^{2}+
k^{2})\\
&=&(a^{2}(\mathbf{k})+b^{2}(\mathbf{k}))(2E_{k}(E_{k}+m))=(a^{2}(\mathbf{k})+b^{2}(\mathbf{k}))(2E_{k}\widehat{E}_{k})
\end{eqnarray*}
we get:

$$\langle s_{
\mathbf{k}}^{*},\alpha_{1}s_{\mathbf{k}}\rangle=\frac{k_{1}}{\sqrt{k^{2}+m^{2}}}\langle
s_{\mathbf{k}}^{*},s_{\mathbf{k}}\rangle\;.$$ Analogously we get:

$$\langle
s_{\mathbf{k}}^{*},\boldsymbol{\alpha}s_{\mathbf{k}}\rangle=\frac{\mathbf{k}}{\sqrt{k^{2}+m^{2}}}\langle
s_{\mathbf{k}}^{*},s_{\mathbf{k}}\rangle\;.$$

By linearity (\ref{ridofalpha}) follows.

\subsection{Proof of Lemma \ref{properties}}\label{appendix3}

$\mathbf{(a)}$

To begin with, we consider the integral

\begin{equation}\label{form}
I(\mathbf{x})=\int\frac{1}{\mid\mathbf{x}-\mathbf{x}^{\prime}\mid^{j}}f(\mathbf{x}^{\prime})d^{3}x^{\prime}
\end{equation}
for bounded, integrable $\parallel f\parallel_{s}$ and j=1;2.

For j=1 it has been proven by Ikebe \cite{ikebe}, that I is
H\"older continuous. We extend this to j=2. Therefore we need to
estimate:

$$I(\mathbf{x}+\mathbf{h})-I(\mathbf{x}-\mathbf{h})$$ for
arbitrary $\mathbf{h}$ with $h\leq\frac{1}{4}$ (We do not need to
focus on $h>\frac{1}{4}$, as $I(\mathbf{x})$ is bounded). We
split the integral into:

\begin{eqnarray}\label{formest}
\lefteqn{\hspace{-1cm}I(\mathbf{x}+\mathbf{h})-I(\mathbf{x}-\mathbf{h})=}\\&=&
\int_{B(\mathbf{x},\sqrt{h})}(\frac{1}{\mid\mathbf{x}+\mathbf{h}-\mathbf{x}^{\prime}\mid^{2}}-\frac{1}{\mid\mathbf{x}-\mathbf{h}-\mathbf{x}^{\prime}\mid^{2}})f(\mathbf{x}^{\prime})d^{3}x^{\prime}\nonumber\\&&+
\int_{B(\mathbf{x},1)\backslash
B(\mathbf{x},\sqrt{h})}(\frac{1}{\mid\mathbf{x}+\mathbf{h}-\mathbf{x}^{\prime}\mid^{2}}-\frac{1}{\mid\mathbf{x}-\mathbf{h}-\mathbf{x}^{\prime}\mid^{2}})f(\mathbf{x}^{\prime})d^{3}x^{\prime}\nonumber\\&&+
\int_{\mathbb{R}^{3}\backslash B(\mathbf{x},1)
}(\frac{1}{\mid\mathbf{x}+\mathbf{h}-\mathbf{x}^{\prime}\mid^{2}}-\frac{1}{\mid\mathbf{x}-\mathbf{h}-\mathbf{x}^{\prime}\mid^{2}})f(\mathbf{x}^{\prime})d^{3}x^{\prime}=:I_{1}+I_{2}+I_{3}\;.
\end{eqnarray}
For $I_{1}$ we have:

$$\parallel
I_{1}\parallel_{s}\leq2\sup_{x\in\mathbb{R}^{3}}\{\parallel
f(\mathbf{x})\parallel_{s}\}\int_{B(\mathbf{x},\sqrt{h})}\frac{1}{\mid\mathbf{x}-\mathbf{x}^{\prime}\mid^{2}}d^{3}x^{\prime}\;.$$
So we can find a constant $M<\infty$, so that

\begin{equation}\label{holder1/2a}
\parallel I_{1}(\mathbf{x},\mathbf{h})\parallel_{s}\leq M\sqrt{h}\hspace{1cm}\forall\mathbf{h}\in\mathbb{R}^{3}\;.
\end{equation}

For $I_{2}$ we have, using
$\mid\sqrt{h}-h\mid\leq\frac{1}{2}\sqrt{h}$:

\begin{eqnarray*}
\parallel I_{2}\parallel_{s}&=&\parallel\int_{B(\mathbf{x},1)\backslash B(0,\sqrt{h})
}(\frac{1}{\mid\mathbf{x}^{\prime}+\mathbf{h}\mid^{2}}-\frac{1}{\mid\mathbf{x}^{\prime}-\mathbf{h}\mid^{2}})f(\mathbf{x}-\mathbf{x}^{\prime})d^{3}x^{\prime}\parallel_{s}
\\&\leq&\sup_{x\in\mathbb{R}^{3}}\{\parallel
f(\mathbf{x})\parallel_{s}\}\int_{B(\mathbf{x},1)\backslash
B(0,\sqrt{h})}\frac{\mid
\mid\mathbf{x}^{\prime}-\mathbf{h}\mid^{2}-\mid\mathbf{x}^{\prime}+\mathbf{h}\mid^{2}\mid}{\mid\mathbf{x}^{\prime}+\mathbf{h}\mid^{2}\mid\mathbf{x}^{\prime}-\mathbf{h}\mid^{2}}d^{3}x^{\prime}
\\&\leq&\sup_{x\in\mathbb{R}^{3}}\{\parallel
f(\mathbf{x}\parallel_{s})\}\int_{B(\mathbf{x},1)\backslash
B(0,\sqrt{h})}\frac{4hx^{\prime}}{\mid\mathbf{x}^{\prime}+\mathbf{h}\mid^{2}\mid\mathbf{x}-\mathbf{h}\mid^{2}}d^{3}x^{\prime}
\\&\leq&\sup_{x\in\mathbb{R}^{3}}\{\parallel
f(\mathbf{x})\parallel_{s}\}\int_{B(\mathbf{x},1)\backslash
B(0,\sqrt{h})}\frac{8h}{x^{\prime3}}d^{3}x^{\prime}\;.
\end{eqnarray*}
So we can find a constant $M<\infty$, so that

\begin{equation}\label{holder1/2b}
\parallel I_{2}(\mathbf{x},\mathbf{h})\parallel_{s}\leq M\sqrt{h}\hspace{1cm}\forall\mathbf{h}\in\mathbb{R}^{3}\;.
\end{equation}

For $I_{3}$ we have, using similar reasoning as above:

\begin{eqnarray*}
\parallel I_{3}\parallel_{s}\leq\int_{\mathbb{R}^{3}\backslash B(\mathbf{0},1)
}\frac{8h}{x^{\prime3}}\parallel
f(\mathbf{x}-\mathbf{x}^{\prime})\parallel_{s} d^{3}x^{\prime}
\leq8h\int\parallel f(\mathbf{x}-\mathbf{x}^{\prime})\parallel_{s}
d^{3}x^{\prime}\;.
\end{eqnarray*}
Since $f$ is absolutely integrable, we can find a constant
$M<\infty$, so that

\begin{equation}\label{holder1/2c}
\parallel I_{3}(\mathbf{x},\mathbf{h})\parallel_{s}\leq Mh\hspace{1cm}\forall\mathbf{h}\in\mathbb{R}^{3}\;.
\end{equation}

We use this estimate on (\ref{LSE}), observing, that
$G_{\mathbf{k}}^{+}$ multiplied by
$A\hspace{-0.2cm}/\widetilde{\varphi}_{\mathbf{k}}^{s}$ is
essentially of the form of the integrals in (\ref{form}).Therfore:

\begin{equation}\label{holder1/2}
\parallel
\widetilde{\varphi}^{s}_{\mathbf{k}}(\mathbf{x}+\mathbf{h})-\widetilde{\varphi}^{s}_{\mathbf{k}}(\mathbf{x})\parallel_{s}\leq
M\sqrt{h}\hspace{1cm}\forall\mathbf{h}\in\mathbb{R}^{3}\;.
\end{equation}

Now we want to focus on integrals of the form (\ref{form}) for j=2
where $f(\mathbf{x})$ satisfies:

\begin{equation}\label{holder1/2f}
\parallel f(\mathbf{x}+\mathbf{h})-f(\mathbf{x})\parallel_{s}\leq M\sqrt{h}\;.
\end{equation}
We do a similar splitting as in (\ref{formest}). Now we have for
$I_{1}$, using (\ref{holder1/2f}):

\begin{eqnarray*}
\parallel I_{1}\parallel_{s}&\leq&\parallel\int_{B(\mathbf{x}
,\sqrt{h})}\frac{1}{\mid\mathbf{x}-\mathbf{x}^{\prime}\mid^{2}}(f(\mathbf{x}^{\prime}+\mathbf{h})-f(\mathbf{x}^{\prime}-\mathbf{h}))d^{3}x^{\prime}\parallel_{s}
\\&\leq&\mid\int_{B(\mathbf{x},\sqrt{h})}\frac{1}{\mid\mathbf{x}-\mathbf{x}^{\prime}\mid^{2}}M\sqrt{h}d^{3}x^{\prime}\mid\;.
\end{eqnarray*}
Thus with an appropriate $\widetilde{M}<\infty$:

\begin{equation}\label{holdera}
\parallel I_{1}(\mathbf{x},\mathbf{h})\parallel_{s}\leq Mh\hspace{1cm}\forall\mathbf{h}\in\mathbb{R}^{3}\;.
\end{equation}

For $I^{2}$ we have:

\begin{eqnarray*}
\parallel I_{2}^{2}\parallel_{s}&=&\parallel\int_{B(0,1)\backslash
B(0,\sqrt{h})}(\frac{1}{\mid\mathbf{x}^{\prime}+\mathbf{h}\mid^{2}}-\frac{1}{\mid\mathbf{x}^{\prime}-\mathbf{h}\mid^{2}})f(\mathbf{x}-\mathbf{x}^{\prime})d^{3}x^{\prime}\parallel_{s}\\&=&\parallel\int_{B(0,1)\backslash
B(0,\sqrt{h})}\frac{\mid\mathbf{x}^{\prime}-\mathbf{h}\mid^{2}-\mid\mathbf{x}^{\prime}+\mathbf{h}\mid^{2}}{\mid\mathbf{x}^{\prime}-\mathbf{h}\mid^{2}\mid\mathbf{x}^{\prime}+\mathbf{h}\mid^{2}}f(\mathbf{x}-\mathbf{x}^{\prime})d^{3}x^{\prime}\parallel_{s}\;.
\end{eqnarray*}
Since the fraction under this integral is point-symmetric to zero,
we can estimate the integral by:

\begin{eqnarray*}
\parallel I_{2}^{2}\parallel_{s}&\leq&\parallel\int_{B(0,1)\backslash
B(0,\sqrt{h})}\mid\frac{\mid\mathbf{x}^{\prime}-\mathbf{h}\mid^{2}-\mid\mathbf{x}^{\prime}+\mathbf{h}\mid^{2}}{\mid\mathbf{x}^{\prime}-\mathbf{h}\mid^{2}\mid\mathbf{x}^{\prime}+\mathbf{h}\mid^{2}}\mid
(f(\mathbf{x}-\mathbf{x}^{\prime})-f(\mathbf{x}+\mathbf{x}^{\prime}))d^{3}x^{\prime}\parallel_{s}
\\&\leq&\parallel\int_{B(0,1)\backslash
B(0,\sqrt{h})}\mid\frac{\mid\mathbf{x}^{\prime}-\mathbf{h}\mid^{2}-\mid\mathbf{x}^{\prime}+\mathbf{h}\mid^{2}}{\mid\mathbf{x}^{\prime}-\mathbf{h}\mid^{2}\mid\mathbf{x}^{\prime}+\mathbf{h}\mid^{2}}\mid
M\sqrt{2x^{\prime}}d^{3}x^{\prime}\parallel_{s}
\\&\leq&\parallel\int_{B(0,1)\backslash
B(0,\sqrt{h})}4\mid\frac{2h}{x^{\prime3}}\mid
M\sqrt{2x^{\prime}}d^{3}x^{\prime}\mid\leq\mid16\pi
M\sqrt{2}\int^{1}_{\sqrt{h}}\mid\frac{2h}{x^{\prime-\frac{1}{2}}}\mid
d^{3}x^{\prime}\parallel_{s}\;.
\end{eqnarray*}
So we can find a $\widetilde{M}<\infty$ with:

\begin{equation}\label{holderc}
\parallel I_{2}^{2}(\mathbf{x},\mathbf{h})\parallel_{s}\leq \widetilde{M}h\hspace{1cm}\forall\mathbf{h}\in\mathbb{R}^{3}\;.
\end{equation}

For $I_{3}$ we do the same estimations as before.

Applying this to (\ref{LSE}) we obtain the H\"older continuity of
degree 1 for $\widetilde{\varphi}_{\mathbf{k}}^{s}$.

\bigskip

$\mathbf{(b)}$

Assume that $\widetilde{\varphi}^{s}_{\mathbf{k}}(\mathbf{x})$
satisfies (\ref{LSE}) and is H\"older continuous of degree 1.
Inserting $\widetilde{\varphi}^{s}_{\mathbf{k}}(\mathbf{x})$ in
the right hand side of (\ref{dgmp}) leads to:

\begin{eqnarray*}
H\widetilde{\varphi}^{s}_{\mathbf{k}}(\mathbf{x})&=&(H_{0}+A\hspace{-0.2cm}/(\mathbf{x}))\left(\varphi_{\mathbf{k}}^{s}(\mathbf{x})-\int
 A\hspace{-0.2cm}/(\mathbf{x^{\prime}})G^{+}_{k}(\mathbf{x}-\mathbf{x^{\prime}})
 \widetilde{\varphi}_{\mathbf{k}}^{s}(\mathbf{x^{\prime}})d^{3}x^{\prime}\right)\\&=&(E_{k}+A\hspace{-0.2cm}/(\mathbf{x}))\varphi_{\mathbf{k}}^{s}(\mathbf{x})-(H_{0}+A\hspace{-0.2cm}/(\mathbf{x}))\int
 A\hspace{-0.2cm}/(\mathbf{x^{\prime}})G^{+}_{k}(\mathbf{x}-\mathbf{x^{\prime}})
 \widetilde{\varphi}_{\mathbf{k}}^{s}(\mathbf{x^{\prime}})d^{3}x^{\prime}\;.
\end{eqnarray*}
For (\ref{dgmp}) this term has to be equal to
$E_{k}\widetilde{\varphi}^{s}_{\mathbf{k}}(\mathbf{x})$. So we
have to prove, that

$$(H_{0}-E_{k})\int
 G^{+}_{k}(\mathbf{x}-\mathbf{x^{\prime}})A\hspace{-0.2cm}/(\mathbf{x^{\prime}})
 \widetilde{\varphi}_{\mathbf{k}}^{s}(\mathbf{x^{\prime}})d^{3}x^{\prime}=A\hspace{-0.2cm}/(\mathbf{x})\widetilde{\varphi}^{s}_{\mathbf{k}}(\mathbf{x})\;.$$
In other words we have to prove, that with $f$ H\"older continuous
of degree $1$:

\begin{equation}\label{povzner}
(H_{0}-E_{k})\int
 G^{+}_{k}(\mathbf{x}-\mathbf{x^{\prime}})f(\mathbf{x}^{\prime})d^{3}x^{\prime}=f(\mathbf{x}^{\prime})\;.
\end{equation}


$G^{+}_{k}$ can be written as \cite{thaller}:

$$G^{+}_{k}(\mathbf{x})=(H_{0}+E_{k})\frac{e^{ikx}}{4\pi
x}=:(H_{0}+E_{k})G_{k}^{KG}$$
with

\begin{equation}\label{GKG}
(H_{0}-E_{k})(H_{0}+E_{k})G_{k}^{KG}=(\triangle-k^{2})G_{k}^{KG}=\delta\;.
\end{equation}
So for (\ref{povzner}) we need to show, that:

\begin{equation}\label{hauff}
(H_{0}-E_{k})(H_{0}+E_{k})\int
 G^{KG}_{k}(\mathbf{x}-\mathbf{x^{\prime}})f(\mathbf{x}^{\prime})d^{3}x^{\prime}=
(\triangle-k^{2})\int
 G^{KG}_{k}(\mathbf{x}-\mathbf{x^{\prime}})f(\mathbf{x}^{\prime})d^{3}x^{\prime}=f(\mathbf{x})\;.
\end{equation}
We define for $\varepsilon>0$ the following function
$G^{\varepsilon}_{k}$:

\begin{eqnarray}\label{gepsilon}
G_{k}^{\varepsilon}(\mathbf{x}):=G^{KG}_{\mathbf{k}}(\mathbf{x})\text{
for } x\geq\varepsilon\hspace{1cm}
G^{\varepsilon}_{k}(\mathbf{x})=G^{KG}_{\mathbf{k}}(\mathbf{x})(1-e^{\frac{x}{\varepsilon-x}})\text{
for } x<\varepsilon\;.
\end{eqnarray}
We denote

\begin{equation}\label{gkprime}
 G_{k}^{\prime}(\mathbf{x})=\nabla
G^{KG}_{k}=\frac{ik\mathbf{x}e^{ikx}}{4\pi
x^{2}}+\frac{\mathbf{x}e^{ikx}}{x^{3}}\;.
\end{equation}
We split the right hand side of (\ref{hauff}) into:

\begin{equation}\label{haufsplit}
(\triangle-k^{2})\int
 (G^{KG}_{k}(\mathbf{x}-\mathbf{x^{\prime}})-G_{k}^{\varepsilon}(\mathbf{x}))f(\mathbf{x}^{\prime})d^{3}x^{\prime}+(\triangle-k^{2})\int
 G_{k}^{\varepsilon}(\mathbf{x})f(\mathbf{x}^{\prime})d^{3}x^{\prime}\;.
\end{equation}
By definition of $G_{k}^{KG}$ (\ref{GKG}) we have outside the Ball
$B(0,\varepsilon)$:

\begin{equation}\label{outball}
(\triangle-k^{2})G_{k}^{\varepsilon}(\mathbf{x})=(\triangle-k^{2})G_{k}^{KG}(\mathbf{x})=0\;.
\end{equation}
So for the first summand we have:

\begin{eqnarray*}
\lefteqn{\hspace{-1cm}\lim_{\varepsilon\rightarrow0}\parallel(\triangle-k^{2})\int
(G^{KG}_{k}(\mathbf{x}-\mathbf{x^{\prime}})-G_{k}^{\varepsilon}(\mathbf{x}-\mathbf{x^{\prime}}))f(\mathbf{x}^{\prime})d^{3}x^{\prime}\parallel_{s}}\\
&=&\lim_{\varepsilon\rightarrow0}\parallel\triangle\int_{B(\mathbf{x},\varepsilon)}
(G^{KG}_{k}(\mathbf{x}-\mathbf{x^{\prime}})-G_{k}^{\varepsilon}(\mathbf{x}-\mathbf{x^{\prime}}))f(\mathbf{x}^{\prime})d^{3}x^{\prime}\parallel_{s}\\
 &=&\lim_{\varepsilon\rightarrow0}\parallel\nabla\int_{B(\mathbf{x},\varepsilon)}
\nabla(G^{KG}_{k}(\mathbf{x}-\mathbf{x^{\prime}})-G_{k}^{\varepsilon}(\mathbf{x}-\mathbf{x^{\prime}}))f(\mathbf{x}^{\prime})d^{3}x^{\prime}\parallel_{s}\\
&=&\lim_{\varepsilon\rightarrow0}\parallel\nabla\int_{B(\mathbf{x},\varepsilon)}
\big(\nabla_{x^{\prime}}(G^{KG}_{k}(\mathbf{x}-\mathbf{x^{\prime}})-G_{k}^{\varepsilon}(\mathbf{x}-\mathbf{x^{\prime}}))\big)f(\mathbf{x}^{\prime})d^{3}x^{\prime}\parallel_{s}\\
&=&\lim_{\varepsilon\rightarrow0}\parallel\nabla\int_{B(\mathbf{x},\varepsilon)}
\big(\nabla_{x^{\prime}}(G^{KG}_{k}(\mathbf{x}^{\prime})-G_{k}^{\varepsilon}(\mathbf{x^{\prime}}))\big)f(\mathbf{x}-\mathbf{x}^{\prime})d^{3}x^{\prime}\parallel_{s}\\
&\leq&\lim_{\varepsilon\rightarrow0}\int_{B(\mathbf{x},\varepsilon)}\parallel
\big(\nabla_{x^{\prime}}(G^{KG}_{k}(\mathbf{x^{\prime}})-G_{k}^{\varepsilon}(\mathbf{x}^{\prime}))\big)\frac{f(\mathbf{x}-\mathbf{x^{\prime}})-f(\mathbf{x}+\mathbf{h}-\mathbf{x^{\prime}})}{h}\parallel_{s}
 d^{r}x^{\prime}\;.
\end{eqnarray*}
As f is H\"older continuous, the last term can be estimated by:

\begin{eqnarray*}
\lefteqn{\hspace{-1cm}\lim_{\varepsilon\rightarrow0}\parallel(\triangle-k^{2})\int
(G^{KG}_{k}(\mathbf{x}-\mathbf{x}^{\prime})-G_{k}^{\varepsilon}(\mathbf{x}-\mathbf{x}^{\prime}))f(\mathbf{x}^{\prime})d^{3}x^{\prime}\parallel_{s}}\\
&\leq&\lim_{\varepsilon\rightarrow0}\int_{B(0,\varepsilon)}\mid
\nabla_{x^{\prime}}(G^{KG}_{k}(\mathbf{x}^{\prime})-G_{k}^{\varepsilon}(\mathbf{x}^{\prime}))M\mid
 d^{3}x^{\prime}\\
 &\leq&\lim_{\varepsilon\rightarrow0}\int_{B(0,\varepsilon)}\mid
\left(G^{\prime}_{k}(\mathbf{x}^{\prime})-G^{\prime}_{k}(\mathbf{x}^{\prime})(1-e^{\frac{x^{2}}{\varepsilon-x}})-G^{\varepsilon}_{k}(\mathbf{x}^{\prime})\frac{-\varepsilon}{(x-\varepsilon)^{2}}\right)M\mid
 d^{3}x^{\prime}=0\;.
\end{eqnarray*}


For the second summand, we use (\ref{outball}) and the mean value theorem

\begin{eqnarray*}
 \lefteqn{\hspace{-1cm}\lim_{\varepsilon\rightarrow0}(\triangle-k^{2})\int
 G_{k}^{\varepsilon}(\mathbf{x}-\mathbf{x}^{\prime})f(\mathbf{x}^{\prime})d^{3}x^{\prime}}\\
&=&\lim_{\varepsilon\rightarrow0}(\triangle-k^{2})\int_{B(\mathbf{x},\varepsilon)}
 G_{k}^{\varepsilon}(\mathbf{x}-\mathbf{x}^{\prime})f(\mathbf{x}^{\prime})d^{3}x^{\prime}\\
&=&\lim_{\varepsilon\rightarrow0}\int_{B(\mathbf{x},\varepsilon)}
 (\triangle-k^{2}) G_{k}^{\varepsilon}(\mathbf{x}-\mathbf{x}^{\prime})f(\mathbf{x}^{\prime})d^{3}x^{\prime}\\
&=&\lim_{\varepsilon\rightarrow0}\int_{B(\mathbf{x},\varepsilon)}
 (\triangle-k^{2}) (e^{-ik\mid\mathbf{x}-\mathbf{x}^{\prime}\mid}G_{k}^{\varepsilon}(\mathbf{x}-\mathbf{x}^{\prime})e^{ik\mid\mathbf{x}-\mathbf{x}^{\prime}\mid})f(\mathbf{x}^{\prime})d^{3}x^{\prime}\\
&=&\lim_{\varepsilon\rightarrow0}\int_{B(\mathbf{x},\varepsilon)}
 (\triangle-k^{2}) (e^{-ik\mid\mathbf{x}-\mathbf{x}^{\prime}\mid}G_{k}^{\varepsilon}(\mathbf{x}-\mathbf{x}^{\prime})e^{ik\mid\mathbf{x}-\mathbf{x}^{\prime}\mid})f(\mathbf{x}^{\prime})d^{3}x^{\prime}\\
&=&\lim_{\varepsilon\rightarrow0}\int_{B(\mathbf{x},\varepsilon)}
 (\triangle+2ik\nabla) \big(e^{-ik\mid\mathbf{x}-\mathbf{x}^{\prime}\mid}G_{k}^{\varepsilon}(\mathbf{x}-\mathbf{x}^{\prime})\big)e^{ik\mid\mathbf{x}-\mathbf{x}^{\prime}\mid}f(\mathbf{x}^{\prime})d^{3}x^{\prime}\\
&=&\lim_{\varepsilon\rightarrow0}\int_{B(\mathbf{x},\varepsilon)}
 \triangle \big(e^{-ik\mid\mathbf{x}-\mathbf{x}^{\prime}\mid}G_{k}^{\varepsilon}(\mathbf{x}-\mathbf{x}^{\prime})\big)e^{ik\mid\mathbf{x}-\mathbf{x}^{\prime}\mid}f(\mathbf{x}^{\prime})d^{3}x^{\prime}\\
&=&\lim_{\varepsilon\rightarrow0}e^{ik\mid\mathbf{x}-\mathbf{x}_{\varepsilon}\mid}f(\mathbf{x}_{\varepsilon})\int_{B(\mathbf{x},\varepsilon)}
 \triangle \big(e^{-ik\mid\mathbf{x}-\mathbf{x}^{\prime}\mid}G_{k}^{\varepsilon}(\mathbf{x}-\mathbf{x}^{\prime})\big)d^{3}x^{\prime}\\
\end{eqnarray*}
where $\mathbf{x}_{\varepsilon}\in B(\mathbf{x},\varepsilon)$
using the positivity of

$$\triangle \big(e^{-ik\mid\mathbf{x}-\mathbf{x}^{\prime}\mid}G_{k}^{\varepsilon}(\mathbf{x}-\mathbf{x}^{\prime})\big)=2\frac{1-e^{\frac{x}{\varepsilon-x}}}{4\pi x^{3}}+2\frac{\varepsilon e^{\frac{x}{\varepsilon-x}}}{(x-\varepsilon)^{2}4\pi x^{2}}+\frac{(\varepsilon^{2}+2x\varepsilon)e^{\frac{x}{\varepsilon-x}})}{(x-\varepsilon)^{4}4\pi x}\geq0\;.$$
Hence with Gauss' theorem and (\ref{gkprime})

\begin{eqnarray*}
 \lefteqn{\hspace{-1cm}\lim_{\varepsilon\rightarrow0}(\triangle-k^{2})\int
 G_{k}^{\varepsilon}(\mathbf{x}-\mathbf{x}^{\prime})f(\mathbf{x}^{\prime})d^{3}x^{\prime}}\\
&=&f(\mathbf{x})\lim_{\varepsilon\rightarrow0}\int_{B(\mathbf{x},\varepsilon)}
 \triangle \big(e^{-ik\mid\mathbf{x}-\mathbf{x}^{\prime}\mid}G_{k}^{\varepsilon}(\mathbf{x}-\mathbf{x}^{\prime})\big)d^{3}x^{\prime}\\
&=&f(\mathbf{x})\lim_{\varepsilon\rightarrow0}\int_{\partial
B(\mathbf{x},\varepsilon)}
 \nabla \big(e^{-ik\mid\mathbf{x}-\mathbf{x}^{\prime}\mid}G_{k}^{\varepsilon}(\mathbf{x}-\mathbf{x}^{\prime})\big)\cdot\mathbf{n}d\Omega\\
&=&f(\mathbf{x})\lim_{\varepsilon\rightarrow0}\int_{\partial
B(\mathbf{x},\varepsilon)}
 \nabla \big( e^{-ik\mid\mathbf{x}-\mathbf{x}^{\prime}\mid}G_{k}^{KG}(\mathbf{x}-\mathbf{x}^{\prime})\big)\cdot\mathbf{n}\mid\mathbf{x}-\mathbf{x}^{\prime}\mid^{\prime2}d\Omega \\
&=&f(\mathbf{x})\lim_{\varepsilon\rightarrow0}\int_{\partial
B(\mathbf{x},\varepsilon)}
 \frac{\mathbf{x}-\mathbf{x}^{\prime}}{4\pi \mid\mathbf{x}-\mathbf{x}^{\prime}\mid^{3}}\cdot\mathbf{n}\mid\mathbf{x}-\mathbf{x}^{\prime}\mid^{2}d\Omega=f(\mathbf{x})
\end{eqnarray*}
and (b) is proved.


\bigskip

We show now, that for any $\mathbf{k}\in \mathbb{R}^{3}$ there
exists a unique solution
$\widetilde{\varphi}^{s}_{\mathbf{k}}(\mathbf{x})$ of (\ref{LSE}).

Using the definition of the $\zeta_{\mathbf{k}}^{s}(\mathbf{x})$
(see \ref{zeta}) in (\ref{LSE}) yields:

\begin{equation}\label{zetalse}
\zeta_{\mathbf{k}}^{s}(\mathbf{x})=v_{\mathbf{k}}(\mathbf{x})-\int
A\hspace{-0.2cm}/(\mathbf{x}^{\prime})G_{\mathbf{k}}^{+}(\mathbf{x}-\mathbf{x}^{\prime})\zeta_{\mathbf{k}}^{s}(\mathbf{x}^{\prime})d^{3}x^{\prime}
\end{equation}
where

\begin{equation}\label{vk}
v_{\mathbf{k}}(\mathbf{x}):=-\int
A\hspace{-0.2cm}/(\mathbf{x}^{\prime})G_{\mathbf{k}}^{+}(\mathbf{x}-\mathbf{x}^{\prime})\varphi_{\mathbf{k}}^{s}(\mathbf{x}^{\prime})d^{3}x^{\prime}\;.
\end{equation}
It suffices to prove, that (\ref{zetalse}) has a unique solution
for any $\mathbf{k}\in \mathbb{R}^{3}$. For the Schr\"odinger
Greens-function, this has already been proven by Ikebe
\cite{ikebe}. We want to proceed in the same way.

Let $\mathcal{B}$ be the Banach space of all continuous functions
tending uniformly to zero as $x\rightarrow\infty$. Due to
(\ref{decayofint}) $v(\mathbf{x})\in\mathcal{B}$. Ikebe uses the Riesz-Schauder theory of completely continuous operators
in a Banach space \cite{riesz}:

If T is a completely continuous operator in $\mathcal{B}$, then
for any given $g\in\mathcal{B}$ the equation

\begin{equation}\label{zetalse2}
f=g+Tf
\end{equation}
has a unique solution in $\mathcal{B}$ if
$\widetilde{f}=T\widetilde{f}$ implies that $\widetilde{f}=0$.

Defining the integral operator $T$ by:

$$Tf(\mathbf{x}):=-\int
A\hspace{-0.2cm}/(\mathbf{x}^{\prime})G_{\mathbf{k}}^{+}(\mathbf{x}-\mathbf{x}^{\prime})f(\mathbf{x}^{\prime})d^{3}x^{\prime}$$
and using $v$ for $g$, (\ref{zetalse2}) is equivalent to
(\ref{zetalse}). Note, that this operator is completely continuous
by the proof of Lemma \ref{properties}(a) following a similar
argumentation as in \cite{ikebe} Lemma 4.2. So it is left to show,
that the integral equation

\begin{equation}\label{ftilde}
\widetilde{f}(\mathbf{x})=-\int
A\hspace{-0.2cm}/(\mathbf{x}^{\prime})G_{\mathbf{k}}^{+}(\mathbf{x}-\mathbf{x}^{\prime})\widetilde{f}(\mathbf{x}^{\prime})d^{3}x^{\prime}
\end{equation}
has the unique solution $\widetilde{f}\equiv0$.








Obviously $\widetilde{f}\equiv0$ is a solution of (\ref{ftilde}).
By virtue of
(\ref{decayofint}) any solution of (\ref{ftilde}) has to be of order $x^{-1}$.
Furthermore $\widetilde{f}$ satisfies

\begin{equation}\label{kleingordon}
(-\Delta-k^{2}+A\hspace{-0.2cm}/)\widetilde{f}=0
\end{equation}
which can be shown by direct calculation.

Following Ikebe, $\widetilde{f}\equiv0$
is the only solution of (\ref{ftilde}).

\bigskip

$\mathbf{(c)}$

(c)i) follows directly from (\ref{decayofint}). For (c)ii) we need
to work more. We exemplarily prove (c)ii) for j=1,2.

Heuristically deriving (\ref{zetalse}) with respect to $k$ will
yield $\partial_{k}\zeta$. We denote the function we get by this
formal method by $\zeta^{\prime s}_{\mathbf{k}}$. Then

\begin{equation}\label{ableitung}
\zeta^{\prime
s}_{\mathbf{k}}(\mathbf{x})=\partial_{k}v_{\mathbf{k}}(\mathbf{x})-\int
A\hspace{-0.2cm}/(\mathbf{x}^{\prime})\partial_{k}G_{\mathbf{k}}^{+}(\mathbf{x}-\mathbf{x}^{\prime})\zeta_{\mathbf{k}}^{s}(\mathbf{x}^{\prime})d^{3}x^{\prime}-\int
A\hspace{-0.2cm}/(\mathbf{x}^{\prime})G_{\mathbf{k}}^{+}(\mathbf{x}-\mathbf{x}^{\prime})\zeta^{\prime
s}_{\mathbf{k}}(\mathbf{x}^{\prime})d^{3}x^{\prime}\;.
\end{equation}

We will now show, that this integral equation has a unique
solution. We define

\begin{eqnarray}\label{dkzeta}
p(\mathbf{x})&:=&\partial_{k}v_{\mathbf{k}}(\mathbf{x})+\int
A\hspace{-0.2cm}/(\mathbf{x}^{\prime})\partial_{k}G_{\mathbf{k}}^{+}(\mathbf{x}-\mathbf{x}^{\prime})\zeta_{\mathbf{k}}^{s}(\mathbf{x}^{\prime})d^{3}x^{\prime}\\
\overline{\zeta}_{\mathbf{k}}^{s}(\mathbf{x})&:=&\zeta^{\prime
s}_{\mathbf{k}}(\mathbf{x})-p(\mathbf{x})
\end{eqnarray}
so $\overline{\zeta}_{\mathbf{k}}^{s}$ satisfies:

\begin{equation*}
\overline{\zeta}_{\mathbf{k}}^{s}(\mathbf{x})=-\int
A\hspace{-0.2cm}/(\mathbf{x}^{\prime})G_{\mathbf{k}}^{+}(\mathbf{x}-\mathbf{x}^{\prime})p(\mathbf{x}^{\prime})d^{3}x^{\prime}-\int
A\hspace{-0.2cm}/(\mathbf{x}^{\prime})G_{\mathbf{k}}^{+}(\mathbf{x}-\mathbf{x}^{\prime})\overline{\zeta}^{\prime
s}_{\mathbf{k}}(\mathbf{x}^{\prime})d^{3}x^{\prime}\;.
\end{equation*}
Since

$$v^{\prime}(\mathbf{x}):=-\int
A\hspace{-0.2cm}/(\mathbf{x}^{\prime})G_{\mathbf{k}}^{+}(\mathbf{x}-\mathbf{x}^{\prime})v^{\prime}(\mathbf{x}^{\prime})d^{3}x^{\prime}\in\mathcal{B}$$
this integral equation again has a unique solution, so does
(\ref{dkzeta}).

 We will now
show, that $\zeta^{\prime}=\partial_{k}\zeta$.

We define the integral of $\zeta^{\prime}$:

\begin{equation}\label{intzeta}
\widetilde{\zeta}^{s}_{k,\vartheta,\varphi}(\mathbf{x}):=\zeta^{s}_{0}(\mathbf{x})+\int_{0}^{k}\zeta^{\prime
s}_{k^{\prime},\vartheta,\varphi}(\mathbf{x})dk\mathbf{\prime}\;.
\end{equation}
Obviously
$\partial_{k}\widetilde{\zeta}_{\mathbf{k}}^{s}=\zeta_{\mathbf{k}}^{s\prime}$
and $\widetilde{\zeta}^{s}_{0}=\zeta_{0}^{s}$. Using
(\ref{zetalse}) and (\ref{dkzeta}) in (\ref{intzeta}) leads to:

\begin{eqnarray*}
\widetilde{\zeta}_{\mathbf{k}}^{s}(\mathbf{x})&=&\zeta^{s}_{0}(\mathbf{x})+\int_{0}^{k}\zeta^{\prime
s}_{k^{\prime},\vartheta,\varphi}(\mathbf{x})dk^{\prime}
\\&=&v_{0}(\mathbf{x})-\int
A\hspace{-0.2cm}/(\mathbf{x}^{\prime})G_{0}^{+}(\mathbf{x}-\mathbf{x}^{\prime})\zeta_{0}^{s}(\mathbf{x}^{\prime})d^{3}x^{\prime}
+\int_{0}^{k}\partial_{k^{\prime}}v_{\mathbf{k}^{\prime}}(\mathbf{x})dk^{\prime}
\\&&-\int\int
A\hspace{-0.2cm}/(\mathbf{x}^{\prime})\partial_{k^{\prime}}G_{\mathbf{k}^{\prime}}^{+}(\mathbf{x}-\mathbf{x}^{\prime})\zeta_{\mathbf{k}^{\prime}}^{s}(\mathbf{x}^{\prime})d^{3}x^{\prime}+\int
A\hspace{-0.2cm}/(\mathbf{x}^{\prime})G_{\mathbf{k}^{\prime}}^{+}(\mathbf{x}-\mathbf{x}^{\prime})\zeta^{\prime
s}_{\mathbf{k}^{\prime}}(\mathbf{x}^{\prime})d^{3}x^{\prime}dk^{\prime}
\\&=&v_{\mathbf{k}}(\mathbf{x})-\int
A\hspace{-0.2cm}/(\mathbf{x}^{\prime})G_{0}^{+}(\mathbf{x}-\mathbf{x}^{\prime})\widetilde{\zeta}_{0}^{s}(\mathbf{x}^{\prime})d^{3}x^{\prime}
\\&&-\int\int
A\hspace{-0.2cm}/(\mathbf{x}^{\prime})\partial_{k^{\prime}}G_{\mathbf{k}^{\prime}}^{+}(\mathbf{x}-\mathbf{x}^{\prime})\widetilde{\zeta}_{\mathbf{k}^{\prime}}^{s}(\mathbf{x}^{\prime})d^{3}x^{\prime}+\int
A\hspace{-0.2cm}/(\mathbf{x}^{\prime})G_{\mathbf{k}^{\prime}}^{+}(\mathbf{x}-\mathbf{x}^{\prime})\zeta^{\prime
s}_{\mathbf{k}^{\prime}}(\mathbf{x}^{\prime})d^{3}x^{\prime}dk^{\prime}
\\&&-\int\int
A\hspace{-0.2cm}/(\mathbf{x}^{\prime})\partial_{k^{\prime}}\big(G_{\mathbf{k}^{\prime}}^{+}(\mathbf{x}-\mathbf{x}^{\prime})\widetilde{\zeta}_{\mathbf{k}^{\prime}}^{s}(\mathbf{x}^{\prime})\big)d^{3}x^{\prime}dk^{\prime}
\\&=&v_{\mathbf{k}}(\mathbf{x})-\int
A\hspace{-0.2cm}/(\mathbf{x}^{\prime})G_{\mathbf{k}}^{+}(\mathbf{x}-\mathbf{x}^{\prime})\widetilde{\zeta}_{\mathbf{k}}^{s}(\mathbf{x}^{\prime})d^{3}x^{\prime}\;.
\end{eqnarray*}
So $\widetilde{\zeta}_{\mathbf{k}}^{s}$ satisfies (\ref{zetalse}).
As the solution is unique, it follows, that
$\widetilde{\zeta}_{\mathbf{k}}^{s}=\zeta_{\mathbf{k}}^{s}$, hence

$$\partial_{k}\zeta_{\mathbf{k}}^{s}=\zeta_{\mathbf{k}}^{s\prime}\;.$$
By (\ref{decayofint}) $\widetilde{\zeta}_{\mathbf{k}}^{s}$ and
$p(\mathbf{x})$ have uniform bound, so

$$\sup_{\mathbf{x}\in\mathbb{R}^{3}}\parallel\partial_{k}\zeta^{s}_{\mathbf{k}}(\mathbf{x})\parallel_{s}<\infty\;.$$

\bigskip

For the second derivative we have:

$$\partial_{k}^{2}\frac{\zeta^{s}}{x+1}=\partial_{k}\frac{\overline{\zeta}^{s}}{x+1}+\partial_{k}\frac{p}{x+1}\;.$$
The proof of the existence and uniqueness of
$\partial_{k}\frac{\overline{\zeta}^{s}}{x}$ is the same as for
$\partial_{k}\zeta^{s}$, furthermore
$\partial_{k}\overline{\zeta}^{s}$ is bounded uniformly in
$\mathbf{x}$.

For $\partial_{k}\frac{p}{x+1}$ we have:

\begin{eqnarray}\label{partialp}
\parallel\partial_{k}\frac{p}{x+1}\parallel_{s}&=&\parallel\frac{1}{x+1}\partial_{k}\big(\partial_{k}v_{\mathbf{k}}(\mathbf{x})-\int
A\hspace{-0.2cm}/(\mathbf{x}^{\prime})\partial_{k}G_{\mathbf{k}}^{+}(\mathbf{x}-\mathbf{x}^{\prime})\zeta_{\mathbf{k}}^{s}(\mathbf{x}^{\prime})d^{3}x^{\prime}\big)\parallel_{s}\nonumber\\
&=&\parallel\frac{1}{x+1}\big(\partial_{k}^{2}v_{\mathbf{k}}(\mathbf{x})-\int
A\hspace{-0.2cm}/(\mathbf{x}^{\prime})\partial_{k}^{2}G_{\mathbf{k}}^{+}(\mathbf{x}-\mathbf{x}^{\prime})\zeta_{\mathbf{k}}^{s}(\mathbf{x}^{\prime})d^{3}x^{\prime}\nonumber\\
&&-\int
A\hspace{-0.2cm}/(\mathbf{x}^{\prime})\partial_{k}G_{\mathbf{k}}^{+}(\mathbf{x}-\mathbf{x}^{\prime})\partial_{k}\zeta_{\mathbf{k}}^{s}(\mathbf{x}^{\prime})d^{3}x^{\prime}\big)\parallel_{s}\;.
\end{eqnarray}
Note, that

\begin{eqnarray}\label{xxxx}
\mid\partial_{k}^{2}G_{\mathbf{k}}^{+}(\mathbf{x})\mid=\mid
x^{2}S_{\mathbf{k}}^{+}(\mathbf{x})+x\partial_{k}S_{\mathbf{k}}^{+}(\mathbf{x})+\partial_{k}^{2}S_{\mathbf{k}}^{+}(\mathbf{x})\mid\leq
M(xk+\frac{k}{x^{2}})\;.
\end{eqnarray}
Observing (\ref{xxxx}) and (\ref{kernel})
$\frac{\partial_{k}^{2}G_{\mathbf{k}}^{+}}{x+1}$ and
$\partial_{k}\zeta^{s}_{\mathbf{k}}$ are bounded uniformly in
$\mathbf{x}$, we have also, that

$$\frac{1}{x+1}\big(\partial_{k}^{2}v_{\mathbf{k}}(\mathbf{x})-\int
A\hspace{-0.2cm}/(\mathbf{x}^{\prime})\partial_{k}G_{\mathbf{k}}^{+}(\mathbf{x}-\mathbf{x}^{\prime})\partial_{k}\zeta_{\mathbf{k}}^{s}(\mathbf{x}^{\prime})d^{3}x^{\prime}\big)$$
is uniformly bounded in $\mathbf{x}$.

For the other summand we have get:

\begin{eqnarray*}
\lefteqn{\hspace{-1cm}\sup_{x,k\in\mathbb{R}^{3}}\parallel\frac{1}{x+1}\int
A\hspace{-0.2cm}/(\mathbf{x}^{\prime})\partial_{k}^{2}G_{\mathbf{k}}^{+}(\mathbf{x}-\mathbf{x}^{\prime})\zeta_{\mathbf{k}}^{s}(\mathbf{x}^{\prime})d^{3}x^{\prime}\parallel_{s}}\\
&\leq&\sup_{x\in\mathbb{R}^{3}}\parallel\frac{1}{x+1}\int
A\hspace{-0.2cm}/(\mathbf{x}^{\prime})(\mathbf{x}-\mathbf{x}^{\prime})\zeta_{\mathbf{k}}^{s}(\mathbf{x}^{\prime})d^{3}x^{\prime}\parallel_{s}\\
&\leq&\sup_{x\in\mathbb{R}^{3}}\parallel\int
A\hspace{-0.2cm}/(\mathbf{x}^{\prime})\frac{M(\mathbf{x}-\mathbf{x}^{\prime})}{(x+1)(x^{\prime}+1)}(x^{\prime}+1)\zeta_{\mathbf{k}}^{s}(\mathbf{x}^{\prime})d^{3}x^{\prime}\parallel_{s}\\
&\leq&\sup_{x\in\mathbb{R}^{3}}\parallel\int
A\hspace{-0.2cm}/(\mathbf{x}^{\prime})M(x^{\prime}+1)\zeta_{\mathbf{k}}^{s}(\mathbf{x}^{\prime})d^{3}x^{\prime}\parallel_{s}<\infty
\end{eqnarray*}
This proves (c)ii).

\bigskip

$\mathbf{(c)iii)}$

The proof of (c)iii) is very similar to the proof of (c)ii). The
only difference is, that we get new functions $p(\mathbf{x})$.

$$p(\mathbf{x})=k^{\mid\gamma\mid-1}D_{\mathbf{k}}^{\gamma}v_{\mathbf{k}}(\mathbf{x}))+\int
A\hspace{-0.2cm}/(\mathbf{x}^{\prime})k^{\mid\gamma\mid-1}D^{\gamma}_{\mathbf{k}}G_{\mathbf{k}}^{+}(\mathbf{x}-\mathbf{x}^{\prime})\zeta_{\mathbf{k}}^{s}(\mathbf{x}^{\prime})d^{3}x^{\prime}\;.$$

To have $p(\mathbf{x})$ in $\mathcal{B}$ one only has to assure,
that $k^{\mid\gamma\mid-1}D_{\mathbf{k}}^{\gamma}k$ is bounded for
$\mid\gamma\mid\leq2$, which follows by direct calculation.
\bigskip

$\mathbf{(d)}$

For potentials satisfying Condition A (\ref{potcond}) the
scattering system $(H,H_{0})$ is asymptotically complete (see
\cite{thaller}), i.e. for any scattering state $\psi$ there exists
a free outgoing asymptotic $\psi_{\text{out}}$ with:

\begin{equation}\label{asymp}
\lim_{t\rightarrow\infty}\parallel\psi(\mathbf{x},t)-\psi_{\text{out}}(\mathbf{x},t)\parallel=0\;.
\end{equation}
We write this, using the Fourier transform
$\widehat{\psi}_{\text{out}}^{s}$ of $\psi_{\text{out}}$:

$$\lim_{t\rightarrow\infty}\parallel\psi(\mathbf{x},t)-\sum_{s=1}^{2}\int(2\pi)^{-\frac{3}{2}}\widehat{\psi}_{out,s}(\mathbf{k})\varphi^{s}_{\mathbf{k}}(\mathbf{x},t)d^{3}k\parallel=0\;.$$
We shall show that

\begin{equation}\label{zerfall}
\lim_{t\rightarrow\infty}\parallel\int(2\pi)^{-\frac{3}{2}}\widehat{\psi}_{out,s}(\mathbf{k})\left(\widetilde{\varphi}^{s}_{\mathbf{k}}(\mathbf{x},t)-\varphi^{s}_{\mathbf{k}}(\mathbf{x},t)\right)d^{3}k\parallel=0\;.
\end{equation}
With that:

\begin{eqnarray*}
\lefteqn{\hspace{-1cm}\lim_{t\rightarrow\infty}\parallel\psi(\mathbf{x},t)-\sum_{s=1}^{2}\int(2\pi)^{-\frac{3}{2}}\widehat{\psi}_{out,s}(\mathbf{k})\widetilde{\varphi}^{s}_{\mathbf{k}}(\mathbf{x},t)d^{3}k\parallel}
\\&=&\lim_{t\rightarrow\infty}\parallel
e^{-iHt}(\psi(\mathbf{x})-\sum_{s=1}^{2}\int(2\pi)^{-\frac{3}{2}}\widehat{\psi}_{out,s}(\mathbf{k})\widetilde{\varphi}^{s}_{\mathbf{k}}(\mathbf{x})d^{3}k)\parallel
\\&=&\parallel
\psi(\mathbf{x})-\sum_{s=1}^{2}\int(2\pi)^{-\frac{3}{2}}
\widehat{\psi}_{out,s}(\mathbf{k})\widetilde{\varphi}^{s}_{\mathbf{k}}(\mathbf{x})d^{3}k\parallel=0\;.
\end{eqnarray*}
which establishes (\ref{hin}). For (\ref{zerfall}) we consider

\begin{eqnarray}\label{s2}
\lefteqn{\hspace{-1cm}\int(2\pi)^{-\frac{3}{2}}\widehat{\psi}_{out,s}(\mathbf{k})\left(\widetilde{\varphi}^{s}_{\mathbf{k}}(\mathbf{x},t)-\varphi^{s}_{\mathbf{k}}(\mathbf{x},t)\right)d^{3}k\nonumber}\\
&=&\int(2\pi)^{-\frac{3}{2}}e^{iE_{k}t}\widehat{\psi}_{out,s}(\mathbf{k})\zeta_{\mathbf{k}}^{s}(\mathbf{x})d^{3}k
\nonumber\\&=&\int(2\pi)^{-\frac{3}{2}}e^{iE_{k}t}\widehat{\psi}_{out,s}(\mathbf{k})v_{\mathbf{k}}^{s}(\mathbf{x})d^{3}k
\nonumber\\&&-\int(2\pi)^{-\frac{3}{2}}e^{iE_{k}t}\widehat{\psi}_{out,s}(\mathbf{k})\int
A\hspace{-0.2cm}/(\mathbf{x}^{\prime)}G_{\mathbf{k}}^{+}(\mathbf{x}-\mathbf{x}^{\prime})\zeta_{\mathbf{k}}^{s}(\mathbf{x}^{\prime})d^{3}x^{\prime}d^{3}k\nonumber\\&=:&\xi_{1}(\mathbf{x})+\xi_{2}(\mathbf{x})\;.
\end{eqnarray}
For the k-integration of $\xi_{1}$ we introduce (\ref{vk}) and
(\ref{kernel}) and then use Lemma \ref{statphas}, setting:

$$\mu=t ;\; a=t^{-1}\mid\mathbf{x}-\mathbf{x}^{\prime}\mid ;\;
\mathbf{y}=t^{-1}\mathbf{x}^{\prime} ;\; k^{\prime}=k;\;
\chi(\mathbf{k}^{\prime})=(2\pi)^{-\frac{3}{2}}\widehat{\psi}_{out,s}(\mathbf{k}^{\prime})\;.$$
Furthermore we recall, that:

\begin{eqnarray*}
\frac{k_{stat}}{\sqrt{k_{stat}^{2}+m^{2}}}+a-y&=&0\\
k_{stat}^{2}&=&(k_{stat}^{2}+m^{2})(y-a)^{2}\\ k_{stat}&=&
m(y-a)\sqrt{k_{stat}^{2}+m^{2}}=
m\frac{x^{\prime}}{t}\sqrt{k_{stat}^{2}+m^{2}}\;.
\end{eqnarray*}
For $\xi_{2}$ we set:

$$\mu=t ;\; a=t^{-1}\mid\mathbf{x}-\mathbf{x}^{\prime}\mid ;\;
\mathbf{y}=0 ;\; k^{\prime}=k;\;
\chi(\mathbf{k}^{\prime})=(2\pi)^{-\frac{3}{2}}\zeta_{\mathbf{k}}^{s}(\mathbf{k}^{\prime})\widehat{\psi}_{out,s}(\mathbf{k}^{\prime})\;.$$
Hence by (\ref{hoerm}) we obtain for (\ref{s2}) that there exists
$M<\infty$ uniform in $\mathbf{y}$ and $a$, such that:

\begin{eqnarray}\label{summesj}
\parallel\xi_{1}(\mathbf{x})+\xi_{2}(\mathbf{x})\parallel_{s}&\leq&
Mt^{-\frac{3}{2}}\mid\int
A\hspace{-0.2cm}/(\mathbf{x}^{\prime})G_{\mathbf{k}}^{+}(\mathbf{x}-\mathbf{x}^{\prime})(1+x^{\prime})d^{3}x^{\prime}\mid\nonumber\\&=:&Mt^{-\frac{3}{2}}G(\mathbf{x})\;.
\end{eqnarray}
The integral $G(\mathbf{x})$ is bounded and goes to zero in the
limit $x\rightarrow\infty$ (see \ref{decayofint}). This we shall
use in the following estimate. For (\ref{zerfall}) we need to
control

$$\lim_{t\rightarrow\infty}\parallel\xi_{1}+\xi_{2}\parallel=\lim_{t\rightarrow\infty}\big(\int\parallel\xi_{1}+\xi_{2}\parallel_{s}^{2}d^{3}x\big)^{\frac{1}{2}}\;.$$
We split this integral into three parts, which are time dependent
by introducing for all $\varepsilon>0$:

\begin{eqnarray*}
\rho_{\varepsilon}(\mathbf{x})&=&\mathbb{I}_{B(0,\varepsilon
t)}\,\,, \text{ the indicator function of the set }B(0,\varepsilon
t)
\\
\widetilde{\rho}_{\varepsilon}(\mathbf{x})&=&\mathbb{I}_{B(0,t)\backslash
B(0,\varepsilon t)}\\
\rho_{\text{out}}(\mathbf{x})&=&\mathbb{I}_{\mathbb{R}^{3}\backslash
B(0,t)}
\end{eqnarray*}
thus splitting our integral into:

\begin{eqnarray}\label{tripart}
\lim_{t\rightarrow\infty}\int\parallel\xi_{1}+\xi_{2}\parallel_{s}^{2}d^{3}x
&=&\lim_{t\rightarrow\infty}\int\rho_{\varepsilon}(\mathbf{x})\parallel\xi_{1}+\xi_{2}\parallel_{s}^{2}d^{3}x
\nonumber\\&&+\lim_{t\rightarrow\infty}\int\widetilde{\rho}_{\varepsilon}(\mathbf{x})\parallel\xi_{1}+\xi_{2}\parallel_{s}^{2}d^{3}x
\nonumber\\&&+\lim_{t\rightarrow\infty}\int\rho_{\text{out}}(\mathbf{x})\parallel\xi_{1}+\xi_{2}\parallel_{s}^{2}d^{3}x
\nonumber\\&=:&I_{1}+I_{2}+I_{3}\;.
\end{eqnarray}
The last part of this integral is the part, that lies outside the
light cone. For large times, all wavefunctions which are solutions
of the free or full Dirac equation will lie inside the light cone.
By virtue of (\ref{spacelike}):

\begin{eqnarray*}
\lim_{t\rightarrow\infty}\parallel\rho_{\text{out}}(\mathbf{x})\psi_{out,s}(\mathbf{x})\parallel&=&0
\\\lim_{t\rightarrow\infty}\parallel\rho_{\text{out}}(\mathbf{x})\int(2\pi)^{-\frac{3}{2}}\widehat{\psi}_{out,s}(\mathbf{k})\varphi_{\mathbf{k}}^{s}(\mathbf{x})d^{3}k\parallel&=&0
\end{eqnarray*}
or by (\ref{asymp}):

$$\lim_{t\rightarrow\infty}\parallel\rho_{\text{out}}(\mathbf{x})\int(2\pi)^{-\frac{3}{2}}\widehat{\psi}_{out,s}(\mathbf{k})\widetilde{\varphi}_{\mathbf{k}}^{s}(\mathbf{x})d^{3}k\parallel=0\;.$$
By (\ref{s2}) it follows, that:

$$I_{3}=\lim_{t\rightarrow\infty}\parallel\rho_{\text{out}}(\mathbf{x})\int\widehat{\psi}_{out,s}(\mathbf{k})\left(\widetilde{\varphi}_{\mathbf{k}}^{s}(\mathbf{x})-\varphi_{\mathbf{k}}^{s}(\mathbf{x})\right)d^{3}k\parallel=0\;.$$
Now we use (\ref{summesj}) on:

$$I_{1}\leq M^{2}\lim_{t\rightarrow\infty}(\sup_{x\leq\varepsilon
t}\{G(x)\})^{2}t^{-3}\frac{4\pi}{3}(\varepsilon
t)^{3}=C\varepsilon^{3}\;.$$ Since $\varepsilon$ is arbitrary,
$I_{1}=0$.

For $I_{2}$ we have:

\begin{eqnarray*}
I_{2}&=&\lim_{t\rightarrow\infty}\mid\int\widetilde{\rho}_{\varepsilon}(\mathbf{x})\parallel\xi_{1}+\xi_{2}\parallel_{s}^{2}d^{3}x\mid
\\&=&\lim_{t\rightarrow\infty}\mid M^{2}\int t^{-3}\widetilde{\rho}_{\varepsilon}(\mathbf{x})G(\mathbf{x})^{2}d^{3}x\mid
\\&\leq&\lim_{t\rightarrow\infty}\sup_{x\geq\varepsilon t}\mid
G(\mathbf{x})^{2}\mid=0
\end{eqnarray*}
and (\ref{hin}) is proved.

We first prove (\ref{her}) for wavefunctions, where
$\psi_{\text{out}}$ is in $L^{1}\cap L^{2}$. The general result
can then be obtained by density arguments.

Therefore we again use the unitarity of the time propagator:

\begin{eqnarray*}
\int(2\pi)^{-\frac{3}{2}}\langle\widetilde{\varphi}^{s}_{\mathbf{k}}(\mathbf{x}),\psi(\mathbf{x})\rangle
d^{3}x
&=&\lim_{t\rightarrow\infty}\int(2\pi)^{-\frac{3}{2}}e^{iHt}\langle\widetilde{\varphi}^{s}_{\mathbf{k}}(\mathbf{x}),e^{-iHt}\psi(\mathbf{x})\rangle
d^{3}x\\&=&\lim_{t\rightarrow\infty}e^{iEt}\int(2\pi)^{-\frac{3}{2}}\langle\widetilde{\varphi}^{s}_{\mathbf{k}}(\mathbf{x}),\psi(\mathbf{x},t)\rangle
d^{3}x\\
&=&\lim_{t\rightarrow\infty}e^{iEt}\int_{B(0,R)}(2\pi)^{-\frac{3}{2}}\langle\widetilde{\varphi}^{s}_{\mathbf{k}}(\mathbf{x}),\psi(\mathbf{x},t)\rangle
d^{3}x
\\&&+\lim_{t\rightarrow\infty}e^{iEt}\int_{\mathbb{R}^{3}\backslash
B(0,R)}(2\pi)^{-\frac{3}{2}}\langle\widetilde{\varphi}^{s}_{\mathbf{k}}(\mathbf{x}),\psi(\mathbf{x},t)\rangle
d^{3}x\;.
\end{eqnarray*}
By asymptotical completeness (\ref{asymp}) we obtain therefore

\begin{eqnarray*}
\int(2\pi)^{-\frac{3}{2}}\langle\widetilde{\varphi}^{s}_{\mathbf{k}}(\mathbf{x}),\psi(\mathbf{x})\rangle
d^{3}x
&=&\lim_{t\rightarrow\infty}e^{iEt}\int_{B(0,R)}(2\pi)^{-\frac{3}{2}}\langle\widetilde{\varphi}^{s}_{\mathbf{k}}(\mathbf{x}),\psi_{\text{out}}(\mathbf{x},t)\rangle
d^{3}x
\\&&+\lim_{t\rightarrow\infty}e^{iEt}\int_{\mathbb{R}^{3}\backslash
B(0,R)}(2\pi)^{-\frac{3}{2}}\langle\widetilde{\varphi}^{s}_{\mathbf{k}}(\mathbf{x}),\psi_{\text{out}}(\mathbf{x},t)\rangle
d^{3}x\;.
\end{eqnarray*}
By the free scattering into cones theorem, the first integral of
the right hand side goes to zero because any freely evolving
wavefunction leaves any bounded region. For the second integral we
write for all $R>0$:

\begin{eqnarray*}
\int(2\pi)^{-\frac{3}{2}}\langle\widetilde{\varphi}^{s}_{\mathbf{k}}(\mathbf{x}),\psi(\mathbf{x})\rangle
d^{3}x
&=&\lim_{R\rightarrow\infty}\lim_{t\rightarrow\infty}e^{iEt}\int_{\mathbb{R}^{3}\backslash
B(0,R)}(2\pi)^{-\frac{3}{2}}\langle\varphi^{s}_{\mathbf{k}}(\mathbf{x}),\psi_{\text{out}}(\mathbf{x},t)\rangle
d^{3}x
\\&&+\lim_{R\rightarrow\infty}\lim_{t\rightarrow\infty}e^{iEt}\int_{\mathbb{R}^{3}\backslash
B(0,R)}(2\pi)^{-\frac{3}{2}}\langle\zeta^{s}_{\mathbf{k}}(\mathbf{x}),\psi_{\text{out}}(\mathbf{x},t)\rangle
d^{3}x\;.
\end{eqnarray*}
Using Lemma \ref{properties}(c)i), the second integral on the
right hand side becomes:

\begin{eqnarray*}
\lefteqn{\hspace{-1cm}\mid\lim_{R\rightarrow\infty}\lim_{t\rightarrow\infty}e^{iEt}\int_{\mathbb{R}^{3}\backslash
B(0,R)}(2\pi)^{-\frac{3}{2}}\langle\zeta^{s}_{\mathbf{k}}(\mathbf{x}),\psi_{\text{out}}(\mathbf{x},t)\rangle
d^{3}x\mid}
\\&\leq&\lim_{R\rightarrow\infty}\frac{M}{R}\parallel\lim_{t\rightarrow\infty}e^{iEt}\int_{\mathbb{R}^{3}\backslash
B(0,R)}(2\pi)^{-\frac{3}{2}}\psi_{\text{out}}(\mathbf{x},t)d^{3}x\parallel
\\&=&\lim_{R\rightarrow\infty}\frac{M}{R}\parallel\int_{\mathbb{R}^{3}\backslash
B(0,R)}(2\pi)^{-\frac{3}{2}}\psi_{\text{out}}(\mathbf{x},0)d^{3}x\parallel=0\;.
\end{eqnarray*}
Therefore:

\begin{eqnarray*}
\lefteqn{\hspace{-1cm}\int(2\pi)^{-\frac{3}{2}}\langle\widetilde{\varphi}^{s}_{\mathbf{k}}(\mathbf{x}),\psi(\mathbf{x})\rangle
d^{3}x}\\
&=&\lim_{R\rightarrow\infty}\lim_{t\rightarrow\infty}e^{iEt}\int_{\mathbb{R}^{3}\backslash
B(0,R)}(2\pi)^{-\frac{3}{2}}\langle\varphi^{s}_{\mathbf{k}}(\mathbf{x}),\psi_{\text{out}}(\mathbf{x},t)\rangle
d^{3}x\\
&=&\lim_{R\rightarrow\infty}\lim_{t\rightarrow\infty}e^{iEt}\int_{\mathbb{R}^{3}\backslash
B(0,R)}(2\pi)^{-\frac{3}{2}}\langle\varphi^{s}_{\mathbf{k}}(\mathbf{x}),\psi_{\text{out}}(\mathbf{x},t)\rangle
d^{3}x\\
&=&\lim_{R\rightarrow\infty}\lim_{t\rightarrow\infty}e^{iEt}\int(2\pi)^{-\frac{3}{2}}\langle\varphi^{s}_{\mathbf{k}}(\mathbf{x}),\psi_{\text{out}}(\mathbf{x},t)\rangle
d^{3}x
\\
&=&\lim_{R\rightarrow\infty}\int(2\pi)^{-\frac{3}{2}}\langle\varphi^{s}_{\mathbf{k}}(\mathbf{x}),\psi_{\text{out}}(\mathbf{x},0)\rangle
d^{3}x =\widehat{\psi}_{out,s}(\mathbf{k})\;.
\end{eqnarray*}
and (\ref{her}) is proved.

\subsection{Proof of Lemma \ref{equiv}}\label{appendix4}



First we want to prove "$\Rightarrow$":

Let $\widehat{\psi}_{\text{out}}(\mathbf{k})\in\mathcal{G}$.
According to (\ref{hin}) we have for any $n\in\mathbb{N}_{0}$:

\begin{eqnarray*}
H^{n}\psi(\mathbf{x})&=&\sum_{s=1}^{2}\int(2\pi)^{-\frac{3}{2}}
  H^{n}\widetilde{\varphi}^{s}_{\mathbf{k}}(\mathbf{x})\widehat{\psi}_{out,s}(\mathbf{k})d^{3}k
\\&=&\sum_{s=1}^{2}\int(2\pi)^{-\frac{3}{2}}
  E_{k}^{n}\widetilde{\varphi}^{s}_{\mathbf{k}}(\mathbf{x})\widehat{\psi}_{out,s}(\mathbf{k})d^{3}k\;.
\end{eqnarray*}
Since $\widehat{\psi}_{\text{out}}(\mathbf{k})$ decays faster than
any polynom, this term is bounded and in
$L^{2}\bigotimes\mathbb{C}^{4}$ for all $n\in\mathbb{N}_{0}$. As
the potential $A\hspace{-0.2cm}/\in C^{\infty}$, also
$$(H-A\hspace{-0.2cm}/-\beta
m)^{n}\psi(\mathbf{x})=\nabla\hspace{-0.25cm}/^{n}\psi(\mathbf{x})$$
is bounded and in $L^{2}\bigotimes\mathbb{C}^{4}$ for all
$n\in\mathbb{N}_{0}$.

Furthermore we have, using (\ref{LSE}) in (\ref{hin}):

\begin{eqnarray*}
H^{n}\psi(\mathbf{x})&=&\sum_{s=1}^{2}\int(2\pi)^{-\frac{3}{2}}
\widetilde{\varphi}^{s}_{\mathbf{k}}(\mathbf{x})E_{k}^{n}\widehat{\psi}_{out,s}(\mathbf{k})d^{3}k
\\&=&\sum_{s=1}^{2}\int(2\pi)^{-\frac{3}{2}}
\varphi^{s}_{\mathbf{k}}(\mathbf{x})E_{k}^{n}\widehat{\psi}_{out,s}(\mathbf{k})d^{3}k
\\&&-\sum_{s=1}^{2}\int(2\pi)^{-\frac{3}{2}} \int
 A\hspace{-0.2cm}/(\mathbf{x^{\prime}})G^{+}_{k}(\mathbf{x}-\mathbf{x^{\prime}})
 \widetilde{\varphi}_{\mathbf{k}}^{s}(\mathbf{x^{\prime}})d^{3}x^{\prime}E_{k}^{n}\widehat{\psi}_{out,s}(\mathbf{k})d^{3}k=:I_{1}+I_{2}\;.
\end{eqnarray*}
$I_{1}$ is the Fourier transform of
$E_{k}^{n}\widehat{\psi}_{out,s}(\mathbf{k})$. As
$E_{k}^{n}\widehat{\psi}_{out,s}(\mathbf{k})\in\mathcal{G}$,
$I_{1}$ lies in $\widehat{G}$.

Next we write for $I_{2}$:

\begin{eqnarray*}
I_{2} &=&-\sum_{s=1}^{2}\int(2\pi)^{-\frac{3}{2}} \int
 A\hspace{-0.2cm}/(\mathbf{x^{\prime}})e^{ikx+ik(\mid\mathbf{x}-\mathbf{x^{\prime}}\mid-x)}\frac{S^{+}_{k}(\mathbf{x}-\mathbf{x^{\prime}})}{\mid\mathbf{x}-\mathbf{x^{\prime}}\mid}
 \widetilde{\varphi}_{\mathbf{k}}^{s}(\mathbf{x^{\prime}})d^{3}x^{\prime}E_{k}^{n}\widehat{\psi}_{out,s}(\mathbf{k})d^{3}kd\Omega
\\&=&-\sum_{s=1}^{2}\int\int_{0}^{\infty}(2\pi)^{-\frac{3}{2}} \int
 A\hspace{-0.2cm}/(\mathbf{x^{\prime}})e^{ikx}F(\mathbf{k},\mathbf{x},\mathbf{x}^{\prime})d^{3}x^{\prime}E_{k}^{n}\widehat{\psi}_{out,s}(\mathbf{k})k^{2}dkd\Omega
\end{eqnarray*}
where

\begin{equation}\label{deff}
F(\mathbf{k},\mathbf{x},\mathbf{x}^{\prime}):=e^{ik(\mid\mathbf{x}-\mathbf{x^{\prime}}\mid-x)}\frac{S^{+}_{k}(\mathbf{x}-\mathbf{x^{\prime}})}{\mid\mathbf{x}-\mathbf{x^{\prime}}\mid}
 \widetilde{\varphi}_{\mathbf{k}}^{s}(\mathbf{x^{\prime}})\;.
\end{equation}
We make now two partial integrations under the k-integral, which
is possible by Fubinis theorem:


\begin{eqnarray*}
I_{2}&=&-\sum_{s=1}^{2}\int(2\pi)^{-\frac{3}{2}}
\int_{0}^{\infty}\int
 A\hspace{-0.2cm}/(\mathbf{x^{\prime}})e^{ikx}F(\mathbf{k},\mathbf{x},\mathbf{x}^{\prime})d^{3}x^{\prime}E_{k}^{n}k^{2}\widehat{\psi}_{out,s}(\mathbf{k})dkd\Omega
\\&=&-\sum_{s=1}^{2}\frac{1}{x^{2}}\int(2\pi)^{-\frac{3}{2}} \int_{0}^{\infty}\int
 A\hspace{-0.2cm}/(\mathbf{x^{\prime}})e^{ikx}\partial_{k}^{2}\big(F(\mathbf{k},\mathbf{x},\mathbf{x}^{\prime})d^{3}x^{\prime}E_{k}^{n}k^{2}\widehat{\psi}_{out,s}(\mathbf{k})\big)dkd\Omega
\\&=&-\sum_{s=1}^{2}\frac{1}{x^{2}}\int\int_{0}^{\infty}\int(2\pi)^{-\frac{3}{2}}
 A\hspace{-0.2cm}/(\mathbf{x^{\prime}})e^{ikx}\partial_{k}^{2}F(\mathbf{k},\mathbf{x},\mathbf{x}^{\prime})E_{k}^{n}k^{2}\widehat{\psi}_{out,s}(\mathbf{k})\\&&\hspace{3cm}+\;2\partial_{k}F(\mathbf{k},\mathbf{x},\mathbf{x}^{\prime})\partial_{k}\big(E_{k}^{n}k^{2}\widehat{\psi}_{out,s}(\mathbf{k})\big)dkd\Omega d^{3}x^{\prime}
\\&&-\sum_{s=1}^{2}\frac{1}{x^{2}}\int_{0}^{\infty}\int(2\pi)^{-\frac{3}{2}} \int
 A\hspace{-0.2cm}/(\mathbf{x^{\prime}})e^{ikx}F(\mathbf{k},\mathbf{x},\mathbf{x}^{\prime})d^{3}x^{\prime}\partial_{k}^{2}\big(E_{k}^{n}k^{2}\widehat{\psi}_{out,s}(\mathbf{k})\big)dkd\Omega\\&=:&I_{3}+I_{4}\;.
\end{eqnarray*}
For $I_{4}$ we can write, using the definition of $F$ (\ref{deff})
and (\ref{LSE})

\begin{eqnarray*}
x^{2}I_{4}&=&\sum_{s=1}^{2}\int(2\pi)^{-\frac{3}{2}}
\widetilde{\varphi}^{s}_{\mathbf{k}}(\mathbf{x})\partial_{k}^{2}\big(E_{k}^{n}k^{2}\widehat{\psi}_{out,s}(\mathbf{k})\big)\frac{1}{k^{2}}d^{3}k\\&&-\sum_{s=1}^{2}\int(2\pi)^{-\frac{3}{2}}
\varphi^{s}_{\mathbf{k}}(\mathbf{x})\partial_{k}^{2}\big(E_{k}^{n}k^{2}\widehat{\psi}_{out,s}(\mathbf{k})\big)\frac{1}{k^{2}}d^{3}k\;.
\end{eqnarray*}
As $\widehat{\psi}_{\text{out}}\in\mathcal{G}$,
$\partial_{k}^{2}\big(E_{k}^{n}k^{2}\widehat{\psi}_{out,s}(\mathbf{k})\big)\frac{1}{k^{2}}$
lies in $L^{2}$ and so does $x^{2}\partial_{k}^{n}I_{4}$ for
$n\in\mathbb{N}_{0}$.

Under the k-integral in $I_{3}$ one more partial integration is
possible.

\begin{eqnarray*}
I_{3}=-\sum_{s=1}^{2}\frac{1}{x^{3}}\int(2\pi)^{-\frac{3}{2}} \int
 A\hspace{-0.2cm}/(\mathbf{x^{\prime}})\widetilde{F}(\mathbf{k},\mathbf{x},\mathbf{x}^{\prime})d^{3}x
\end{eqnarray*}
where
$$\widetilde{F}(\mathbf{k},\mathbf{x},\mathbf{x}^{\prime}):=\partial_{k}\big(\partial_{k}^{2}F(\mathbf{k},\mathbf{x},\mathbf{x}^{\prime})E_{k}^{n}k^{2}\widehat{\psi}_{out,s}(\mathbf{k})+2\partial_{k}F(\mathbf{k},\mathbf{x},\mathbf{x}^{\prime})\partial_{k}\big(E_{k}^{n}k^{2}\widehat{\psi}_{out,s}(\mathbf{k})\big)\big)\;.$$

Due to Lemma \ref{properties} (c)
$\partial_{k}^{n}\widetilde{\varphi}_{\mathbf{k}}(\mathbf{x}^{\prime)}\leq
Mx^{\prime}$. Furthermore we have, that

$$\mid\partial_{k}e^{ik(\mid\mathbf{x}-\mathbf{x}^{\prime}\mid-x)}\mid=\mid(\mid\mathbf{x}-\mathbf{x}^{\prime}\mid-x)e^{ik(\mid\mathbf{x}-\mathbf{x}^{\prime}\mid-x)}\mid\leq
x^{\prime}\mid
e^{ik(\mid\mathbf{x}-\mathbf{x}^{\prime}\mid-x)}\mid\;.$$ It
follows, that (remember the definition of $F$ (\ref{deff}))

$$\parallel \widetilde{F}(\mathbf{x},\mathbf{x})\parallel_{s}\leq
M_{2}\frac{x^{\prime3}}{\mid\mathbf{x}-\mathbf{x}^{\prime}\mid}\;.$$
So due to (\ref{decayofint}), with Condition B (\ref{potcond2}) on
the potential, the integral

$$\int
 A\hspace{-0.2cm}/(\mathbf{x^{\prime}})\widetilde{F}(\mathbf{k},\mathbf{x},\mathbf{x}^{\prime})d^{3}x^{\prime}$$
decays as fast as or faster than $x^{-1}$, so $x^{4}I_{3}$ is
bounded. It follows, that $x^{2}I_{3}$ lies in $L^{2}$ for
$n\in\mathbb{N}_{0}$.

The proof, that $x\partial_{x}^{n}\psi\in L^{2}$ is similar as
above, just with one partial integration less. It follows, that
$\psi\in\widehat{\mathcal{G}}$.


It is left to prove "$\Leftarrow$":

By Lemma \ref{properties}(b) it follows, that

\begin{eqnarray*}
E_{k}\widehat{\psi}_{out,s}(\mathbf{k})&=&H\widehat{\psi}_{out,s}(\mathbf{k})=\int(2\pi)^{-\frac{3}{2}}\langle\widetilde{\varphi}^{s}_{\mathbf{k}}(\mathbf{x}),H\psi(\mathbf{x})\rangle
d^{3}x
\\&=&\int(2\pi)^{-\frac{3}{2}}\langle\widetilde{\varphi}^{s}_{\mathbf{k}}(\mathbf{x}),(H_{0}+A\hspace{-0.2cm}/)\psi(\mathbf{x})\rangle
d^{3}x\;.
\end{eqnarray*}
For $\psi\in\mathcal{\widehat{G}}$, the right hand side is
integrable, so $E_{k}\widehat{\psi}_{out,s}(\mathbf{k})$ is
bounded. As $A\hspace{-0.2cm}/\in C^{\infty}$, this can be
repeated, so $E_{k}^{n}\widehat{\psi}_{out,s}(\mathbf{k})$ is
bounded for any $n\in\mathbb{N}$.

Since $E_{k}=\sqrt{k^{2}+m^{2}}\geq k$, it follows, that
$$k^{n}\widehat{\psi}_{out,s}(\mathbf{k})<\infty\;.$$
Equivalently we get:

\begin{eqnarray*}
E_{k}^{n}\partial_{k}^{j}\widehat{\psi}_{out,s}(\mathbf{k})&=&\int(2\pi)^{-\frac{3}{2}}\langle\partial_{k}^{j}\widetilde{\varphi}^{s}_{\mathbf{k}}(\mathbf{x}),H^{n}\psi(\mathbf{x})\rangle
d^{3}x\\
E_{k}^{n}k^{\mid\gamma\mid-1}D_{\mathbf{k}}^{\gamma}\widehat{\psi}_{out,s}(\mathbf{k})&=&\int(2\pi)^{-\frac{3}{2}}\langle
k^{\mid\gamma\mid-1}D_{\mathbf{k}}^{\gamma}\widetilde{\varphi}^{s}_{\mathbf{k}}(\mathbf{x}),H^{n}\psi(\mathbf{x})\rangle
d^{3}x
\end{eqnarray*}
With (c) of Lemma \ref{properties} it follows, that for
$\psi\in\widehat{\mathcal{G}}$ these terms are bounded for
$j=1,2$, $n\in\mathbb{N}_{0}$ and $\mid\gamma\mid\leq2$. So
$\widehat{\psi}_{out,s}(\mathbf{k})\in\mathcal{G}$.

\end{document}